\documentclass[twocolumn]{aastex63}

\newcommand{\IRAS}{\textit{IRAS}}
\newcommand{\ISO}{\textit{ISO}}
\newcommand{\AKARI}{\textit{AKARI}}
\newcommand{\Spitzer}{\textit{Spitzer}}
\newcommand{\JWST}{\textit{JWST}}
\newcommand{\WISE}{\textit{WISE}}
\newcommand{\Herschel}{\textit{Herschel}}
\newcommand{\Swift}{\textit{Swift}}
\newcommand{\unitpw}[2]{\ensuremath{\mathrm{#1}^{#2}}}
\newcommand{\per}[1]{\unitpw{#1}{-1}}
\newcommand{\persq}[1]{\unitpw{#1}{-2}}
\newcommand{\percb}[1]{\unitpw{#1}{-3}}
\newcommand{\kmps}{\ensuremath{\mathrm{km}\,\per{s}}}
\newcommand{\ergps}{\ensuremath{\mathrm{erg}\,\per{s}}}
\newcommand{\Vsys}{\ensuremath{V_\mathrm{sys}}}

\newcommand{\HH}{\ensuremath{\mathrm{H_2}}}
\newcommand{\nHH}{\ensuremath{n_\HH}}
\newcommand{\ncrit}{\ensuremath{n_\mathrm{crit}}}
\newcommand{\Tkin}{\ensuremath{T_\mathrm{kin}}}
\newcommand{\Tex}{\ensuremath{T_\mathrm{ex}}}
\newcommand{\Tb}{\ensuremath{T_\mathrm{b}}}
\newcommand{\Tbemi}{\ensuremath{T_\mathrm{b,emi}}}
\newcommand{\Tbabs}{\ensuremath{T_\mathrm{b,abs}}}
\newcommand{\Tdust}{\ensuremath{T_\mathrm{dust}}}
\newcommand{\TBG}{\ensuremath{T_\mathrm{BG}}}
\newcommand{\NHH}{\ensuremath{N_\HH}}
\newcommand{\NH}{\ensuremath{N_\mathrm{H}}}
\newcommand{\NCO}{\ensuremath{N_\mathrm{CO}}}
\newcommand{\NHHsubmm}{\ensuremath{N_\mathrm{\HH,\,submm}}}
\newcommand{\NCOsubmm}{\ensuremath{N_\mathrm{CO,\,submm}}}
\newcommand{\Nco}{\NCO} 
\newcommand{\Tco}{\ensuremath{T_\mathrm{CO}}}
\newcommand{\vturb}{\ensuremath{v_\mathrm{turb}}}
\newcommand{\NHHNIR}{\ensuremath{N_\mathrm{\HH,\,NIR}}}
\newcommand{\NCONIR}{\ensuremath{N_\mathrm{CO,\,NIR}}}
\newcommand{\Lsun}{\ensuremath{L_\sun}}
\newcommand{\LIR}{\ensuremath{L_\mathrm{IR}}}
\newcommand{\LAGN}{\ensuremath{L_\mathrm{AGN}}}
\newcommand{\LHX}{\ensuremath{L_\mathrm{2\text{--}10\,keV}}}
\newcommand{\LvHX}{\ensuremath{L_\mathrm{14\text{--}195\,keV}}}
\newcommand{\LHCNvib}{\ensuremath{L_\mathrm{HCN\text{-}vib}}}
\newcommand{\fAGN}{\ensuremath{f_\mathrm{AGN}}}
\newcommand{\fAGNMIR}{\ensuremath{f_\mathrm{AGN,\,MIR}}}
\newcommand{\fAGNNIRs}{\ensuremath{f_\mathrm{AGN,\,2.5\text{--}4\,\mu m}}}
\newcommand{\fAGNNIRl}{\ensuremath{f_\mathrm{AGN,\,4\text{--}5\,\mu m}}}
\newcommand{\EWPAH}{\ensuremath{\mathrm{EW_{3.3\,PAH}}}}
\newcommand{\tauice}{\ensuremath{\tau_{3.1}}}
\newcommand{\taudust}{\ensuremath{\tau_{3.4}}}
\newcommand{\tauCO}{\ensuremath{\tau_\mathrm{CO,NIR}}}
\newcommand{\AKARIcolor}{\ensuremath{f_{4.3}/f_{2.8}}}

\shorttitle{Nucleus of IRAS 17208$-$0014}
\shortauthors{Baba et al.}
\watermark{} 

\begin{document}

\title{Extremely Buried Nucleus of IRAS 17208$-$0014 Observed at Sub-Millimeter and Near-Infrared Wavelengths}

\correspondingauthor{Shunsuke Baba}
\email{shunsuke.baba@astrophysics.jp}

\author[0000-0002-9850-6290]{Shunsuke Baba}
\altaffiliation{JSPS Fellow (PD)}
\affiliation{National Astronomical Observatory of Japan (NAOJ),
2-21-1 Osawa, Mitaka, Tokyo 181-8588, Japan}
\affil{Graduate School of Science and Engineering, Kagoshima University, 1-21-35 Korimoto, Kagoshima, Kagoshima 890-0065, Japan}

\author[0000-0001-6186-8792]{Masatoshi Imanishi}
\affiliation{National Astronomical Observatory of Japan (NAOJ),
2-21-1 Osawa, Mitaka, Tokyo 181-8588, Japan}
\affiliation{Department of Astronomy, School of Science, The Graduate University for Advanced Studies (SOKENDAI),
2-21-1 Osawa, Mitaka, Tokyo 181-8588, Japan}

\author[0000-0001-9452-0813]{Takuma Izumi}
\altaffiliation{NAOJ Fellow}
\affiliation{National Astronomical Observatory of Japan (NAOJ),
2-21-1 Osawa, Mitaka, Tokyo 181-8588, Japan}
\affiliation{Department of Astronomy, School of Science, The Graduate University for Advanced Studies (SOKENDAI),
2-21-1 Osawa, Mitaka, Tokyo 181-8588, Japan}

\author[0000-0002-6808-2052]{Taiki Kawamuro}
\altaffiliation{FONDECYT fellow}
\affiliation{N\'{u}cleo de Astronom\'{i}a de la Facultad de Ingenier\'{i}a, Universidad Diego Portales,
Av. Ej\'{e}rcito Libertador 441, Santiago, Chile}

\author[0000-0002-5678-1008]{Dieu D. Nguyen}
\affiliation{Department of Physics, International University - Quarter 6, Linh Trung Ward, Thu Duc City, Ho Chi Minh City, Vietnam}
\affiliation{Vietnam National University - Quarter 6, Linh Trung Ward, Thu Duc City, Ho Chi Minh City, Vietnam}

\author[0000-0002-6660-9375]{Takao Nakagawa}
\affiliation{Institute of Space and Astronautical Science (ISAS),
Japan Aerospace Exploration Agency (JAXA),
3-1-1 Yoshinodai, Chuo-ku, Sagamihara, Kanagawa 252-5210, Japan}

\author{Naoki Isobe}
\affiliation{Institute of Space and Astronautical Science (ISAS),
Japan Aerospace Exploration Agency (JAXA),
3-1-1 Yoshinodai, Chuo-ku, Sagamihara, Kanagawa 252-5210, Japan}

\author{Shusuke Onishi}
\affiliation{Institute of Space and Astronautical Science (ISAS),
Japan Aerospace Exploration Agency (JAXA),
3-1-1 Yoshinodai, Chuo-ku, Sagamihara, Kanagawa 252-5210, Japan}
\affiliation{Department of Physics, Graduate School of Science,
The University of Tokyo,
7-3-1 Hongo, Bunkyo-ku, Tokyo 113-0033, Japan}

\author[0000-0002-5012-6707]{Kosei Matsumoto}
\affiliation{Institute of Space and Astronautical Science (ISAS),
Japan Aerospace Exploration Agency (JAXA),
3-1-1 Yoshinodai, Chuo-ku, Sagamihara, Kanagawa 252-5210, Japan}
\affiliation{Department of Physics, Graduate School of Science,
The University of Tokyo,
7-3-1 Hongo, Bunkyo-ku, Tokyo 113-0033, Japan}
\affiliation{Sterrenkundig Observatorium, Universiteit Gent, Krijgslaan 281, B-9000 Gent, Belgium}



\begin{abstract} 
The ultraluminous infrared galaxy IRAS 17208$-$0014 is a late-stage merger that hosts a buried active galactic nucleus (AGN).
To investigate its nuclear structure, we performed high spatial resolution ($\sim0\farcs04\sim32\,\mathrm{pc}$) Atacama Large Millimeter/submillimeter Array (ALMA) observations in Band 9 ($\sim$450\,\micron\ or $\sim$660\,GHz), along with near-infrared \AKARI\ spectroscopy in 2.5--5.0\,\micron.
The Band 9 dust continuum peaks at the AGN location, and toward this position CO($J$=6--5) and CS($J$=14--13) are detected in absorption.
Comparison with non-local thermal equilibrium calculations indicates that, within the central beam ($r\sim20\,\mathrm{pc}$), there exists a concentrated component that is dense ($10^7\,\percb{cm}$) and warm ($>$200\,K) and has a large column density ($\NHH>10^{23}\,\persq{cm}$).
The \AKARI\ spectrum shows deep and broad CO ro-vibrational absorption at 4.67\,\micron.
Its band profile is well reproduced with a similarly dense and large column but hotter ($\sim$1000\,K) gas.
The region observed through absorption in the near-infrared is highly likely in the nuclear direction, as in the sub-millimeter, but with a narrower beam including a region closer to the nucleus.
The central component is considered to possess a hot structure where vibrationally excited HCN emission originates.
The most plausible heating source for the gas is X-rays from the AGN.
The \AKARI\ spectrum does not show other AGN signs in 2.5--4\,\micron, but this absence may be usual for AGNs buried in a hot mid-infrared core.
Besides, based on our ALMA observations, we relate various nuclear structures of IRAS 17208$-$0014 that have been proposed in the literature.
\end{abstract}




\section{Introduction}
\label{sec:intro}

Active galactic nuclei (AGNs) are the central regions of galaxies that release radiative and kinetic energy, owing to mass accretion onto supermassive black holes (SMBHs).
Despite their compactness compared to the host galaxies, the intense activity in AGNs affects star formation in the host galaxies by heating or expelling the interstellar medium via radiation, winds, and jets \citep[see][for a review]{Fabian12}.
The questions, concerning how SMBHs are fueled, how AGNs occur, and what the characteristics of their surrounding environment are, are important for understanding galaxy evolution.

It has been proposed that major galaxy mergers in the local universe play a significant role in triggering AGN activity, particularly in the case of powerful AGNs.
Nearby AGNs are frequently observed in luminous ($\LIR=10^{11\text{--}12}\,\Lsun$) and ultraluminous ($\LIR>10^{12}\,\Lsun$) infrared (IR) galaxies \citep[(U)LIRGs;][]{Sanders&Mirabel96}.
Most of these AGNs show signatures of past or ongoing interactions with other galaxies, such as double nuclei, tidal tails, bridges, and disturbed morphologies, in optical and near-IR images \citep[e.g.,][]{Murphy+96,Duc+97,Scoville+00,Borne+00,Cui+01,Colina+01,Bushouse+02,Veilleux+02,Veilleux+06}.
Numerical simulations have shown that the gravitational torque generated during the mergers of disk galaxies funnels gas inward \citep{Barnes&Hernquist91,Barnes&Hernquist96} causing an intense starburst in nuclear regions \citep{Mihos&Hernquist96}, leading to a rapid feeding of the SMBHs, until the AGN feedback expels the gas from inner regions \citep{Di-Matteo+05}.
These processes that take place during a galaxy merger are hierarchically modeled as a unified evolutionary scenario \citep{Hopkins+06}.
This model has been simulated by including a vast range of merger properties in a cosmological framework, and has successfully reproduced the observed AGN luminosity functions, fractions, and clustering as a function of redshift and luminosity \citep{Hopkins+08}.
Consistent with theoretical predictions, a concentration of material along the merger sequence has been confirmed by a decrease in the bulge radius observed in the $H$-band \citep{Haan+11} as well as a decrease in the size of the CO(1--0) emission \citep{Yamashita+17}.
High dust temperatures (i.e., steep mid-IR slope) and deep silicate dust absorption at 9.7\,\micron\ in the late stages of a merger also imply a high mass concentration \citep{Stierwalt+13}.
The fraction of Compton-thick (CT; $\NH>10^{24}\,\persq{cm}$) AGNs in late-merger galaxies is higher than that in early-merger galaxies, peaking when the nuclei of merging galaxies are at a projected distance of $\lesssim$10\,kpc \citep{Ricci+17_fraction, Koss+18}.
In post-merger galaxies, the excess of X-ray selected AGNs over non-interacting control galaxies is poorly constrained because of heavy AGN obscuration.
However, there is an excess of mid-IR AGNs in the redshift range $z=0.02\text{--}0.16$, which is statistically significant \citep[i.e., a factor of $\sim$17 excess in post-mergers compared to non-mergers;][]{Secrest+20}.
A mid-IR AGN excess in mergers in the redshift range $z=0\text{--}0.6$ has also been confirmed by another study \citep{Gao+20}.
This study further showed that the merger fraction in mid-IR AGNs is higher than that in non-AGNs, and that it increases with the stellar mass and AGN power.
This suggests that mergers play a significant role in triggering AGNs, especially in the case of luminous AGNs \citep{Gao+20}.

The final stage of coalescence, during which the mass accretion rate of the SMBHs increases rapidly and the AGN feedback becomes more efficient, is a critical phase in the merger sequence.
However, observing the site of coalescence is challenging because the nucleus is deeply buried in dusty gas during this stage.
If the column density of the matter obscuring the nucleus is in the CT regime, then even the hard X-ray photons arising from the AGN can become nearly extinct.
Moreover, the commonly observed millimeter (mm) and sub-millimeter (sub-mm) molecular lines may become optically thick and can undergo self-absorption \citep[see e.g.,][]{Sakamoto+09,Rangwala+15,Scoville+17,Wheeler+20}.

To overcome this difficulty, HCN, which is a standard dense gas tracer, has been observed in the vibrationally excited state $v_2=1$ (HCN-vib) over the last decade \citep{Sakamoto+10,Imanishi&Nakanishi13,Aalto+15_Mrk231,Aalto+15_4ULIRGs,Imanishi+16_IRAS20551,Imanishi+16_14galaxies,Martin+16,Falstad+19,Aalto+19}.
\citet{Aalto+15_4ULIRGs} observed four (U)LIRGs that show intense activity in their compact nuclei.
They also found that the $J=4\text{--}3$ and 3--2 lines of both HCN and HCO$^+$ are severely self-absorbed; in contrast, the $J=4\text{--}3$ and 3--2 $l=1\mathrm{f}$ lines of HCN-vib are prominent and single-peaked.
The authors interpreted these observations to be caused by a hot mid-IR core having a brightness temperature high enough to vibrationally pump up HCN molecules ($\Tb(14\micron)>100$\,K) and a cooler gas envelope (extending up to tens of pc) surrounding the core, where the photons at the line centers of the ground-state HCN are absorbed.
\citet{Falstad+19} found that at least some of the galaxies that exhibit bright HCN-vib emission have fast and collimated outflows \citep[e.g., Arp 220;][]{Barcos-Munoz+18}.
Based on this, they proposed that the core from which the HCN-vib lines are emerging is passing through an intermediate stage before AGN feedback is fully developed.
Unfortunately, there are only a few studies that can directly resolve the structure of such cores \citep{Martin+16,Aalto+19}.
HCN-vib is indeed an excellent tracer of the nuclear region of CT AGNs because it is less susceptible to self-absorption.
However, it has some disadvantages such as it is less abundant, and thus, relatively faint; moreover, its doublet $l=1\mathrm{f}$ and 1e is affected by blending with HCO$^+$ and HCN $v=0$, respectively, as observed by \citet{Martin+16}.
Notably, the velocity difference between HCN-vib $l=1\mathrm{f}$ and HCO$^+$ $v=0$ is $\sim$400\,\kmps, and that between HCN-vib $l=1\mathrm{e}$ and HCN $v=0$ is $\sim$40\,\kmps, for all $J$ levels up to $J=9\text{--}8$.

By contrast, we propose to study the structure of a deeply buried gas cocoon embedded in a late-stage merger by observing it in CO($J$=6--5).\footnote{Hereafter, we usually write this line simply as CO(6--5) for readability.}
We aim to observe the CO(6--5) absorption against the dust continuum emission along the line of sight toward the nucleus, and the CO(6--5) emission along the line of sight away from the nucleus.
By inspecting the morphology of CO(6--5) and comparing its emission and absorption intensities with those of the underlying dust emission, we hope to infer the structure in the nuclear region of the merger.
The abundance of CO is more than four orders of magnitude than that of HCN in the X-ray-dominated region (XDR) \citep{Meijerink&Spaans05}, making it ideal for high-resolution observations.
Owing to its high excitation level, CO(6--5) is suitable for probing the moderately warm and dense gas in the central region rather than the extended gas in the host galaxy.
The upper and lower energy levels are $E_6=116$\,K and $E_5=83$\,K, respectively.
The reason for detecting  CO(6--5) in both absorption and emission is to obtain complementary information.

The target galaxy in our study is IRAS 17208$-$0014 (hereafter, referred to as IRAS 17208), which is one of the most luminous ULIRGs in the local universe \citep[$\LIR=10^{12.4}\,\Lsun$;][]{Sanders+03}.
It is a late-stage merger \citep{Haan+11} that has been suggested to harbor a CT AGN \citep{Gonzalez-Martin+09}.
IRAS 17208 exhibits bright HCN-vib emission \citep{Aalto+15_4ULIRGs}, thus establishing itself as an ideal candidate to study the highly obscured interiors of mergers (see Section \ref{sec:IRAS17208} for details).
In this work, we present the high-angular-resolution ($\sim0\farcs04\sim32$\,pc) CO(6--5) data of IRAS 17208 obtained with the Atacama Large Millimeter/submillimeter Array (ALMA).

Another advantage of observing a CO pure rotational transition in absorption is that it can be compared with the CO ro-vibrational absorption in the near-IR region ($v=1\leftarrow0$, $\Delta J=\pm1$, band center 4.67\,\micron, line separations $\sim$0.01\,\micron).
This band can probe the immediate vicinity of AGNs with an effectively high spatial resolution because the size of the background continuum source is compact, which is dominated by the dust sublimation region ($\Tdust\sim1500$\,K).
For instance, the interferometric observations of the nearby type-2 AGN NGC 1068 revealed a small sublimation radius of 0.24\,pc \citep{GRAVITYcollab20}.
In addition to the spatial resolution, dozens of $J$ level transitions can be observed simultaneously in this band, unlike in the case of pure rotational transitions, making it easy to gauge the physical state of the gas.

The near-IR CO band has been detected in several obscured AGNs \citep{Spoon+04,Spoon+05,Shirahata+13,Baba+18,Onishi+21}.
\citet{Shirahata+13} observed the obscured AGN IRAS 08572$+$3915 NW with the 8.2-m Subaru telescope, and detected 18 blueshifted ($-160$\,\kmps) lines of different rotational levels that were spectroscopically resolved.
The corresponding population diagram as a function of $J$ revealed that the absorption mainly originated from the gas in local thermal equilibrium (LTE) at a temperature of 270\,K with a total molecular column density $\NHH=3\times10^{22}\,\persq{cm}$.
The authors proposed that the high radial velocity, gas temperature, and column density observed are best explained by a molecular cloud moving near the AGN and heated by X-ray photons.
\citet{Baba+18} analyzed the CO absorption bands detected in 10 nearby ULIRGs, which were observed with the space telescopes \AKARI\ and \Spitzer.
Although the spectral resolution used in their study was not sufficient to resolve the rotational lines, they found high temperatures and column densities similar to those observed by \citet{Shirahata+13} based on band profile fitting.

Thus, previous observations of the near-IR CO absorption band suggest that it is useful for probing molecular gas near obscured AGNs.
In this paper, in addition to the ALMA sub-mm mapping, we present the \AKARI\ near-IR spectroscopic observations of IRAS 17208, and compare the data from these two telescopes to understand the structure of the nuclear region.
The \AKARI\ spectrum of IRAS 17208 has been previously published by \citet{Inami+18} as part of the spectral atlas of 145 galaxies.
However, in this work, we performed data reduction with special care to increase the signal-to-noise ratio (S/N) of the observations.
Improving the \AKARI\ spectrum is also useful for discussing the nature of the nucleus in IRAS 17208 using the near-IR AGN/starburst diagnostics \citep{Imanishi+08,Imanishi+10,Inami+18}.
Consequently, we also analyzed the 3.3\,\micron\ polycyclic aromatic hydrocarbon (PAH) emission, 3.1\,\micron\ H$_2$O ice absorption, 3.4\,\micron\ aliphatic carbon absorption, and the near-IR continuum color.

In this paper, we discuss the structure of the deeply buried nucleus of IRAS 17208, based on  observations with ALMA and \AKARI.
This paper is organized as follows.
Section \ref{sec:IRAS17208} summarizes the properties of IRAS 17208 reported in the literature.
Section \ref{sec:observations} describes the details of ALMA and \AKARI\ observations and the data reduction performed.
In Section \ref{sec:models}, the data from the two observatories are compared with theoretical models to investigate the properties of gas, and the corresponding findings are discussed comprehensively in Section \ref{sec:discussion}.
Finally, our conclusions are summarized in Section \ref{sec:summary}.

\section{IRAS 17208$-$0014}
\label{sec:IRAS17208}

IRAS 17208 (or IRAS F17207$-$0014) is the most luminous ULIRG ($\LIR=10^{12.4}\,\Lsun$) in the southern sky among the galaxies in the \IRAS\ Revised Bright Galaxy Sample \citep{Sanders+03}.
Its redshift is $z=0.0428$ \citep{Saunders+00}, corresponding to a luminosity distance of $D_\mathrm{L}=189$\,Mpc and an angular scale of $1\arcsec=844$\,pc.\footnote{A standard cosmology with $H_0=70\,\kmps\,\per{Mpc}$, $\Omega_\mathrm{M}=0.3$, and $\Omega_\Lambda=0.7$ is adopted.}
Owing to its negative declination, high luminosity, and moderate redshift, IRAS 17208 is one of the most suitable ULIRGs for high-spatial-resolution ALMA observations.

The optical and near-IR images of IRAS 17208 show two long tidal tails toward  the northwest and southeast, and a disturbed morphology in the central region \citep{Melnick&Mirabel90,Murphy+96,Duc+97,Scoville+00}.
\citet{Haan+11} and \citet{Stierwalt+13} classified this galaxy as a late-stage merger based on its optical and near-IR images, respectively.
\citet{Medling+14} conducted adaptive-optics assisted near-IR integral field spectroscopy to reveal that the nuclei of the two progenitor galaxies are separated by 0\farcs24 (or 200\,pc in projection), which are contained in overlapping stellar disks (with radii 200 and 410\,pc).

The activity of IRAS 17208 has been determined to be starburst-dominated based on its spectral properties in optical \citep[from line ratios;][]{Veilleux+95,Yuan+10}, near-IR \citep[from spectral slope, 3.3\,\micron\ PAH emission, and 3.4\,\micron\ dust absorption;][]{Risaliti+06}, mid-IR \citep[from 7.7\,\micron\ PAH emission or 5--8\,\micron\ spectral shape;][]{Lutz+99,Nardini+09,Nardini+10}, hard X-ray \citep[from luminosity, spectral index, and/or 6.4\,keV Fe K$\alpha$ line;][]{Franceschini+03,Iwasawa+11}, and radio \citep[from far-IR-to-radio ratio;][]{Momjian+03}.
However, numerous studies have claimed that the explanation is more complicated.
\citet{Duc+97} created three diagnostic diagrams of optical line ratios, and classified the galaxy as a low-ionization nuclear emission-line region (LINER) based on one diagram and as a starburst based on the other two.
\citet{Arribas&Colina03} found from optical integral field spectroscopy that the optical nucleus is not coincident with the dynamically determined true nucleus, owing to dust extinction; consequently, this changes the classification of the true nucleus from \ion{H}{2} to a LINER.
\citet{Stierwalt+13} and \citet{Inami+18} classified the galaxy to be a composite of starburst and AGN activity based on the equivalent width of the 6.2\,\micron\ PAH emission and a combination of the equivalent width of the 3.3\,\micron\ PAH emission and near-IR color, respectively.
\citet{Gonzalez-Martin+09} analyzed the X-ray data from both \textit{Chandra} and \textit{XMM-Newton}, and concluded that IRAS 17208 is a CT AGN candidate based on the spectral index, flux ratio of hard X-ray to [\ion{O}{3}], and equivalent width of Fe K$\alpha$.
This classification is in contrast with that of \citet{Franceschini+03} and \citet{Iwasawa+11} cited above, who used data from only one of the two aforementioned observatories.
\citet{Imanishi+06_HCN} found that the HCN(1--0) to HCO$^+$(1--0) ratio in the central 3\arcsec\ is in between the typical values observed in AGN- and starburst-dominated galaxies.

The contribution of a possibly hidden AGN in IRAS 17208 to the bolometric luminosity has been evaluated by various studies.
\citet{Gonzalez-Martin+09} estimated the intrinsic hard X-ray luminosity to be $\LHX\sim10^{43}\,\ergps$, which results in an AGN fraction of $\fAGN\sim0.1\%$.
However, this estimate could have a large systematic uncertainty owing to the Compton-thickness correction.
From IR diagnostics, much higher AGN fractions have been derived in the range of $\fAGN=5\text{--}21\%$ \citep{Veilleux+09,Veilleux+13,Rupke&Veilleux13,Gowardhan+18}.
By modeling the observed HCN-vib line, \citet{Aalto+15_4ULIRGs} calculated the AGN luminosity \LAGN\ to be $8\times10^{11}\,\Lsun$ ($\fAGN\sim30\%$).
This \LAGN\ value is supported by \citet{Garcia-Burillo+15} as a viable driver of a molecular outflow observed in the nucleus of IRAS 17208.
From mid-IR observations, comparably high AGN fractions have been obtained using spectral decomposition techniques:  31\% in 5--15\,\micron\ \citep{Alonso-Herrero+16} and 17\% in 4--20\,\micron\ \citep{Leja+18}.
Given these findings, it is highly likely that IRAS 17208 possesses a heavily obscured AGN that has a significant contribution to the bolometric luminosity of the galaxy.

The considerable amount of nuclear material in IRAS 17208 has been evidenced not only by the heavy visual extinction and X-ray obscuration of the central region, but also by the self-absorption of sub-mm lines.
\citet{Aalto+15_4ULIRGs} observed the IRAS 17208 nucleus with ALMA at a resolution of $\sim$0\farcs4, and detected deeply self-absorbed HCN(4--3) and HCO$^+$(4--3) lines.
We expect that the CO(6--5) line can be detected in absorption if observed at an angular resolution that is an order of magnitude higher.\footnote{The critical density of CO(6--5) is $\sim3\times10^4\,\percb{cm}$ \citep{Yang+10}, and those of HCN(4--3) and HCO$^+$(4--3) are $\sim1\times10^7\,\percb{cm}$ and $\sim2\times10^6\,\percb{cm}$, respectively \citep{Izumi+16}.}

Note that the \AKARI\ spectrum of IRAS 17208 presented by \citet{Inami+18} was noisy, as it was damaged by hot pixels.
Consequently, the presence of CO ro-vibrational absorption was unclear in the spectrum.
However, the ULIRG UGC 5101, which shows HCN-vib emission similar to IRAS 17208 \citep{Falstad+19}, exhibits deep CO absorption \citep{Baba+18}.
Therefore, we expect to detect significant CO absorption in IRAS 17208 as well, if data reduction is performed carefully.

\section{Observations and Results}
\label{sec:observations}

\subsection{ALMA}

\subsubsection{Observations and Data Reduction}

The nucleus of IRAS 17208 was observed with ALMA Band 9 on 2017 Aug 9 as part of our Cycle 4 program (2016.1.01223.S; PI: S. Baba). 
In total, 45 antennas were used, and the projected baselines spanned from 19.6\,m to 3.36\,km.
The maximum recoverable scale (MRS) was $\sim$0\farcs46,\footnote{Evaluated by the ALMA Cycle 4 approved formula: $0.983\times\lambda/L5$, where $\lambda$ is the wavelength and $L5$ is the fifth percentile baseline length (199\,m).} and the half power beam width of the primary beam was $\sim$9\arcsec.
Of the four spectral windows (each 1.875\,GHz wide), one was centered on the CO($J$=6--5) line, another was adjusted to include CS($J$=14--13), which might be serendipitously detected, and the rest were used to image the continuum emission.
The total on-source time was 26.7\,min.
The bandpass, phase, and flux calibrators were J1517$-$2422, J1743$-$0350, and J1751$+$0939, respectively.
The accuracy of the absolute flux calibration for Band 9 is expected to be at the 20\% level according to the ALMA Cycle 4 Proposer's Guide.

The data reduction and subsequent analysis were conducted with CASA version 4.7.2 \citep{CASA}.
The bandpass, phase, and absolute flux were calibrated by executing the scripts provided by ALMA along with the raw data.
Channels with too high system temperatures due to the presence of atmospheric lines were flagged by the script and excluded from further analysis.
One example of these atmospheric lines is at a sky frequency of 658.035\,GHz.
We found that several channels adjacent to the ones flagged at this frequency also had fairly high system temperatures, and hence we manually added flags to them using the \texttt{flagdata} task in CASA.

The continuum flux level was determined from the line-free region (observed frequency 658.2--661.7\,GHz) and subtracted from the visibility using the \texttt{uvcontsub} task assuming a constant for the fit.
Continuum-subtracted and unsubtracted image cubes were reconstructed from all the spectral windows by applying the \texttt{tclean} task.
Briggs weighting with a robustness parameter of 0.5 was adopted.
The pixel scale was set to 5\,mas.
To improve the S/N, the original spectral channels were binned into a common optical velocity width of 29.35\,\kmps.
The resulting rms noise of the data cube was $\sim$4.0\,mJy\,\per{beam}.
The full width at half maximum (FWHM) of the obtained synthesized beam was $45\times31\,\unitpw{mas}{2}$ or $38\times26\,\unitpw{pc}{2}$.

\subsubsection{Spectra}

\begin{figure}[t]
\plotone{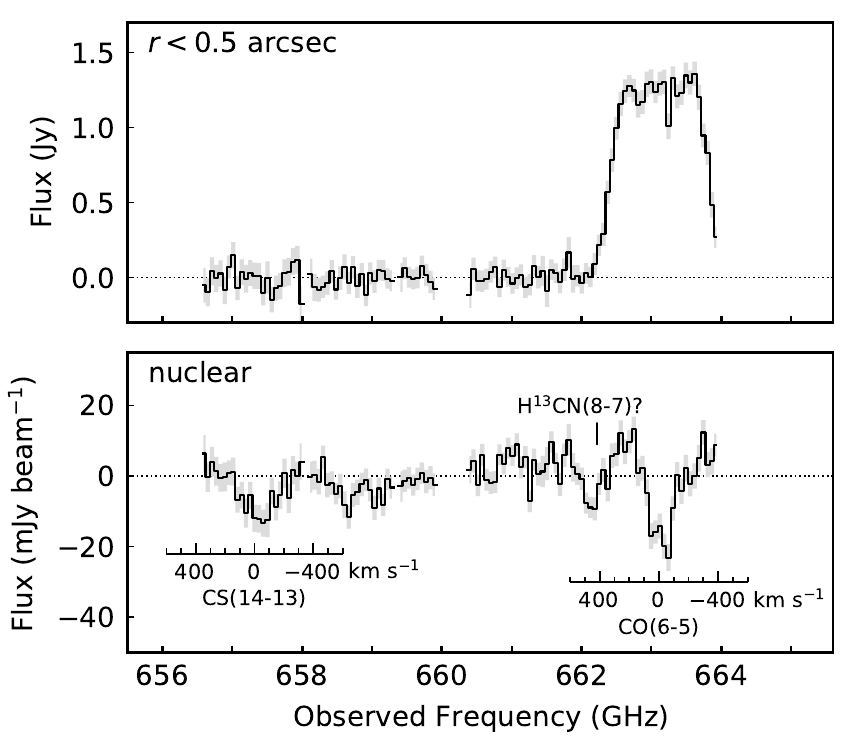}
\caption{
ALMA Band 9 line spectra of IRAS 17208 with different spatial scales.
Top panel: area-integrated spectrum within 0\farcs5 from the 436\,\micron\ continuum peak.
The most blueshifted ($<-390$\,\kmps) part of the emission is outside the observed range.
The gray shaded area represents statistical uncertainties ($\sim$0.08\,Jy).
The blank around 660\,GHz corresponds to a gap between spectral windows, and the discontinuities at 658.1 and 659.3\,GHz correspond to regions flagged for severe atmospheric absorption.
Bottom panel: Beam-sized ($\sim$0\farcs04) spectrum extracted at the nucleus.
The gray shaded area represents statistical uncertainties ($\sim4\,\mathrm{mJy}\,\per{beam}\sim8\,\mathrm{K}$).
The velocity scale with respect to \Vsys\ is drawn for CO(6--5) and CS(14--13).
The position of H$^{13}$CN(8--7) at \Vsys\ is indicated with a short vertical bar.
\label{fig:submm-spec}}
\end{figure}

Band 9 line spectra of different spatial scales were created from the cleaned data cube.
The upper panel of Figure \ref{fig:submm-spec} shows a spectrum spatially integrated within 0\farcs5 from the nucleus (continuum peak, Section \ref{sec:continuum}).
The CO(6--5) emission is clearly observed.
Unfortunately, however, its most blueshifted ($<-390$\,\kmps\ relative to the systemic velocity $\Vsys(\mathrm{LSR, optical})=12824\pm15\,\kmps$; \citealt{Garcia-Burillo+15}\footnote{Note that the beam size of the observation by \citet{Garcia-Burillo+15} was 0\farcs5, so the \Vsys\ adopted here is a composite of the two nuclei of IRAS 17208.}) part did not fit into a single spectral window and was missed.
Other emission lines are not detected with this angular size.

The lower panel of the same figure presents a beam-sized spectrum extracted at the nucleus.
It is clearly shown that CO(6--5) is detected as continuum absorption as expected.
The absorption depth relative to the continuum level reaches $-18$\,mJy\,\per{beam}, and the FWHM is $\sim$200\,\kmps.
In addition, CS(14--13) is detected in absorption as well.
These two lines could be slightly blueshifted ($\sim-50$\,\kmps) with respect to \Vsys\ (Section \ref{sec:disks}).
Another absorption line at $\nu_\mathrm{obs}=662.2\,$GHz could be H$^{13}$CN(8--7), but we do not investigate it further in this paper because the identification is somewhat questionable.

\subsubsection{Dust Continuum}
\label{sec:continuum}

\begin{figure}[t]
\plotone{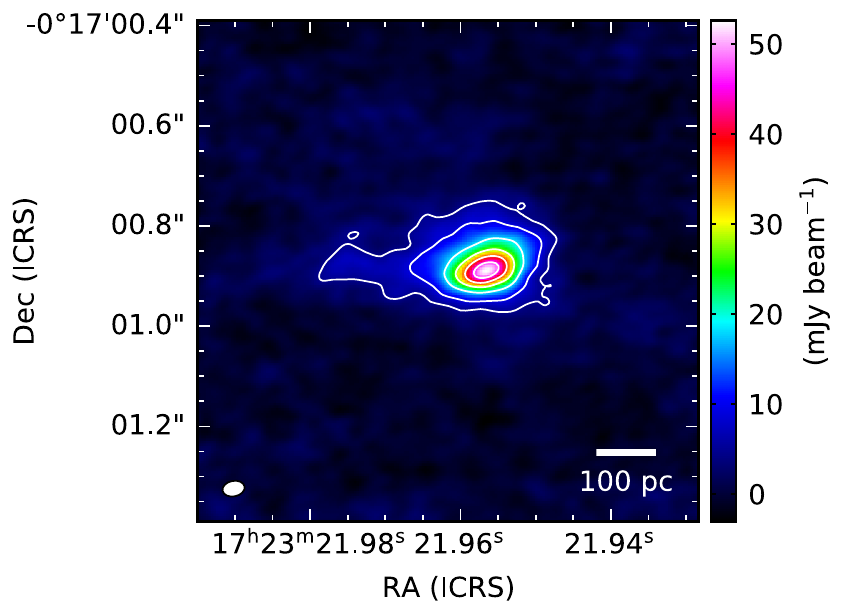}
\caption{Dust continuum emission map at $\lambda_\mathrm{rest}=436\,\micron$ in the central $1\arcsec\times1\arcsec$ region of IRAS 17208.
The contours indicate $5\sigma$, $10\sigma$, $20\sigma$, ..., and $60\sigma$ levels, where $1\sigma=0.92$\,mJy\,\per{beam}.
The white ellipse at the bottom left corner represents the synthesized beam size ($45\times31\,\unitpw{mas}{2}$, $\mathrm{PA}=101\arcdeg$).
Primary beam correction was applied.
\label{fig:continuum}}
\end{figure}

The image of the continuum emission at $\lambda_\mathrm{rest}=436\,\micron$ was obtained by averaging the line-free region of the continuum-unsubtracted data cube using the \texttt{immoments} task with \texttt{moments=-1}.
Figure \ref{fig:continuum} shows the result.
At this wavelength, the contribution of a radio jet, even if it exists, is minor.
Therefore, we attribute the continuum entirely to the thermal emission from dust.
The peak intensity of the continuum is 52.6\,mJy\,\per{beam}, which corresponds to a brightness temperature of 106\,K.\footnote{Brightness temperatures are expressed with the Rayleigh--Jeans approximation.}
The peak position is $\mathrm{(\alpha_{ICRS},\,\delta_{ICRS})=(17^h23^m21\fs957,\,-00\arcdeg17\arcmin00\farcs89)}$, which is close to the nuclear dynamical center determined by the kinematics of CO(2--1) observed with PdBI\footnote{Plateau de Bure interferometer.} \citep[$\sim$0\farcs5 resolution;][]{Garcia-Burillo+15}.
A faint elongation is observed to the east of the continuum map.
Of the two stellar disks revealed by \citet{Medling+14}, the western disk is brighter than the eastern one.
We consider that our continuum peak and elongation correspond to the western and eastern disks, respectively.
\citet{Garcia-Burillo+15} suggest that the molecular outflow they observed is launched from the western nuclei and that an obscured AGN is a convincing driving agent for it.
Thus, we regard our dust continuum peak as the location of the AGN hereafter.
The flux density encompassed within the central 0\farcs5 radius region is $0.60\pm0.12$\,Jy.
The error includes the systematic uncertainty in the flux calibration and is dominated by it.
Meanwhile, the total flux density of IRAS 17208 at 436\,\micron\ can be estimated to be 1.29\,Jy from a logarithmic interpolation between \Herschel\ SPIRE photometric fluxes at 350 and 500\,\micron\ obtained with a much larger aperture size of $\gtrsim$30\arcsec\ \citep{Pearson+16}.
This indicates that $\sim$50\% of the galactic dust emission at this wavelength is concentrated in the nuclear region.

\subsubsection{CO($J$=6--5)}
\label{sec:CO6-5}

\begin{figure}[t]
\plotone{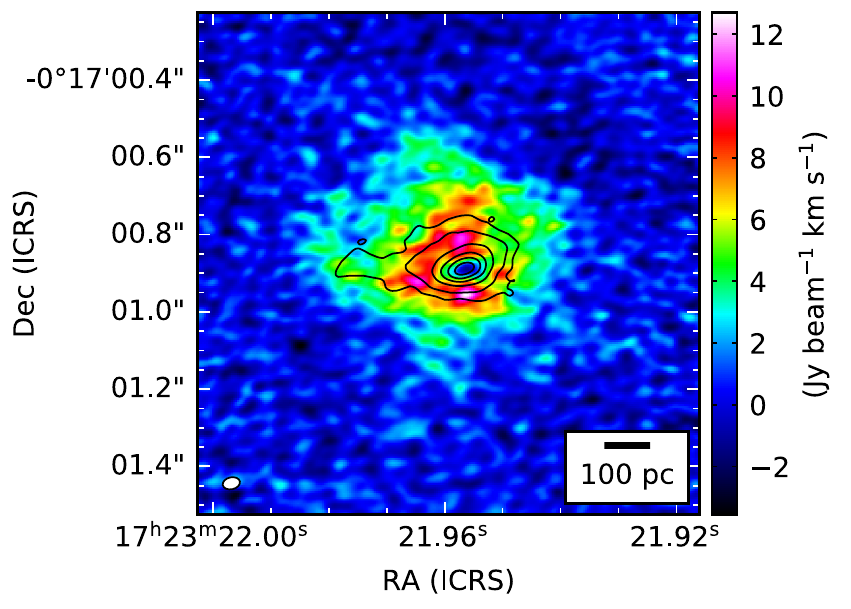}
\caption{CO(6--5) integrated intensity (moment 0) map in the central $1\farcs3\times1\farcs3$ region of IRAS 17208 (color scale), with the 436\,\micron\ continuum map overlaid (the contour levels are the same as in Figure \ref{fig:continuum}).
The velocity range used for the moment is $-$390 to 640\,\kmps.
The rms noise in the blank sky is 0.96\,Jy\,\per{beam}\,\kmps.
The white ellipse at the bottom left corner represents the synthesized beam size of the CO(6--5) map ($45\times31\,\unitpw{mas}{2}$, $\mathrm{PA}=102\arcdeg$).
Primary beam correction was applied.
\label{fig:CO6-5}}
\end{figure}

The obtained line data cube of CO (6--5) is shown as channel maps in Appendix \ref{sec:cm} (Figure \ref{fig:cm}).
From that cube, an integrated intensity (moment 0) map was created by applying the \texttt{ia.moment} task.
The used velocity range was $-$390 to 640\,\kmps\ relative to \Vsys.
The result is presented in Figure \ref{fig:CO6-5} with the contours of the continuum map overlaid for comparison.
A minor fraction of the CO flux in the most blueshifted end was not observed as shown in the upper panel of Figure \ref{fig:submm-spec}.
This is because the change in line-of-sight velocity due to rotation was too large for one spectral window.
Assuming that the emission profile is symmetric, the missed fraction is estimated to be less than 2\%.
The brightest area in the map is shaped like the letter ``C'', and the gap in it corresponds to the missed part (see also below for the velocity map).
The integrated line flux within the central 0\farcs5 radius region is $(8.6\pm1.7)\times10^2$\,Jy\,\kmps\ (i.e., $(1.89\pm0.38)\times10^{-17}$\,W\,\persq{m}).
The error includes the systematic uncertainty in the flux calibration and is dominated by it.
Our CO(6--5) flux accounts for 62$\pm$13\% of the total flux measured with the \Herschel\ SPIRE spectrometer \citep{Pearson+16}, and this percentage is comparable with the recovered fraction of the continuum flux.

Figure \ref{fig:CO6-5} confirms that the CO(6--5) line is observed as absorption at the nucleus.
It is only at the nucleus that CO (6--5) is detected as absorption in multiple consecutive velocity channels (Appendix \ref{sec:cm}).
We interpret that the bright dust emission from the AGN is being absorbed by the foreground gas in the surrounding obscuring structure.
Given the high excitation level and wide FWHM ($\sim$200\,\kmps, Figure \ref{fig:submm-spec}), this absorption does not seem to originate from cold molecular clouds in the host galaxy.
\citet{Rose+20} reported CO(1--0) and CO(2--1) absorption against the bright AGN of Hydra A, resulting from at least 12 individual clouds, but their line widths were of the order of 4\,\kmps\ only.
Our result is noteworthy because, to the best of our knowledge, there is only one other example where a high-$J$ CO line has been detected in such broad and deep continuum absorption:
in a recent high-resolution ($\sim$70\,pc) observation of Arp 220, which is a heavily obscured late-stage merger similar to IRAS 17208, CO(6--5) was detected as continuum absorption in consecutive velocity channels of a total range of $\sim$150\,\kmps\ toward the two nuclei of the galaxy \citep{Sakamoto+21}.
At the nucleus of the radio galaxy NGC 1052, \citet{Kameno+20} observed CO(3--2) and (2--1) in absorption.
However, these lines are at much lower excitation levels, and their equivalent widths are 0.26 times smaller than those observed in our work.
In Section \ref{sec:RADEX}, we attempt to reproduce the intensities of the CO(6--5) emission and absorption we observed in IRAS 17208 with the help of numerical simulations.

\begin{figure}[t]
\plotone{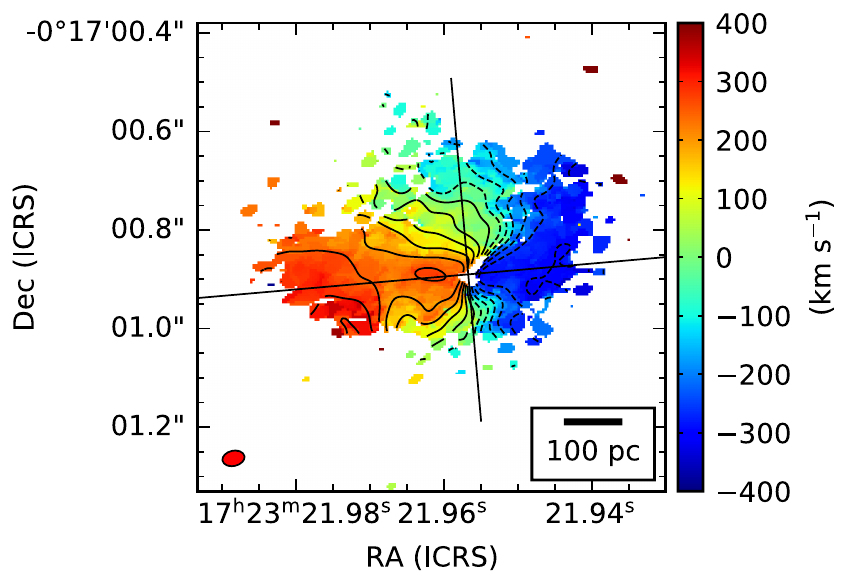}
\caption{CO(6--5) intensity-weighted velocity (moment 1) map in the central $0\farcs9\times0\farcs9$ region of IRAS 17208.
The velocity range used for the moment is $-$390 to 640\,\kmps.
The contours are drawn in steps of $\pm$50\,\kmps\ from 0\,\kmps.
The red-filled ellipse at the bottom left corner represents the synthesized beam size of the CO(6--5) map ($45\times31\,\unitpw{mas}{2}$, $\mathrm{PA}=102\arcdeg$).
The crossed lines show the major and minor axes adopted for the overall velocity gradient ($\mathrm{PA}=95\arcdeg$ and 5\arcdeg, respectively).
Primary beam correction and a 5-$\sigma$ clipping were applied.
\label{fig:velocity}}
\end{figure}

Figure \ref{fig:velocity} shows the CO(6--5) intensity-weighted mean velocity (moment 1) map.
The overall velocity gradient is along the east--west direction, with a position angle of $\mathrm{PA}\sim95\arcdeg$.
This is roughly consistent with the CO(2--1) kinematic axis ($\mathrm{PA}=113\arcdeg$) observed by \citet{Garcia-Burillo+15}.
The immediate vicinity of the nucleus, where the CO(6--5) line is in absorption, is masked because a 5-$\sigma$ clipping was applied.
The large-scale velocity gradient determined from the most red/blueshifted regions is at $\mathrm{PA}\sim95\arcdeg$, probably reflecting the combination of the two overlapping nuclear disks (Section \ref{sec:disks}).
The corresponding PV diagrams are presented in Appendix \ref{sec:PV}.

\subsubsection{CS($J$=14--13)}
\label{sec:CS14-13}

\begin{figure}[t]
\plotone{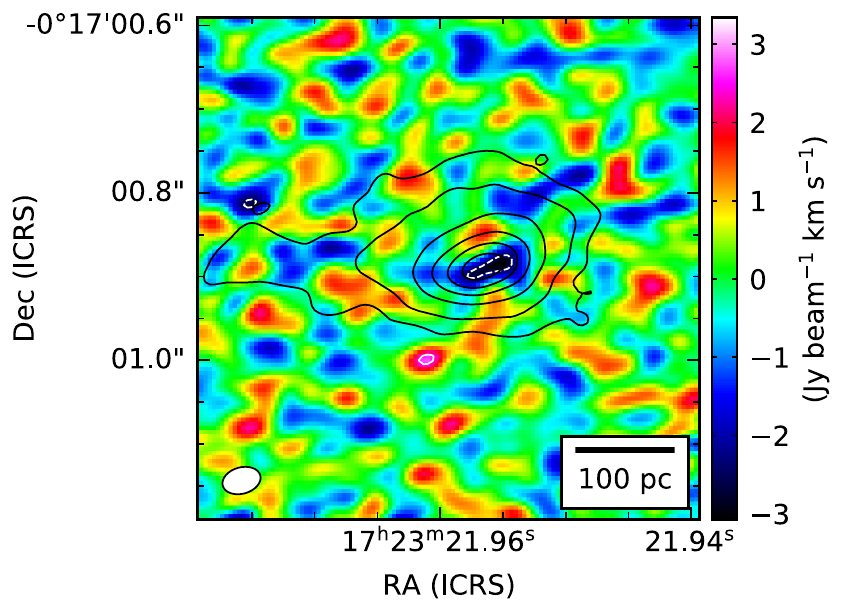}
\caption{
CS(14--13) intensity (moment 0) map in the central region of IRAS 17208 (color scale).
The velocity range used for the moment is $\pm$380\,\kmps.
The white contours indicate $\pm3\sigma$ levels, where $1\sigma=0.83\,\mathrm{Jy\,\per{beam}\,\kmps}$, while black ones are the same as in Figure \ref{fig:continuum} (436\,\micron\ continuum).
The white ellipse represents the synthesized beam size ($47\times32\,\unitpw{mas}{2}$, $\mathrm{PA}=106\arcdeg$).
Primary beam correction was applied.
\label{fig:CS14-13}}
\end{figure}

As shown in the lower panel of Figure \ref{fig:submm-spec}, CS(14--13) is detected in absorption in the beam-sized nuclear spectrum.
Figure \ref{fig:CS14-13} shows  the intensity (moment 0) map of CS(14--13).
This map confirms that the line was detected in absorption at the nucleus.
The significance is $-3.7\sigma$ relative to the rms noise of the blank sky signals.
Given the coincidence between the peak positions of the line and continuum, we regard this absorption to be a real detection.
We attempted to estimate the beam-deconvolved size of the absorption region by fitting a two-dimensional Gaussian using the \texttt{imfit} task, but it was found to be consistent with a point source and no lower limit on the size was obtained.
We confirmed based on channel maps that, in the region away from the nucleus, CS(14--13) was not detected either in emission or in absorption.
These results are referred to in the comparison with models in Section \ref{sec:RADEX}.

\subsection{\AKARI}

\subsubsection{Observations and Data Reduction}
\label{sec:AKARIreduction}

IRAS 17208 was observed twice with the \AKARI\ Infrared Camera (IRC) in the post-cryogenic phase of the satellite.
The observation IDs were 3370001.1 and 3370001.2, respectively, and the observation dates were 2009 Mar 10 and 11, respectively.
In both the observations, the galaxy was acquired within the $1\arcmin\times1\arcmin$ aperture named Np, and its light of wavelength 2.5--5.0\,\micron\ was dispersed with a germanium grism, resulting in a spectral resolution of $\sim$160 at the band center of the CO ro-vibrational transitions.
Owing to the slitless spectroscopy used, the spatial resolution was limited by the size of the point spread function (PSF; $\sim7\arcsec\sim6\,\mathrm{kpc}$).

From each observation, a two-dimensional (2D) spectral image was extracted with the IRC spectroscopic Toolkit version 20181203 in the standard manner, that is, using the \verb|irc_specred| command \citep{Ohyama+07, Baba+16, Baba+19}.
The extraction width to generate a one-dimensional (1D) spectrum was set to five pixels (7\farcs3) to match the PSF size.
Several percent of the pixels in the extraction region turned out to be hot pixels.
However, this is common for observations during the post-cryogenic phase \citep{Onaka+10}.
We corrected for the flux missed by the hot pixels based on the assumption that the spatial profile within the extraction aperture is independent of wavelength.
First, the 2D image was integrated along the dispersion axis to obtain the spatial profile.
Next, at each wavelength, the missed flux was estimated from the flux properly measured at the valid pixels, the ratio expected from the profile, and the positions of the hot and valid pixels.
The uncertainty in this correction was estimated using a bootstrap method,\footnote{We generated the spatial profile $10^3$ times by randomly resampling the 2D image with replacement, and then, evaluated the standard deviation of the resulting correction factors.} and then, propagated to the uncertainty in the flux density.

Each 1D spectrum was extracted using the \verb|plot_spec_with_image| command.
To cover the CO absorption more widely, we turned on the \verb|no_wl_limit| option, and recorded the flux density at wavelengths outside the nominal range (2.5--5.0\,\micron).
At wavelengths longer than 5.0\,\micron, because they are one octave beyond the shorter end of the spectrum, the first-order spectrum overlaps with the second-order spectrum.
To correct for this contamination, we turned on the \verb|NG_2nd_cal| option, which decomposed and removed the contribution from the second-order spectrum as per the methods described by \citet{Baba+16,Baba+19}.
The wavelength-dependent backgrounds were subtracted using the \verb|bg_sub| option.
Based on the output of the \verb|plot_spec_with_image| command, every four data points were binned by discarding the points of poor significance ($\mathrm{S/N}<3$) to improve the overall S/N.
Finally, the two binned spectra, which were generated from the two observations, were averaged.
The final spectrum is shown in Figure \ref{fig:IR-spec}a, where the absolute flux values are scaled as described in the next subsection.

The CO $v=1\leftarrow0$ $\Delta J=\pm1$ ro-vibrartional absorption band is clearly detected at 4.67\,\micron, and seen to have a substantial depth.
The intrinsic lines associated with different rotational levels are not resolved because of insufficient spectral resolution, and the band is observed as a double branched profile that represents $\Delta J=+1$ and $-1$.
The peak-to-peak separation between the two branches is also significantly wide.
Note that, at 4.65 and 4.69\,\micron, the \ion{H}{1} Pf$\beta$ and \HH\ 0--0 $S(9)$ lines may be superposed on the spectrum.
The spectrum also exhibits several emission (Br$\alpha$ at 4.05\,\micron, Br$\beta$ at 2.62\,\micron, and PAH at 3.3\,\micron) and absorption features (CO$_2$ gas or ice at 4.26\,\micron\ and H$_2$O ice at 3.1\,\micron).

\subsubsection{Continuum Normalization}

\begin{figure*}[t]
\plotone{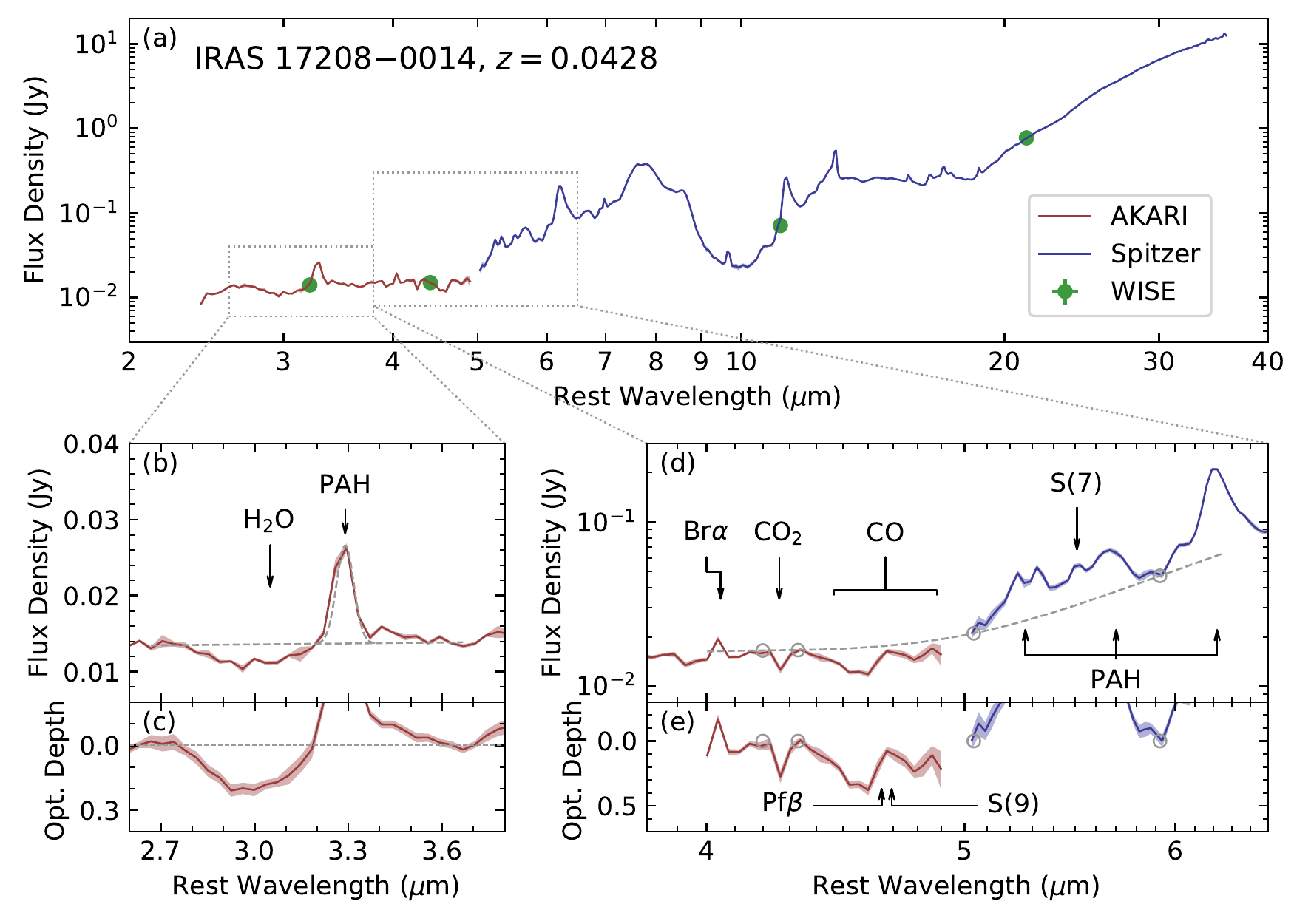}
\caption{(a) \AKARI/IRC and \Spitzer/IRS spectra of IRAS 17208 (red and blue solid lines, respectively).
The two spectra were rescaled to fit the \WISE\ photometric flux densities (green filled circles).
The flux uncertainty for each spectrum is indicated by a lighter color.
However, it is almost indistinguishable from the spectrum except for at $\sim$5 and $\sim$10\,\micron.
(b) \AKARI/IRC spectrum zoomed-in around 3.3\,\micron\, showing the PAH emission and H$_2$O ice absorption.
The dashed curve and dashed straight line show the fitted Gaussian function and linear baseline, respectively.
(c) Optical depth of the normalized absorption spectrum after smoothing with a 0.03-\micron\ boxcar kernel.
(d) \AKARI/IRC and \Spitzer/IRS spectra zoomed-in at a wavelength range of 3.8\,\micron\ to 6.5\,\micron\, showing CO absorption at 4.67\,\micron.
The dashed curve shows the adopted continuum over the CO band, which was obtained by performing cubic spline interpolation between the four pivots (open circles).
The emission and absorption features of the spectra are annotated.
(e) Optical depth of the normalized absorption spectra.
The wavelengths of the \ion{H}{1} Pf$\beta$ and \HH\ 0--0 $S(9)$ lines are indicated.
\label{fig:IR-spec}}
\end{figure*}

The CO band is observed at the longer wavelength end of the \AKARI\ spectrum.
Therefore, to robustly determine a continuum level, we used a spectrum obtained with \Spitzer/IRS similarly to \citet{Baba+18}.
A calibrated low-resolution spectrum with a wavelength range longer than 5.2\,\micron\ was obtained from the Combined Atlas of Sources with \Spitzer\ IRS Spectra \citep[CASSIS;][]{CASSIS}.\footnote{The Combined Atlas of Sources with Spitzer IRS Spectra (CASSIS) is a product of the IRS instrument team supported by NASA and JPL. CASSIS is supported by the ``Programme National de Physique Stellaire'' (PNPS) of CNRS/INSU co-funded by CEA and CNES and through the ``Programme National Physique et Chimie du Milieu Interstellaire'' (PCMI) of CNRS/INSU with INC/INP co-funded by CEA and CNES.}
The spectral segments obtained with the SL and LL slits (shorter and longer than 14\,\micron, respectively) were stitched together by scaling them appropriately, such that they smoothly connected with each other.
To mitigate the difference between the aperture sizes of \AKARI\ and \Spitzer, the \AKARI\ spectrum was rescaled to match the \WISE\ $W1$ and $W2$ flux densities, and the \Spitzer\ spectrum was rescaled to match the \WISE\ $W3$ and $W4$ flux densities.
The results of profile-fit photometry from the AllWISE catalog were used; the PSF size of \WISE\ \citep[$>6\arcsec$;][]{Wright+10} is larger than the nucleus.
The final merged spectrum is shown in Figure \ref{fig:IR-spec}a.

We defined the continuum level as a cubic spline curve that passes through four pivots: two about the CO$_2$ absorption at 4.26\,\micron\, one before the PAH emission at 5.27 \,\micron\, and the last after the PAH emission at 5.70\,\micron\ \citep{Smith+07}.
The adopted pivots and continuum are shown in Figure \ref{fig:IR-spec}d, and the resulting normalized spectrum is presented in Figure \ref{fig:IR-spec}e.
We observe that the optical depth of the CO ro-vibrational band reaches 0.36 after being blurred with the \AKARI\ resolution.
We analyze this spectrum in more detail in Section \ref{sec:Cami}.

\subsubsection{Other Spectral Features}
\label{sec:others}

In addition to the CO band, we measured five more quantities to evaluate the AGN/starburst diagnostics proposed by \citet{Imanishi+08,Imanishi+10} and \citet{Inami+18}.
These are: (1) the equivalent width of the 3.3\,\micron\ PAH emission (\EWPAH); (2) the optical depth of 3.1\,\micron\ H$_2$O ice (\tauice); (3) the optical depth of 3.4\,\micron\ aliphatic carbon absorption (\taudust); (4) the continuum color between 2.8 and 4.3\,\micron\ represented by the flux ratio (\AKARIcolor); and (5) that represented by the power-law slope ($\Gamma$; $f_\nu\propto\lambda^\Gamma$).

The strength of the PAH emission was measured in the same manner as \citet{Inami+18}, i.e., by fitting a Gaussian and linear baseline to the spectrum (see Figure \ref{fig:IR-spec}b).
Note that, at this wavelength range, the spectral shape is complex, as they include the 3.4\,\micron\ PAH sub-feature, H$_2$O absorption, and aliphatic carbon absorption, which were carefully analyzed by \citet{Doi+19}.
In this work, we did not analyze these features in detail, but instead, employed a simpler approach to maintain consistency with \citet{Inami+18}.
The wavelength ranges used for fitting the linear baseline were 2.65--2.75\,\micron\ and 3.55--3.70\,\micron, which correspond to the short wavelength tail of the H$_2$O absorption and the long wavelength tail of the PAH sub-feature and aliphatic carbon absorption, respectively \citep{Doi+19}.
The Gaussian was fitted in the range of 3.2--3.4\,\micron, thus avoiding the PAH sub-feature.
Finally, we obtained $\EWPAH=0.065\pm0.006\,\micron$, which gives a higher S/N than \citet{Inami+18} owing to our rigorous data reduction.
We estimated $\tauice$ and $\taudust$ based on the same baseline by smoothing the optical depth spectrum with a 0.03-\micron\ boxcar kernel (see Figure \ref{fig:IR-spec}c), which yielded $\tauice=0.21\pm0.03$ and $\taudust<0.06$.
The ratio \AKARIcolor\ was calculated to be $1.14\pm0.05$; based on the assumption of a power-law continuum, the slope was estimated to be $\Gamma=0.31\pm0.10$.
We use these results in Section \ref{sec:nature} for further analysis.

\section{Comparison with Theoretical Models}
\label{sec:models}

In this section, we compare the observational results from ALMA and \AKARI\ with the theoretical models to estimate the properties of the gas observed in the sub-mm and near-IR wavelengths.

\subsection{ALMA}
\label{sec:RADEX}

To constrain the properties of the gas observed with ALMA, we compared the observed brightness of CO($J$=6--5) with non-LTE radiative transfer simulations using the RADEX code \citep{RADEX}.
RADEX solves statistical equilibrium problems, and determines optical depths based on the escape probability approximation, assuming a homogeneous one-phase medium of a certain geometry.
We first discuss the region detected in emission, followed by that detected in absorption.
We also compare the absorption depth of CS($J$=14--13) with that of CO($J$=6--5).

\subsubsection{Emission}

The brightest part of the CO(6--5) emission lies approximately between 0\farcs05--0\farcs10 ($\sim$40--80\,pc) from the nucleus (Figure \ref{fig:CO6-5}).
The integrated intensity within the annulus is on average 7.7\,Jy\,\per{beam}\,\kmps, with a standard deviation of 24\% to it.
The FWHM of the line is typically 300\,\kmps\ and yields a mean brightness temperature of $\Tbemi=48$\,K.
We compared this value with that obtained from our RADEX simulations.

We assumed a homogeneous gas cloud with number density \nHH, kinetic temperature \Tkin, and column density \NCO, and calculated the brightness by varying these three parameters.
Given the high fraction of our ALMA flux compared to the \Herschel\ flux ($\sim$60\%, Section \ref{sec:CO6-5}), the gas detected with the former is expected to have similar properties to the gas observed with the latter.
Hence, we decided the parameter space for the RADEX calculation by referring to a result from \Herschel.
\citet{Kamenetzky+14} modeled the CO spectral line energy distribution (SLED) acquired with ground-based single-dish telescopes (for $J_\mathrm{up}=1$, 2, 3) and \Herschel\ (for $J_\mathrm{up}=4$ and higher) using RADEX, and found that the CO(6--5) flux is dominated by a warm component that has $\log\nHH=4.7$ and $\log\Tkin=2.5$.\footnote{Throughout this paper, the logarithms of number density, temperature, and column density are given in units of \percb{cm}, K, and \persq{cm}, respectively.}
In this study, we expect that the gas interferometrically detected with ALMA is likely to be denser and warmer than the warm component detected by \citet{Kamenetzky+14}.
Therefore, we varied \nHH\ in the range of $10^{4\text{--}8}$\,\percb{cm} and \Tkin\ in the range of 200--1000\,K, and searched for the value of \NCO\ required to attain the observed \Tbemi.
We assumed a static slab geometry, used the cosmic microwave background as the source of background radiation ($\TBG=2.73$\,K), and fixed the FWHM of the line at 300\,\kmps.
For our grid calculations, we used pyradex, which is a Python wrapper for RADEX.

\begin{figure*}[t]
\plotone{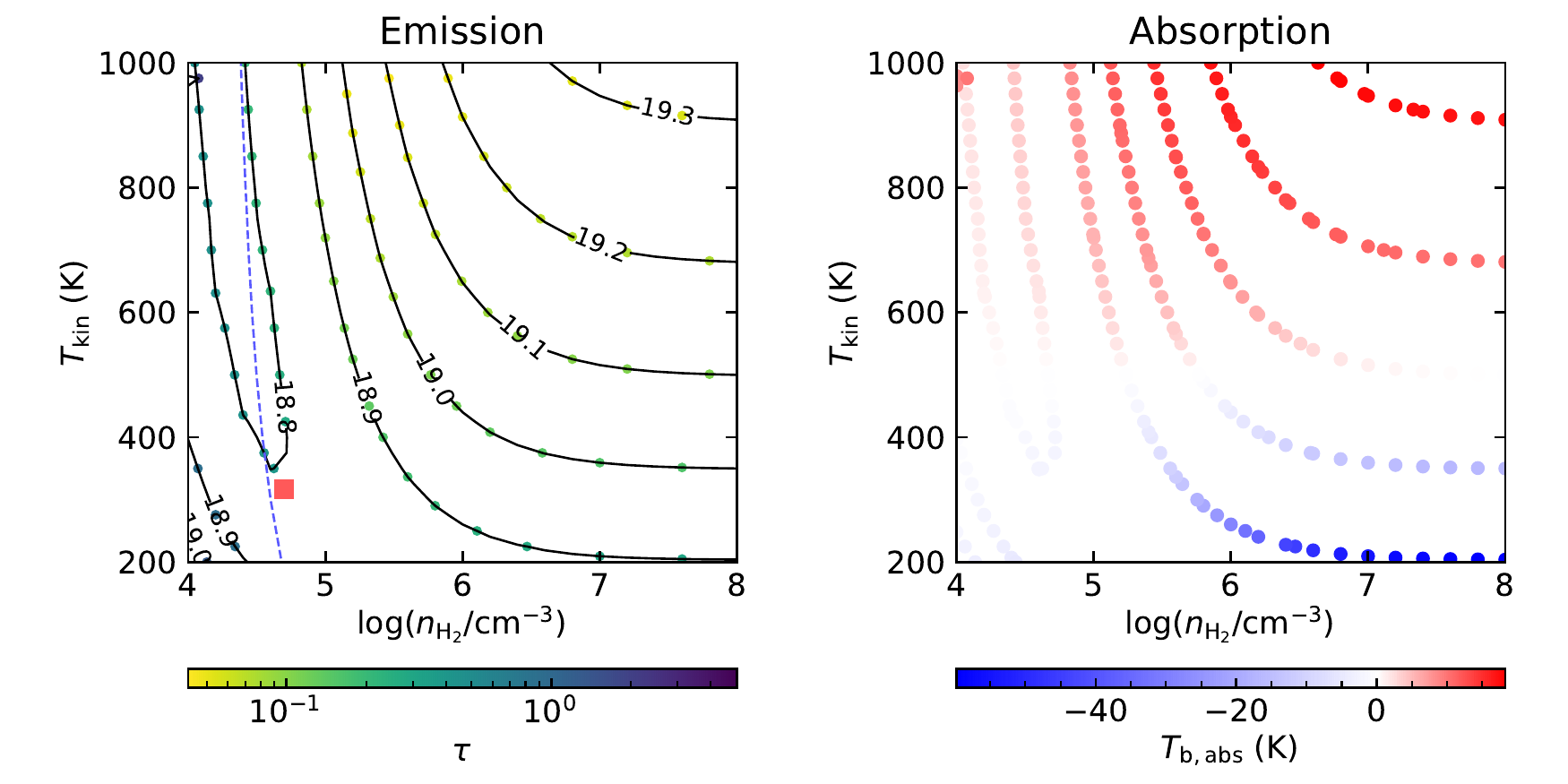}
\caption{Results of the RADEX calculations.
Left panel:
The gas quantities for which the CO(6--5) brightness temperature matches the value observed in emission (48\,K).
The abscissa and ordinate represent the \HH\ number density (\nHH) and gas kinetic temperature (\Tkin), respectively.
The labels of the contours indicate the logarithm of the total CO column density (\NCO) in units of \persq{cm}.
The colored dots on the contours denote the optical depth ($\tau$) for each combination of the parameters $(\nHH, \Tkin, \NCO)$.
The red square indicates \nHH\ and \Tkin\ of the warm SLED component obtained by \citet{Kamenetzky+14}.
The blue dashed line represents the critical density (\ncrit) in the optically thin limit.
Here, \ncrit\ is calculated as $\ncrit=A_{ij}/\sum_{k\ne i}C_{ik}$, where $A_{ij}$ and $C_{ij}$ are the Einstein $A$ coefficient and collision rate, respectively; both values are quoted from the Leiden Atomic and Molecular Database \citep[LAMDA;][]{LAMDA}.
Right panel:
The brightness temperature expected toward the line of sight of the nucleus corresponding to each parameter combination shown in the left panel, based on the assumption of Case (a) (Figure \ref{fig:assumption}).
A factor of 0.5 is multiplied to \NCO\ to consider only the gas in front of the background source.
The background radiation temperature is set to $\TBG=500$\,K, and the effect of the beam filling factor $f$ is not included.
\label{fig:RADEX}}
\end{figure*}

The left panel of Figure \ref{fig:RADEX} presents the results of our emission simulations.
It indicates that at least $\log\NCO=18.8$ is required to achieve $\Tbemi=48$\,K.
This lower limit is found near the critical density \ncrit\ \citep[\(\sim3\times10^4\,\percb{cm}\);][]{Yang+10}, which is conducive to both radiative and collisional excitations, and corresponds to an efficiently populated upper level.
Collisional transitions become dominant as \nHH\ increases.
The CO population moves to levels higher than $J=6$, owing to a \Tkin\ much higher than the corresponding energy level ($E_6 = 116$\,K).
The brightness for a given \NCO\ thus decreases; that is, a higher \NCO\ is needed to attain the same brightness.
In the high-\nHH\ limit, the CO population is no longer density-dependent, and the contours become horizontal.
Thus, the figure suggests that $\log\NCO$ does not exceed 19.3, even when a very high kinetic temperature, $\Tkin=1000$\,K, is assumed.
Hence, $\log\NCO$ is constrained by 0.5\,dex.
Assuming an abundance ratio of $[\mathrm{CO}]/[\HH]=10^{-4}$, this corresponds to $\NHH\sim10^{23}\,\persq{cm}$, suggesting the existence of a considerable amount of gas in the circumnuclear region.
Hereafter, we use $[\mathrm{CO}]/[\HH]=10^{-4}$ as a fiducial value, unless otherwise noted.
At \nHH\ lower than \ncrit, which is also lower than the density obtained by \citet{Kamenetzky+14}, collisional excitation is ineffective.
The excitation temperature between levels $J=6$ and 5 is much lower than the kinetic temperature, and the source function is weak.
Thus, a high optical depth is required.
Unfortunately, we cannot further constrain the parameters because, currently, there are no available observations of other excitation levels at a comparable or better angular resolution.

\subsubsection{Absorption}
\label{sec:RADEX_abs}

The nuclear ($r\sim20$\,pc) spectrum at the dust peak (Figure \ref{fig:submm-spec}, lower panel) shows that the CO (6--5) absorption has a peak of $-$18\,mJy\,\per{beam} (corresponding to $-$36\,K) and a FWHM of 200\,\kmps.
First, we examined whether this observed absorption strength can be consistently explained by the parameters obtained from the emission calculations.

\begin{figure}[t]
\plotone{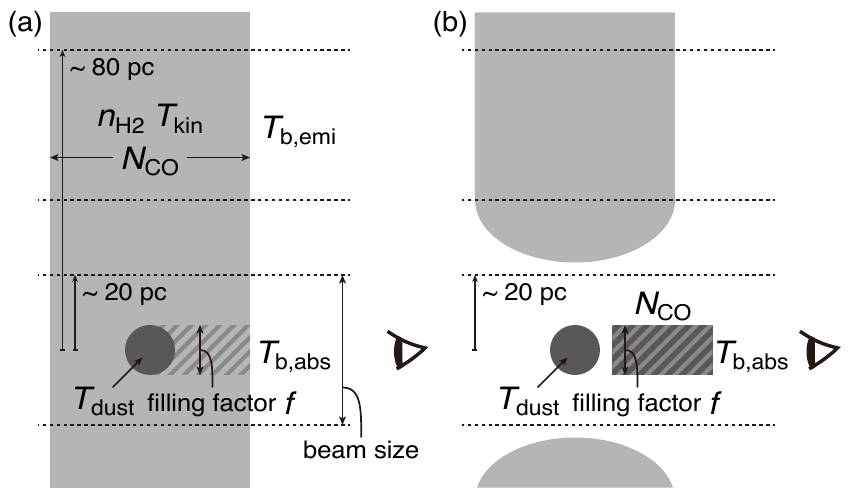}
\caption{
Geometries assumed in the RADEX calculation for the CO(6--5) absorption in the central beam ($r\sim20$\,pc).
Case (a): The case to verify whether the intensity of emission and absorption can be explained simultaneously.
Note, however, that this geometry turns out to be too simple (see text).
The beam is filled with the same gas as seen in emission in the off-nuclear region ($r\sim40$--80\,pc).
The background source, which is smaller than the beam, is buried in the disk, and the gas in front of it (hatched region) contributes as absorption.
The disk is actually inclined, but here it is drawn straight for clarity.
Case (b): The case to estimate the lower limit of the column density toward the nucleus.
There is a dense gas in front of the background source with the same beam filling factor, but it is distinct from the disk in the off-nucleus region.
The disk is actually inclined but drawn straight as in Case (a).
\label{fig:assumption}}
\end{figure}

Figure \ref{fig:assumption}(a) illustrates the geometry assumed here (Case (a)).
A dust continuum source with temperature \Tdust\ is buried in the gas cloud of \nHH, \Tkin, and \NCO\ obtained in the previous subsection.
Since the velocity map (Figure \ref{fig:velocity}) showed a rotation pattern, we assumed that the cloud is an inclined disk.
The size of the continuum source is considered to be smaller than the observation beam ($\sim$30\,pc).
This is because \Tkin\ is now thought to be higher than 320\,K suggested from \Herschel\ \citep{Kamenetzky+14}, and if \Tdust\ were equal to the observed brightness temperature (106\,K), the gas would not be observed in absorption.
Therefore, it is expected that the intrinsic \Tdust\ is higher and that the continuum brightness is diluted from \Tdust\ due to the beam filling factor $f$ of the source being smaller than unity.
Even without the \Herschel\ results, the detection of CS(14--13) in absorption toward the nucleus supports this prediction.
To explain the CS absorption depth while keeping the CS-to-CO abundance ratio in a reasonable range (see below), the $J=13$ level of CS must be occupied with a high percentage.
The level energy of CS at $J=13$ is 214\,K, and the gas should be as warm as or warmer than this temperature.
In order for the line to be absorbed, there must be a hotter background source behind it.
From the 2D Gaussian fitting to the CS absorption region, it was determined to be a point source (Section \ref{sec:CS14-13}).
This also supports $f<1$.
Since no upper limit on the size was obtained from the fitting, we did not assume a specific value for $f$ (and hence \Tdust) in the following comparisons.

We calculated the CO absorption brightness \Tbabs\ using RADEX from each set of parameters obtained from the emission (Figure \ref{fig:RADEX}, left panel).
The temperature of the background radiation, $\TBG=\Tdust$, was set to 500\,K as an initial guess.
In this case, the beam filling factor $f$ should be $\sim$0.2 based on the ratio to the continuum peak brightness (106\,K).
In addition, the column density \NCO\ was multiplied by 0.5 to account only for the part in front of the continuum source, and the velocity FWHM was changed to 200\,\kmps\ as measured in the nuclear spectrum (Figure \ref{fig:submm-spec}, lower panel).

The obtained \Tbabs, corresponding to each point in the left panel of Figure \ref{fig:RADEX}, is shown using a color scale in the right panel of the same figure.
Note that the effect of the beam filling factor $f$ is not included here.
Note also that the CO (6--5) becomes an absorption line when $\Tkin<\TBG=500$\,K.
The minimum value of $\Tbabs$ ($-60$\,K) is obtained in the collision-dominated region.
This value is deeper than the observation.
In practice, however, the absolute value of \Tbabs\ that can be observed should be smaller, owing to the effect of $f$.
If only a simple dilution is assumed, then \Tbabs\ needs to be multiplied by $f$ to yield $\sim-$12\,K.
If the contribution of the emission in the region offset from the continuum source is included, then \Tbabs\ is averaged with $\Tbemi$ weighted by $(1-f)$ to yield $\sim$26\,K.
None of these temperatures match the observed absorption depth.
This is true even if a different \TBG\ is used.

The above discussion on the intensities of emission and absorption is based on the assumption of $\Tkin>320$\,K suggested from \Herschel\ \citep{Kamenetzky+14} and the implication of the CS(14--13) detection.
If based solely on our observations of CO(6–-5), the parameters $\Tdust=106$\,K ($f=1$), $\Tkin\sim65$\,K, $\nHH\sim10^6\,\percb{cm}$, and $\log\NCO\sim19.4$ can explain the observed \Tbemi\ and \Tbabs\ simultaneously.
However, if we adopt $\Tkin>320$\,K (\Herschel\ SLED warm component) or $>200$\,K (CS $J=13$ energy level), then there is no possible solution.

These results demonstrate that the Case (a) is too simplified to explain the observed emission and absorption features.
One natural interpretation is that the gas in the central beam ($r\sim20$\,pc) is more concentrated than in the annular region ($r\sim40$--80\,pc).

\begin{figure}[t]
\plotone{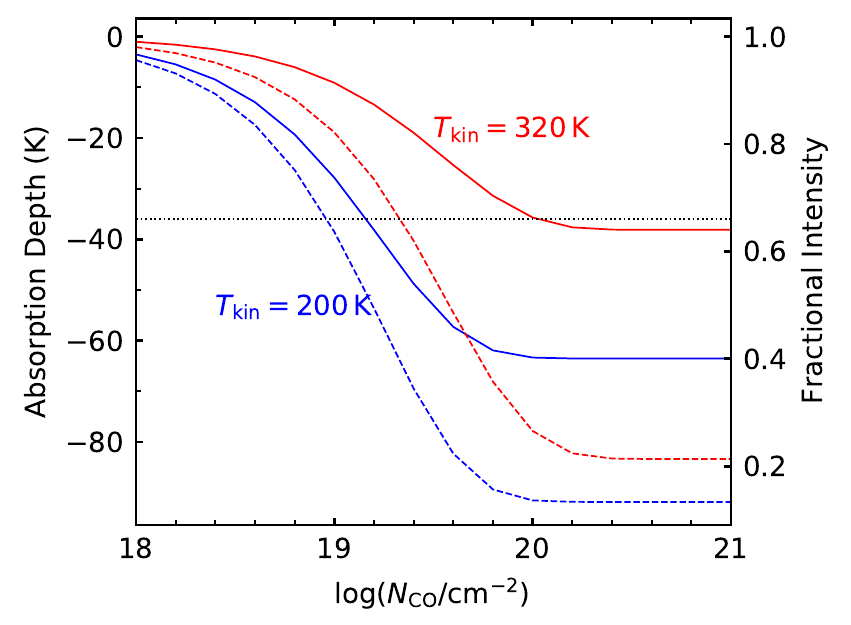}
\caption{
CO(6-5) absorption intensities calculated as a function of CO column density using RADEX, based on the assumption of Case (b) (Figure \ref{fig:assumption}).
The left and right vertical axes represent the intensity as brightness temperature and fraction relative to the continuum level, respectively.
The volume density of the gas is fixed at $\nHH=1\times10^7\,\percb{cm}$, and the kinetic temperature is set to $\Tkin=200$ or 320\,K (blue and red lines, respectively).
The temperature of the background source is assumed to be $\TBG=500$ or 1500\,K (solid and dashed lines, respectively).
The effect of dilution due to the beam filling factor $f<1$ is taken into account in a specific form that $f$ is the same for the continuum source and the line-absorbing gas.
The horizontal dotted line indicates the intensity observed.
\label{fig:coreNco}}
\end{figure}

Next, we attempted to constrain the column density in the central region by comparing only the nuclear absorption with the RADEX calculation, without considering the consistency with the CO(6--5) emission in the off-nuclear region.
Figure \ref{fig:assumption}(b) shows the geometry assumed here (Case (b)).
We first assumed that only a gas cloud of approximately the same size as the continuum source is in the beam.
We also considered that the absorption of CO(6--5) and CS(14--13) occur in this same cloud.
Based on the critical density\footnote{Calculated in the optically thin limit at 200\,K using $A_{ij}$ and $C_{ij}$ quoted from LAMDA \citep{LAMDA}.} and lower level energy of CS(14--13), the number density and kinetic temperature of the gas were fixed at $\nHH=1\times10^7\,\percb{cm}$ and $\Tkin=200$\,K, respectively.
The solid blue line in Figure \ref{fig:coreNco} shows the absorption intensity calculated as a function of the column density \NCO.
The temperature of the background source is set to $\TBG=500$\,K, and beam dilution ($f=0.2$) is also taken into account.
Under this condition, the observed intensity of $-36$\,K is reached at $\log\Nco=19.2$.
Here, since the compensation for the absorption due to emission from gas not in front of the continuum source is ignored, this column density gives a lower limit.

This lower limit hardly depends on the assumed \nHH\ because it is more than two orders of magnitude higher than the critical density of CO(6--5).
When \Tkin\ is increased (e.g., 320\,K suggested by \citealt{Kamenetzky+14}), the CO $J=5$ level (83\,K) becomes less populated, so a larger column density is required (Figure \ref{fig:coreNco}, red solid line).
On the other hand, when the temperature of the background light source is higher, the absorption becomes deeper efficiently, and the required column density becomes smaller (dashed lines).
If \TBG\ is raised to the dust sublimation temperature of 1500\,K, for $\Tkin=200$\,K and 320\,K, $\log\NCO=19.0$ and 19.3, respectively.

In summary, from the spectrum toward the nucleus, the column density of the dense gas can be constrained to be $\log\Nco>19$.
Since the constraint in the off-nucleus region is $\log\Nco=18.8$--19.3,  it is not proved by these alone that the nuclear component is in excess in terms of column density.
However, if we make the assumption that \Tkin\ at the center is higher than outside, then for $\Tkin=320$\,K, $\log\Nco$ does not exceed 19.0 in the off-nuclear region (Figure \ref{fig:RADEX}, left), while it is higher than 19.3 in the center.
This results in an excess of $>$0.3\,dex, which is larger than the scatter of the off-nuclear emission intensities (24\%).
Unfortunately, it is difficult at present to more tightly constrain the column density in front of the sub-mm continuum source.
Further observations in other bands and/or higher resolution are essential to determine the gas temperature and the intrinsic background brightness.

The constrained \NCO\ toward the nucleus can be converted to the hydrogen column density of $\NHHsubmm>10^{23}\,\persq{cm}$.
The implication of this lower limit is discussed in Section \ref{sec:distribution}.

In terms of CS(14--13), the depth of absorption in the nuclear spectrum is $-$12\,mJy\,\per{beam} ($-$24\,K).
From this, we estimated the column density of CS as in Figure \ref{fig:coreNco} for CO.
As a result, $\log N_\mathrm{CS}=16.5$ and 16.8 were obtained for $\Tkin=200$ and 320\,K, respectively, at sufficiently high number densities.
These column densities are 2.5 orders of magnitude smaller than those obtained for CO.
In other words, the depths of absorption of CO and CS can be explained by the abundance ratio of $[\mathrm{CS}]/[\mathrm{CO}]\sim10^{-2.5}$.
This abundance is not in conflict with that predicted for high-density XDR or high-density and low-radiation-field photodissociation region (PDR).
In the latter case, however, the obtained column density cannot be reproduced (see Section \ref{sec:distribution}).

In the SLED analysis of IRAS 17208 by \citet{Kamenetzky+14}, the cool component has $\log\nHH=4.1$, $\log\Tkin=1.1$, $\log\Nco=20.1$.
Owing to its large column density, this component would be admittedly able to explain the depth of CO(6--5) if it were distributed in front of the background source.
However, it is too tenuous and cold to explain the depth of CS(14--13), whose critical density and energy level are much higher than those of CO(6--5).
We found that if this absorption were to be reproduced by the cool component, an abnormally high abundance $[\mathrm{CS}]/[\mathrm{CO}]=0.4$ would be required.
Conversely, if CS(14--13) is originated from warm gas, then the absorption of CO, which is more abundant, should also arise from the same gas.
Therefore, we attribute the CO(6--5) and CS(14--13) absorption entirely to the warm gas.

\subsection{\AKARI}
\label{sec:Cami}

In this subsection, we quantify the properties of the gas responsible for the near-IR CO $v=1\leftarrow0$ $\Delta J=\pm1$ ro-vibrational absorption.
The band profile observed with \AKARI\ is fitted with a gas model, which we discuss in detail here.

\subsubsection{Subtraction of Pf$\beta$ and $S(9)$}

The \ion{H}{1} Pf$\beta$ and \HH\ 0--0 $S(9)$ emission lines may be overlapping in the CO band.
Thus, we first estimated and then subtracted the fluxes of these lines from the CO band spectrum (Figure \ref{fig:IR-spec}e).
The Pf$\beta$ and $S(9)$ emission lines are considered to originate from star-forming regions outside the AGN: \ion{H}{2} and photodissociation regions, respectively.
The flux of Pf$\beta$ was converted from that of Br$\alpha$ at 4.05\,\micron, based on the assumption of Case B of \citet{Baker&Menzel}.
The Br$\alpha$ flux was measured to be $(3.6\pm0.4)\times10^{-21}$\,W\,\persq{cm}, and the Pf$\beta$/Br$\alpha$ ratio in Case B is 0.20 for $n_e\sim10^{2\text{--}7}$\,\unitpw{cm}{-3} and $T_e\sim(3\text{--}30)\times10^3$\,K \citep{Strey&Hummer95}.
The Pf$\beta$ flux was thus calculated to be $(7.2\pm0.8)\times10^{-22}$\,W\,\persq{cm}.
The flux of $S(9)$ was extrapolated from the fluxes of $S(3)$ at 9.66\,\micron\ and $S(7)$ at 5.51\,\micron\, assuming level populations according to LTE.
The $S(3)$ and $S(7)$ fluxes were measured to be $(2.8\pm0.5)\times10^{-21}$\,W\,\persq{cm} and $(2.2\pm0.8)\times10^{-21}$\,W\,\persq{cm}, respectively, yielding an excitation temperature of $1077\pm96$\,K.
From these fluxes and temperature, the $S(9)$ flux was derived to be $(4.5\pm2.7)\times10^{-22}$\,W\,\persq{cm}.
During the subtraction of the line fluxes, the Gaussian profiles of the width limited by the spectral resolution were assumed, which were then binned as in the observed spectrum.
The change in the flux density owing to this subtraction is at most 6\%.

\subsubsection{Comparison with an LTE Gas Model}

We analyzed the CO absorption observed with \AKARI\ by fitting it with the plane parallel LTE gas model developed by \citet{Cami02}, following the way of \citet{Baba+18}.
In \citet{Baba+18}, a one-temperature homogeneous CO gas was assumed, and it was fitted with three free parameters: total column density \Nco, excitation temperature \Tex, and velocity turbulence \vturb.
Here, \Tex\ is equivalent to the kinetic temperature \Tkin.
The S/N of the spectrum in the present study was not sufficient to constrain all three parameters.
Thus, we fixed \vturb\ to be the width of the CO(6--5) absorption line detected in the ALMA observations.
With this \vturb, the modeled spectrum was intrinsically spiky with the different rotational lines clearly separated, but during the fitting, the spectrum was blurred by the resolution of IRC.
We adopted 1500\,K blackbody radiation as the background continuum to represent the thermal emission from the dust sublimation layer around the nucleus, as in \citet{Baba+18}.
Note that this background temperature is higher than that adopted for the comparison of the RADEX calculations with the ALMA data (see Section \ref{sec:RADEX_abs}).
This is owing to the difference in the wavelength regimes in the two cases, as the near-IR continuum source should be smaller and hotter than the sub-mm source.
Here, the area covering fraction of the absorber to the continuum source was assumed to be unity.
We performed non-linear least-squares minimization based on the Levenberg--Marquardt algorithm, using the IDL package MPFIT \citep{MPFIT}.

\begin{figure*}[t]
\plotone{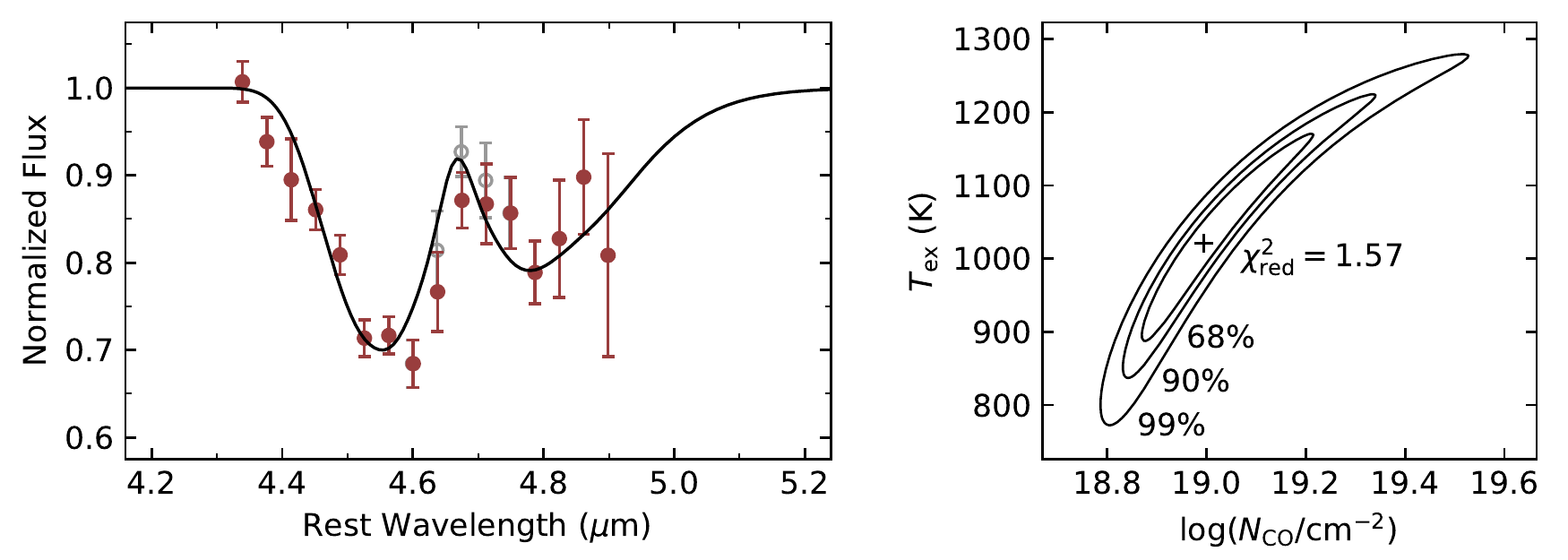}
\caption{Left panel: The observed (filled red circles) and best-fitted (solid black curve) spectrum of the CO ro-vibrational absorption.
Only the data points within the fitting wavelength range are plotted.
The open gray circles represent the fluxes before the subtraction of Pf$\beta$ and $S(9)$;
these are not used for gas model fitting, but shown for reference.
Right panel: Confidence regions for the two parameters of interest \Nco\ and \Tex.
The contours represent the 68\%, 90\%, and 99\% confidence levels ($\Delta\chi^2=2.28$, 4.61, and 9.21, respectively).
The best-fitted values are indicated by the plus sign.
\label{fig:BestFit}}
\end{figure*}

The minimum chi-square value was obtained at $\NCONIR=1.0\times10^{19}$\,\persq{cm} and $\Tex=1021$\,K, with a reduced chi-square of $\chi^2_\mathrm{red}\equiv\chi^2/\mathrm{dof}=1.57$.
The best-fitted gas model is shown in Figure \ref{fig:BestFit} along with the two-parameter confidence regions.
The observed absorption spectrum is fitted fairly well by the single-component gas model.
There is a degeneracy between the two parameters.
This is because, as the gas temperature approaches the background temperature, the emission from the gas becomes more significant.
Thus, a higher column density is required to absorb the continuum significantly, to compensate for the gas emission and maintain the apparent absorption depth.
Despite this degeneracy, it is suggested that the CO gas has $\NCONIR\sim(0.7\text{--}2)\times10^{19}$\,\persq{cm}.
We discuss the origin of this column density in Section \ref{sec:distribution} by comparing it with the ALMA results.

\section{Discussion}
\label{sec:discussion}

\subsection{Gas Distribution}
\label{sec:distribution}

The comparison between our ALMA observations and RADEX simulations shows that the observed emission and absorption intensities cannot be reproduced simultaneously by a simple uniform gas cloud.
One of the primary explanations for the observations is that, within the central beam ($\sim$30\,pc), there is a concentrated gas distribution that contributes to the absorption more than that expected from the brightness of the immediate surrounding region.
The column density toward the nucleus has an excess relative to the surroundings under the assumption of a higher kinetic temperature at the center.

It is natural to consider that a dense obscuring structure distinct from the circumnuclear medium exists in the central $\sim$10\,pc scale region.
\citet{Aalto+15_4ULIRGs} argued that, based on the bright HCN-vib(4--3) emission in IRAS 17208, there is a small core ($r\sim20$\,pc) that is opaque in the mid-IR having a 14\,\micron\ brightness temperature high enough to excite HCN molecules vibrationally.
In the prototypical Seyfert 2 galaxy NGC 1068, an AGN torus has been imaged in spatially resolved observations, with a size of $\sim$5\,pc in HCN and HCO$^+$ lines \citep{Imanishi+18,Imanishi+20} and $\sim$30\,pc in CO lines \citep{Garcia-Burillo+19}.
Thus, it is quite possible that a compact absorber is buried within the beam size of our observations.

The molecular gas column density toward the nucleus obtained in the sub-mm region, $\NHHsubmm>10^{23}\,\persq{cm}$, is a lower limit and does not contradict to that required from the Compton thickness of the AGN ($\NH>10^{24}\,\persq{cm}$).
In addition, the latter includes neutral gas in the dust-free regions  \citep{Merloni+14,Davies+15,Ruffa+18,Ichikawa+19,Kawakatu+20,Tanimoto+20}.
\citet{Ruffa+18} studied the obscured ULIRG IRAS F00183$-$7111, and found that the molecular column density derived from CO(1--0) is $\sim$12\% of the cold gas column density estimated from the X-ray spectral fitting.
Based on a comparison between the fractions of Seyfert 2 galaxies and X-ray absorbed AGNs in the \Swift-BAT AGN survey, \citet{Davies+15} argued that a neutral gas region begins to form inside the dust sublimation layer when the X-ray luminosity at 14--195\,keV (\LvHX) exceeds $10^{44}\,\ergps$.
Following the relation $\log\LAGN=1.12\log\LvHX-4.23$ \citep{Winter+12}, this threshold luminosity corresponds to an AGN bolometric luminosity of $\log\LAGN=45$.
The AGN luminosity of IRAS 17208 was suggested by \citet{Garcia-Burillo+15} to be $\log\LAGN=45.4$, which lies in the high-luminosity regime according to \citet{Davies+15}.
Therefore, a significant fraction of the X-ray absorbing column may be in the neutral phase, although IRAS 17208 is not a typical Seyfert galaxy, and it is not much evident whether its obscurer has the same structure.
To summarize, the gas observed in the CO(6--5) absorption is probably not dominant in the total obscuring column density toward the AGN.

The abundance of CO is also a major uncertainty.
\citet{Wada+16,Wada+18} performed numerical simulations of the non-equilibrium chemistry in the XDR around AGNs, by considering the Seyfert 2 Circinus galaxy as an example.
Their model indicated that, at high column densities ($\NHH=10^{23\text{--}24}\,\persq{cm}$), the abundance ratio [CO]/[\HH] has a large point-to-point scatter in the range of $10^{-5}$--$10^{-4}$.
If a lower abundance is adopted, the \NCOsubmm\ obtained by by our ALMA observation alone is sufficient to make the AGN CT without including neutral gas.
It has been observationally shown that CO dissociates immediately around an AGN, as seen in both the Circinus galaxy \citep{Izumi+18} and NGC 7469 \citep{Izumi+20}, based on the observed high [\ion{C}{1}]/CO ratios.

The comparison of our \AKARI\ spectrum and gas models shows that the observed CO ro-vibrational absorption band can be well represented by the LTE condition with a high column density ($\NHHNIR\sim10^{23}\,\persq{cm}$) and a high temperature ($\Tex\sim1000\,\mathrm{K}$).
Unlike the CO(6--5) line, this band contains multiple rotational levels.
Hence, the column density and temperature can be simultaneously obtained.
The obtained \NHHNIR\ is a lower limit because the area covering fraction of the absorber could be smaller than unity, and the 5\,\micron\ continuum flux could contain not only the AGN contribution but also the star-formation contribution.
Although the gas number density is not directly obtained from the model fitting, to establish LTE up to $J\sim30$,\footnote{The corresponding lines, $R(30)$ and $P(30)$, appear at 4.46 and 4.97\,\micron, respectively.} which corresponds to the observed width of the band, it must be higher than the critical density for these levels \citep[\(\nHH\gtrsim3\times10^6\,\percb{cm}\);][]{Yang+10}.

\begin{figure}[t]
\plotone{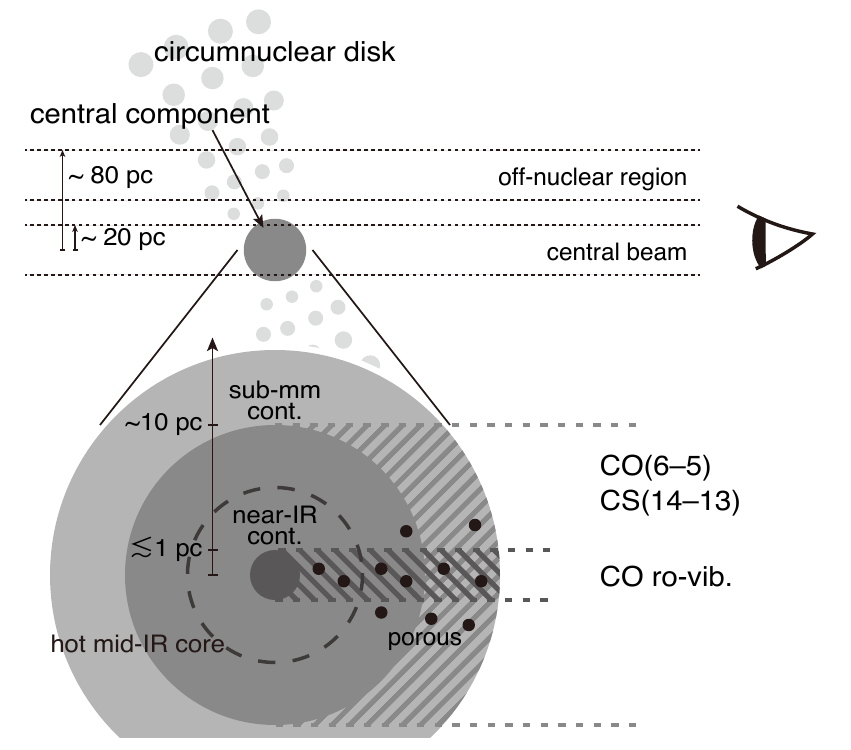}
\caption{
Schematic of the inferred geometry of the central region.
In the annular off-nuclear region ($r\sim40$--80\,pc), where CO(6--5) is observed in emission, there is a circumnuclear disk that is inclined to the line of sight and composed of clumpy gas.
Within the beam toward the nucleus ($r\sim20\,\mathrm{pc}$), there is a centrally concentrated component that contains sub-mm and near-IR continuum sources.
The former is more extended than the latter.
The CO(6--5) and CS(14--13) absorption occurs in front of the sub-mm continuum source (light-gray hatched).
The intrinsic absorption strength is diluted by the larger beam size.
The CO ro-vibrational absorption occurs in front of the near-IR continuum source (dark-gray hatched), probing deeper and hotter regions than the sub-mm absorption.
Because the near-IR continuum light reaches us, the concentrated component should be porous or have gaps.
The representative scale of the hot mid-IR core, from which HCN-vib emission arises, is expected to be between the sizes of the sub-mm and near-IR continuum sources.
\label{fig:geometry}}
\end{figure}

We summarize the gas properties near the nucleus estimated from the absorption in the sub-mm and near-IR regions in Table \ref{tab:submm-vs-NIR}.
Both the sub-mm and near-IR observations suggest similar column and volume densities, but the temperature is higher in the near-IR.
We interpret this result to mean that the region observed in the near-IR is in the nuclear direction, as in the sub-mm, but with a narrower pencil beam than in the sub-mm, containing a hotter region closer to the nucleus.
Figure \ref{fig:geometry} explains this interpretation.
The near-IR continuum should be dominated by a hotter and more compact region than in the sub-mm case.
Reflecting the difference in the temperature and size of the background sources, the near-IR absorption probes deeper into the hotter regions than the sub-mm one.
Here, given that the near-IR light reaches the outside, the concentrated component is considered to be porous or have gaps.
This interpretation is consistent with the fact that the ratio of column density to number density in Table \ref{tab:submm-vs-NIR} is much smaller than the spatial scale we are considering.
Since the line of sight to the near-IR source also crosses the outer region, it may be more accurate to include multiple temperature components in the fitting of the CO ro-vibrational absorption spectrum.
However, doing it with the current \AKARI\ data is difficult due to insufficient wavelength resolution ($R\sim160$).
Such studies will be possible with the \textit{James Webb Space Telescope} (\JWST).

\begin{deluxetable}{ccc}
\tablecaption{Gas properties estimated in the central 30\,pc\label{tab:submm-vs-NIR}}
\tablewidth{0pt}
\tablehead{\colhead{} & \colhead{sub-mm} & \colhead{near-IR}}
\startdata
\HH\ column density & $>10^{23}\,\persq{cm}$ & $>10^{23}\,\persq{cm}$ \\
Temperature         & $\gtrsim200$\,K & $\sim1000$\,K \\
Density             & $\gtrsim1\times10^7\,\percb{cm}$ & $\gtrsim3\times10^6\,\percb{cm}$ \\
\enddata
\end{deluxetable}

The central component, where the sub-mm CO(6--5) and CS(14--13) absorption and near-IR CO ro-vibrational absorption occur, may correspond to the hot mid-IR core that is responsible for HCN-vib emission.
The representative scale of the hot mid-IR core is expected to be between the sizes of the sub-mm and near-IR continuum sources according to the order of the wavelengths.
On the other hand, the gas observed in the off-nuclear region in the CO(6--5) emission shows a rotational velocity pattern and corresponds to a circumnuclear disk (hundreds pc radius).
Here, the ratio of the column density to the number density, shown in the left panel of Figure \ref{fig:RADEX}, yields a physical thickness of $<$1\,pc, which is considerably shorter than the orbital radius.
This implies that the disk is highly clumpy.

The indicated column density is large and the temperature is high, from both the sub-mm and near-IR observations.
The heating source that can most naturally explain such warm gas with a large column density is X-ray photons (i.e., XDRs), which have a high penetrating power compared to UV photons (PDRs) or shock waves, as discussed by \citet{Baba+18}.
If UV photons or shock waves were the heating source, then the gas temperature would decrease to $10^2$\,K before reaching $\NHH\sim10^{22}\,\persq{cm}$.
If the gas were smoothly distributed, then at most points the column density to the nearest OB star would be smaller than that to the AGN, so the stellar contribution to the heating could be large.
However, the gas is now considered to consist of clumps with large column densities.
In such a highly clumpy gas, heating by stars outside the clumps is unlikely to be effective.
Gas heating by cosmic rays works effectively only at moderate densities \citep[$10^3\,\percb{cm}$;][]{Meijerink+06}.
In the SLED analysis of \citet{Kamenetzky+14}, similar high temperatures ($10^3$\,K) with even larger column densities ($\Nco\sim10^{19.5}\,\persq{cm}$) were found for the warm component in the starburst nuclei of NGC 253 and M83.
However, the number densities in these galaxies are in the order of $10^3\,\percb{cm}$ and more than three orders of magnitude lower than the dense gas we observed through absorption.
Thus, UV photons plus cosmic rays from supernovae seem to be unlikely as a heating source.
From the high probability of X-ray heating, the absorption region is expected to be located in the XDR around the AGN.
\citet{Meijerink&Spaans05} has constructed models of chemistry in XDR.
The abundance ratio $[\mathrm{CS}]/[\mathrm{CO}]\sim10^{-2.5}$ found in Section \ref{sec:RADEX_abs} is consistent with the predictions of their high-density cases.
Note that \citet{Meijerink&Spaans05} has modeled the chemistry of PDRs as well, and the abundance can also be explained by their high-density and low-radiation-field PDR.
In this case, however, again, the large column densities we obtained are not reproducible.

At the end of Section \ref{sec:RADEX_abs}, we argue that the sub-mm absorption in the nuclear spectrum is not due to the cool component suggested from the SLED \citep{Kamenetzky+14} and that such cold gas is not distributed in front of the dust continuum source.
On the other hand, 0\farcs5-beam observations by \citet{Garcia-Burillo+15} show that CO(2--1), which should be dominated by cold gas, has a strong peak at the center.
This fact is seemingly inconsistent with the above idea.
However, the FWHM of the CO(2--1) peak is 1\arcsec\ (800\,pc), which is 10 times larger than the range we are discussing.
It is likely that CO(2--1) is intrinsically distributed non-uniformly within the peak and is not much present at the position of the dust source.

\subsection{AGN contribution fraction}
\label{sec:fAGN}

The contribution fraction of the AGN in IRAS 17208 is controversial, as mentioned in Section \ref{sec:IRAS17208}.
\citet{Gonzalez-Martin+09} identified IRAS 17208 as a CT AGN candidate based on X-ray observations, but their estimate of the hard X-ray luminosity after obscuration correction accounts for only $\fAGN=0.1\%$ of the bolometric luminosity.
On the other hand, \citet{Garcia-Burillo+15} proposed $\fAGN=30\%$ based on the observation of the molecular outflow.
As for the contribution at mid-IR wavelengths, $\fAGNMIR=17\text{--}31\%$ has been reported from spectral diagnostics \citep{Alonso-Herrero+16,Leja+18}.
Note that the stellar contribution to the luminosity is thought to arise from the circum-nuclear scale ($>$100\,pc, Section \ref{sec:disks}).

The near-IR CO absorption we observed is thought to be imprinted in the AGN-origin background.
The normalized spectrum (Figure \ref{fig:BestFit}) shows that the depth of the absorption is up to 30\%.
If the star-forming component of the continuum is more than 70\%, the absorption will not reach this depth even if it is saturated.
Therefore, we propose an AGN fraction at 4--5\,\micron\ of $\fAGNNIRl>30\%$.
This reasonably large contribution is not straightforwardly in agreement with the estimate from X-rays, but we speculate that this may be because the X-ray luminosity is not accurately estimated due to the large uncertainties in the obscuration correction.
In addition to this, the AGN fraction at 4--5\,\micron\ can be much higher than the AGN bolometric fraction because AGN-heated hot dust emission becomes very bright at these wavelengths.
\citet{Risaliti+10} argued that the ratio of 4\,\micron\ to the total luminosity is 100 times higher in AGNs than in starbursts and that, if AGN is present, it dominates the 4-\micron\ emission in most cases, even when its contribution to the bolometric luminosity is tiny.
Thus, although the AGN bolometric fraction may be low in IRAS 17208, we can reasonably assume that the CO rotational vibrational absorption occurs against the AGN component.

\subsection{Properties of the Hot Core}
\label{sec:nature}

As explained in Section \ref{sec:IRAS17208}, IRAS 17208 is a late-stage merger whose nuclear activity is not well understood.
Some studies have classified this galaxy to be starburst-dominated, while others, particularly IR studies, have claimed a significant contribution from a deeply buried AGN.
To resolve this issue, we compared the indicators obtained from the \AKARI\ spectrum (see Section \ref{sec:others}) with the diagnostics from other studies.
The 3.3\,\micron\ PAH emission is excited by UV photons, and thus, is an indicator of the star-formation activity.
If thermal emission from hot dust heated by the AGN dominates the near-IR continuum, then the PAH emission is diluted, which decreases the \EWPAH\ and turns the color red.
If the AGN is buried in dust, then its contribution is more suppressed compared to that of the foreground star-forming regions, in which case the \EWPAH\ is not much affected.
However, in this case, the color remains red, and because the energy source is more centrally concentrated than dust, the ice and dust absorption features become deeper than what can be achieved with a mixed dust and source geometry \citep{Imanishi&Maloney03,Imanishi+06_tau}.
Based on this logic, \EWPAH, \tauice, \taudust, and the continuum color have been used as indicators of an (obscured) AGN.

\citet{Inami+18} analyzed the \AKARI\ spectra of local (U)LIRGs, and proposed a classification scheme that uses \EWPAH\ and \AKARIcolor: $\EWPAH\ge0.06\,\micron$ for starbursts; $\EWPAH<0.06\,\micron$ and $\AKARIcolor\ge1.0$ for AGNs; and $\EWPAH<0.06\,\micron$ and $\AKARIcolor<1.0$ for composites.
The authors found $\EWPAH<0.049\,\micron$ and $\AKARIcolor=0.77\pm0.05$ for IRAS 17208, and thus, classified the galaxy to be a composite source.
However, the \AKARI\ spectrum they used was noisy, suffering from hot pixels.
In this study, we carefully reduced the \AKARI\ data to correct for the hot pixels (see Section \ref{sec:AKARIreduction}), and obtained $\EWPAH=0.065\pm0.006\,\micron$ and $\AKARIcolor=1.14\pm0.05$.
With these updated values, IRAS 17208 falls on the boundary between the starburst- and AGN-dominated domains.

The criterion proposed by \citet{Imanishi+08,Imanishi+10} on \EWPAH, which is indicative of a luminous AGN, is stricter than that proposed by \citet{Inami+18}, namely, $\EWPAH<0.04\,\micron$.
Therefore, IRAS 17208 fails to satisfy the criterion of \citet{Imanishi+08,Imanishi+10}.
The authors also suggested that $\tauice>0.3$, $\taudust>0.2$, and $\Gamma>1$ can be used as a signature of an obscured AGN.
However, none of these criteria are satisfied, as in this work, we measured $\tauice=0.21\pm0.03$, $\taudust<0.06$, and $\Gamma=0.31\pm0.10$.

Thus, the near-IR parameters measured in this work, \EWPAH, \tauice, \taudust, and \AKARIcolor\ (or $\Gamma$), do not provide a clear sign of an (obscured) AGN, in contrast to the moderate AGN fractions suggested in the mid-IR (Section \ref{sec:fAGN}).
We interpret the lack of AGN signatures as a result of AGN emission being severely attenuated at 2.5--4\,\micron\ due to a large amount of dusty gas and being overwhelmed by the stellar contribution.
\citet{Ichikawa+14} performed spectral decomposition for nearby IR galaxies by using \AKARI\ spectra.
For the obscured AGNs with $\tauice>0.3$ in their sample (namely, IRAS 10494$+$4424, IRAS 17028$+$5817, MCG $+$08-23-097, and Zw 453.062),\footnote{We exclude IRAS F07353$+$2903 because of its large uncertainty in $\tauice$.} the AGN fraction at 2.5--4\,\micron\ is in the range of $\fAGNNIRs=5\text{--}10\%$.\footnote{We calculated \fAGNNIRs\ and \fAGNNIRl\ from the fluxes and temperatures of the stellar, \ion{H}{2}, and dust blackbody components, along with the flux of the 3.3\,\micron\ emission.}
Therefore, it is implied that \fAGNNIRs\ of IRAS 17208 would be smaller than these galaxies ($<$5\%).
Note that this does not contradict the fraction at 4--5\,\micron\ we determined from the CO band ($\fAGNNIRl>30\%$, Section \ref{sec:fAGN}).
This is because \fAGN\ rapidly increases at longer wavelengths owing to the steepness of the extinction curve and the AGN-heated dust emission.
In fact, the AGN fraction of the obscured AGNs in \citet{Ichikawa+14} in the 4--5\,\micron\ range is as high as $\fAGNNIRl=27\text{--}40\%$.
From the AGN fractions in the two ranges and the corresponding observed fluxes of the continuum (Figure \ref{fig:IR-spec}), the slope of the AGN spectral energy distribution (SED) in IRAS 17208 can be roughly estimated to be $\Gamma^\prime>4$ around these wavelengths, where $f_{\nu,\,\mathrm{AGN}}\propto\lambda^{\Gamma^\prime}$.
In the SED models of \citet{Gonzalez-Alfonso&Sakamoto19}, calculated for an AGN buried in a spherically distributed medium that represents a hot mid-IR core, the slope at 4\,\micron\ is as steep as $\Gamma^\prime=4$ with a column density of $\NHH=10^{23}\,\persq{cm}$, which is consistent with the column obtained from our observations.

\begin{figure*}[t]
\plotone{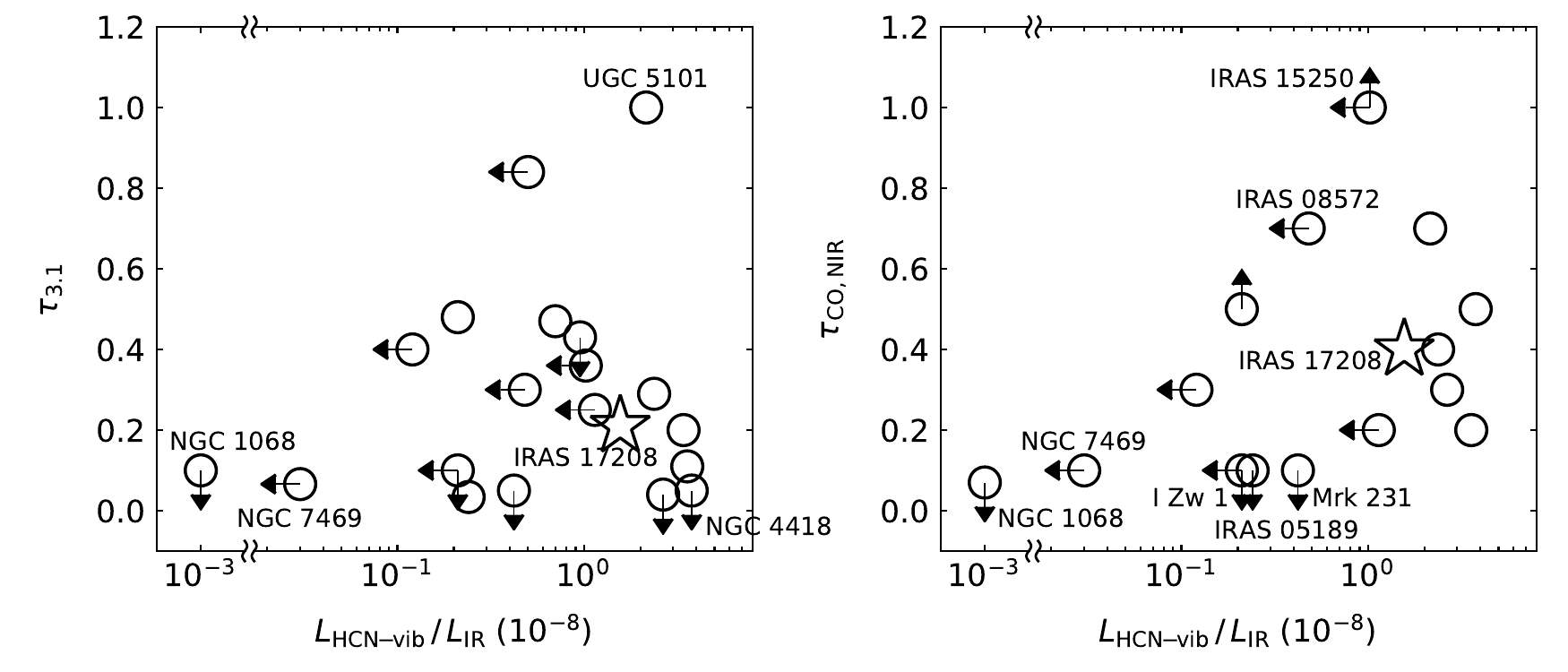}
\caption{Left panel: Optical depth of the 3.1\,\micron\ H$_2$O ice absorption (\tauice) as a function of the HCN-vib luminosity relative to the IR luminosity \citep[\LHCNvib/\LIR;][]{Falstad+19}.
Right panel: Same as left panel but for the optical depth of the near-IR 4.67\,\micron\ CO ro-vibrational absorption band (\tauCO).
The \tauCO\ values of the galaxies were observed with \AKARI/IRC ($R\sim120)$, except for NGC 1068, which was observed with \ISO/SWS ($R\sim2500$).
In both panels, IRAS 17208 is highlighted with a star symbol.
\label{fig:HCN-ice-CO}}
\end{figure*}

Low \tauice\ and high \tauCO\ may be common characteristics of heavily obscured galactic nuclei that exhibit bright HCN-vib emission.
We compare the HCN-vib $J=3\text{--}2$ line luminosity (\LHCNvib) in the literature (\citealt{Falstad+19} and references therein; \citealt{Imanishi+20}), with these two optical depths.
The collected values are tabulated in Table \ref{tab:HCN-vib}, and displayed in Figure \ref{fig:HCN-ice-CO}.
Note that the HCN-vib sample is inhomogeneous probably not without selection bias and that the presence of AGN is not definite in some galaxies.
It can be seen that, when \LHCNvib\ normalized by \LIR\ is brighter than $10^{-8}$, which is the threshold as a CON adopted in \citet{Falstad+19},\footnote{Note that a revised criterion for CON using HCN-vib surface brightness has been proposed by \citet{Falstad+21}. We use a convenient method here.} \tauice\ is lower than 0.3 except for UGC 5101, and \tauCO\ always exceeds 0.2.
Alternatively, when \LHCNvib\ normalized by \LIR\ is fainter, some galaxies have higher \tauice\ and/or insignificant \tauCO.
These results imply that, in extreme cases, the signatures of an obscured AGN are rarely observed in 2.5--4\,\micron, even if it exists (e.g., the Seyfert 2 galaxy NGC 4418).
These results also strengthen the interpretation that the near-IR CO ro-vibrational absorption takes place in the hot mid-IR core, which is responsible for the HCN-vib emission and CO(6--5) absorption observed in this study.

\begin{figure}[t]
\plotone{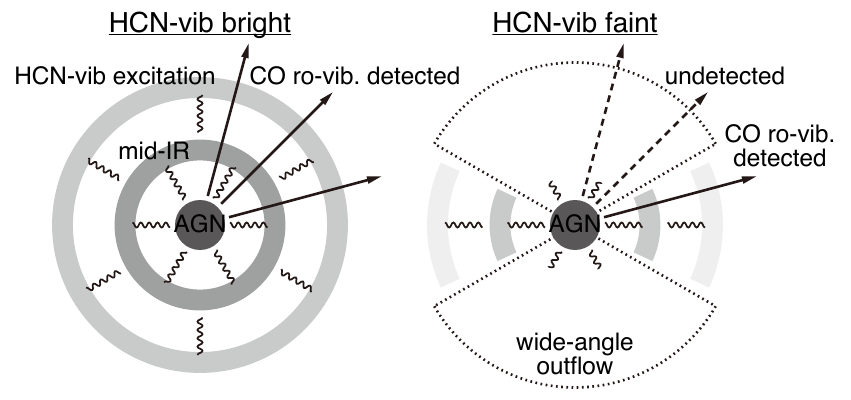}
\caption{Schematics of the structure of a galactic nucleus.
Left panel: The nucleus of a galaxy that exhibits bright HCN-vib emission.
The AGN is obscured from almost every direction, and a hot mid-IR core is formed because of the greenhouse effect.
The HCN molecules are vibrationally excited by 14\,\micron\ photons from the core.
The CO ro-vibrational absorption is frequently detected, regardless of the line of sight, because the covering factor of the obscuring medium is close to unity.
Right panel: The nucleus of a galaxy that does not exhibit bright HCN-vib emission.
The AGN is obscured in only some directions.
The greenhouse effect does not work effectively, and there is no hot mid-IR layer.
As an example, the case in which the hot mid-IR core has been disrupted by wide-angle outflows is shown.
The probability of detecting CO absorption is low because the covering factor of the obscuring medium is small.
\label{fig:HCNvib-bright-faint}}
\end{figure}

Interestingly, \tauCO\ is high in HCN-vib bright galaxies but shows diverse values in faint galaxies, even in obscured ones, as indicated by non-detections in archetypical type-2 AGNs (NGC 1068 and IRAS 05189$-$2524) and detections in LINER-like obscured AGNs (IRAS 08572$+$3915 and IRAS 15250$+$3609).
This may be due to a geometrical effect of the obscuring structure, which is illustrated in Figure \ref{fig:HCNvib-bright-faint}.
First, in HCN-vib luminous systems, the obscuring medium probably covers almost all the solid angles such that the mid-IR radiation field, which is responsible for the HCN pumping, is enhanced by the ``greenhouse effect'' \citep{Gonzalez-Alfonso&Sakamoto19}.
Because the covering factor is extemely close to unity, the CO absorption would be detected regardless of the direction of the line of sight (left panel of Figure \ref{fig:HCNvib-bright-faint}).
Second, if HCN-vib is weak in obscured AGNs, the obscuring medium is likely far from spherical.
For example, disturbed or toroidal morphologies could appear before the hot core is completely formed or after it is disrupted by the onset of AGN winds.
The small covering factor decreases the occurrence rate of the CO absorption (right panel of Figure \ref{fig:HCNvib-bright-faint}).
The presence/absence of CO absorption is perhaps related to the extent to which the nuclear obscuring structure is anisotropic or disturbed.

Again, we consider that the deep and broad CO ro-vibrational absorption probes the gas in the XDR around the AGN.
This is because high \Nco\ along with high \Tco\ can be best explained by X-ray heating.
In addition, an area covering fraction close to unity is indicative of an absorbing medium lying just in front of the background continuum source, rather than being distributed randomly and away from the source.
Note that IRAS 17208 is a late-stage merger that has an outflow that is probably AGN-driven \citep{Garcia-Burillo+15}.
Thus, it may be possible to trace the state of a critical phase in the merger evolutionary scenario with the CO band, where the SMBH rapidly grows and the AGN feedback becomes effective.
To investigate the motion of the CO gas, we first have to determine the blueshift with respect to the systemic velocity.
Consequently, it is important to resolve the different $J$ lines with a resolution of $R\sim1000$.
This type of observations will be possible with \JWST.

\begin{deluxetable*}{lhDDDDc}
\tablecaption{Comparison of $\LHCNvib/\LIR$, \tauice, and \tauCO\label{tab:HCN-vib}}
\tablewidth{0pt}
\tablehead{
\colhead{Object} & \nocolhead{Type} & \multicolumn2c{$z$} & \multicolumn2c{$\LHCNvib/\LIR$} & \multicolumn2c{\tauice} & \multicolumn2c{\tauCO} & \colhead{Ref.} \\
\colhead{}       & \nocolhead{}     & \multicolumn2c{}    & \multicolumn2c{($10^{-8}$)}     & \multicolumn2c{}        & \multicolumn2c{}       & \colhead{}
}
\decimalcolnumbers
\startdata
IRAS 17208$-$0014                & \textsc{Hii} & 0.0428 &  1.56\tablenotemark{a} &  0.21                  &  0.4                     & 1,1 \\
\hline
NGC 4418                         & Sy2          & 0.0071 &  3.77                  & <0.05                  &  0.5                     & 2,8 \\
Zw 049.057 (CGCG 049-057)        & \textsc{Hii} & 0.0130 &  3.57                  &  0.11                  &  0.2                     & 2,8 \\
IRAS 12224$-$0624                & L            & 0.0264 &  3.41\tablenotemark{a} &  0.20                  & \nodata\tablenotemark{d} & 2,6 \\
IC 860                           & \textsc{Hii} & 0.0130 &  2.65                  & <0.04                  &  0.3                     & 2,6 \\
Arp 220 W                        & L            & 0.0181 &  2.37                  &  0.29\tablenotemark{b} &  0.4\tablenotemark{b}    & 2,8 \\
Arp 220 E                        & {          } & {    } &  0.51                  & {                    } & {                      } &     \\
UGC 5101                         & L            & 0.0394 &  2.15                  &  1.0                   &  0.7                     & 3,7 \\
IRAS 22491$-$1808 E              & \textsc{Hii} & 0.0778 &  0.95                  & <0.43\tablenotemark{b} & \nodata\tablenotemark{e} & 2,6 \\
IRAS 12112$+$0305 NE             & L            & 0.0733 &  0.70                  &  0.47\tablenotemark{b} & \nodata\tablenotemark{e} & 2,6 \\
Mrk 231 (UGC 8058)               & Sy1          & 0.0422 &  0.42                  & <0.05                  & <0.1\tablenotemark{c}    & 2,6 \\
IRAS 05189$-$2524                & Sy2          & 0.0428 &  0.24\tablenotemark{a} &  0.034                 & <0.1\tablenotemark{c}    & 4,7 \\
IRAS 20551$-$4250 (ESO 286-IG19) & \textsc{Hii} & 0.0430 &  0.21                  &  0.48                  & >0.5                     & 2,8 \\
Mrk 273                          & Sy2          & 0.0378 & <1.14                  &  0.25                  &  0.2                     & 2,7 \\
IRAS 15250$+$3609                & L            & 0.0552 & <1.02                  &  0.36                  & >1.0                     & 2,8 \\
IRAS 20414$-$1651                & \textsc{Hii} & 0.0871 & <0.50                  &  0.84                  & \nodata\tablenotemark{e} & 2,8 \\
IRAS 08572$+$3915 NW             & L            & 0.0584 & <0.48                  &  0.3                   &  0.7                     & 3,7 \\
I Zw 1 (PG 0050$+$124)           & Sy1          & 0.0610 & <0.21                  & <0.1\tablenotemark{c}  & <0.1                     & 5,9 \\
IRAS 13120$-$5453                & Sy2          & 0.0308 & <0.12\tablenotemark{a} &  0.4\tablenotemark{c}  &  0.3\tablenotemark{c}    & 6,6 \\
NGC 7469                         & Sy1          & 0.0164 & <0.03                  &  0.066                 &  0.1\tablenotemark{c}    & 2,6 \\
\hline
NGC 1068                         & Sy2          & 0.0038 &  0.001                 & <0.1\tablenotemark{c}  & <0.07                    & 10,10
\enddata
\tablecomments{
Column 1: object name.
          Names used in the references are also listed.
Column 2: redshift.
Column 3: ratio of the line luminosity of the HCN-vib $J=3\text{--}2$ line to the total IR luminosity listed in \citet{Falstad+19}, except for NGC 1068, where the value is taken from \citet{Imanishi+20}.
          If only the $J=4\text{--}3$ line is available, then \LHCNvib is scaled down by a factor of 2.5, as in \citet{Falstad+19}.
Column 4: optical depth of the 3.1\,\micron\ H$_2$O ice absorption.
Column 5: optical depth of the 4.67\,\micron\ CO absorption when observed with \AKARI/IRC ($R\sim120)$, except for NGC 1068, which was observed with \ISO/SWS ($R\sim2500$).
Column 6: references for Columns 4 and 5.
    (1) This work;
    (2) \citet{Yamada+13};
    (3) \citet{Imanishi+08};
    (4) \citet{Doi+19};
    (5) \citet{Kim+15};
    (6) \citet{Inami+18};
    (7) \citet{Baba+18};
    (8) \citet{Imanishi+10};
    (9) IRC point source spectral catalogue;
    (10) \citet{Lutz+04}.}
\tablenotetext{a}{Scaled down from the data of the $J=4\text{--}3$ line by a factor of 2.5, as in \citet{Falstad+19}}
\tablenotetext{b}{Multiple nuclei are not resolved by \AKARI.}
\tablenotetext{c}{Estimated by us from the spectrum presented in the reference.}
\tablenotetext{d}{Noisy spectrum to measure \tauCO.}
\tablenotetext{e}{CO is not covered by \AKARI\ because of extremely high $z$.}
\end{deluxetable*}

\subsection{Comparison with Nuclear Structures in the Literature}
\label{sec:disks}

\begin{figure}[t]
\plotone{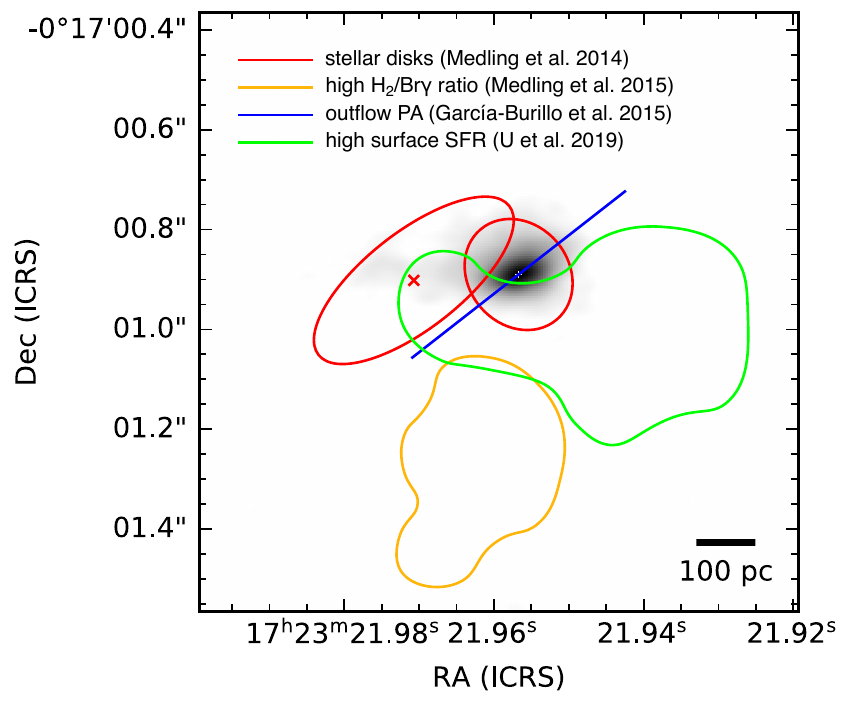}
\caption{Comparison of the different nuclear structures of IRAS 17208.
The underlying gray image is the 436\,\micron\ dust intensity map (see Figure \ref{fig:continuum}).
The white plus symbol marks the dust peak.
The red ellipses represent the effective radii and orientations of the stellar disks observed by \citet{Medling+14} from the near-IR continuum.
The position of the W disk is aligned with our dust peak.
The blue line indicates the axis of the molecular outflow derived by \citet{Garcia-Burillo+15}.
The orange and green curves roughly denote the regions of high \HH(2.12\,\micron)/Br$\gamma$ ratio observed by \citet{Medling+15} and high surface star-formation rate measured by \citet{U+19}, respectively.
The locations of these regions were aligned based on the position of the W disk.
\label{fig:NuclearStructures}}
\end{figure}

Our observations provide a clear picture of the distribution of the dust and gas in the central 100\,pc of IRAS 17208.
We use it as a positional reference for comparison with various nuclear structures reported in the literature: stellar disks, outflows, and starburst regions.
The positions of these structures are shown in Figure \ref{fig:NuclearStructures}.

\citet{Medling+14} revealed two overlapping nuclear disks with a projected separation of 200\,pc from near-IR integral field spectroscopy performed with OSIRIS\footnote{The OH-Suppressing Infra-Red Imaging Spectrograph.} on the W. M. Keck II telescope.
The angular resolution of their observations was similar to that of our ALMA observation.
The western disk is brighter and smaller than the eastern disk in both the continuum and Br$\gamma$ fluxes.
\citet{Aalto+15_4ULIRGs} interpreted that the HCN-vib emission is primarily associated with the W nucleus.
Unfortunately, we cannot directly compare the dust peak position of our observation with the disk positions observed by \citet{Medling+14}, because the OSIRIS data do not provide absolute coordinates.
Thus, we attribute our dust peak to the W nucleus, and the faint elongated emission to the E disk.

The kinematic major axes of the W and E disks measured by \citet{Medling+14} are oriented at $\mathrm{PA}=40\arcdeg$ and $128\arcdeg$, respectively.
In both the disks, the western side is the approaching side.
The direction of the large-scale gradient of our CO(6--5) velocity map (see Figure \ref{fig:velocity}) is approximately $\mathrm{PA}\sim95\arcdeg$.
This probably reflects the velocity field of the entire system comprising the two disks.
Unfortunately, it is currently difficult to reliably decouple the kinematics of the two disks.

\citet{U+19} also analyzed the Keck/OSIRIS integral-field spectroscopic data and calculated the star formation rate for each point from the Br$\gamma$ flux after correcting for dust effects.
They found that the region with a high surface star formation rate is located southwest of the W nucleus.
The size of the region is several hundred parsecs.
The stellar contribution to the bolometric luminosity of IRAS 17208 is thought to originate from this region.

The existence of an outflow is of particular interest.
In the ionized phase, a broad outflowing component that has velocities up to $-$300\,\kmps\ was found from the H$\alpha$ line \citep{Arribas+14}.
In addition, neutral gas outflows blueshifted by up to $-$650\,\kmps\ were identified from \ion{Na}{1} D absorption lines \citep{Rupke&Veilleux13}.
Moving to the molecular phase, in the observations by \citet{Medling+14} and \citet{Medling+15}, the 2.12\,\micron\ \HH\ line flux shows a morphology that is different from that of the continuum or the Br$\gamma$ emission.
It extends south from the system to approximately 400\,pc with a \HH/Br$\gamma$ ratio as high as 4--8.
\citet{Medling+15} attributed this to strong shocks at the base of a starburst-driven wind, which is launched from either of the two nuclei at about 400\,\kmps.
An outflow that appears to be AGN-driven has also been found, by \citet{Garcia-Burillo+15} from their CO(2--1) observations.
They measured the outflow direction as $\mathrm{PA}\sim308\arcdeg$ and tentatively identified the W nucleus as the source of the outflow because the direction and the disk axis are almost perpendicular.
We also propose that the outflow is launched from the W disk, because our high spatial resolution observations show that the AGN is almost certainly buried in the W disk.
The outflow component found in CO(2--1) has two knots located at $r\sim$160\,kpc in the southwest and northeast, and blue- and red-shifted by $\sim$500\,\kmps, respectively.
The CO(6--5) and CS(14--13) lines in our beam-sized ($r\sim20$\,pc) nuclear spectrum (Figure \ref{fig:submm-spec}, lower panel) appear to be slightly blueshift by about $-50$\,\kmps\ from \Vsys.
Unfortunately, this small shift is not very reliable due to the uncertainty in the difference between \Vsys, which was determined with a 0\farcs5 beam not separating the two nuclei \citep{Garcia-Burillo+15}, and the actual systemic velocity of the W nucleus.
If the blueshift is real, it might be associated with the known outflows although the spatial and velocity scales are quite different.
At this time, their detailed relation is unknown including if it exists.

\section{Summary}
\label{sec:summary}

In this paper, we have presented the results of high-spatial-resolution ($\sim0\farcs04\sim32\,\mathrm{pc}$) ALMA observations of CO($J$=6--5), the 436\,\micron\ continuum, and serendipitously detected CS($J$=14--13) of a late-stage merger IRAS 17208, as well as \AKARI\ spectroscopic observations in the 2.5--5.0\,\micron wavelength range.

The dust continuum peak has a high brightness temperature of 106\,K and coincides with the dynamical center reported previously in the literature, which can be regarded as the location of the CT AGN in IRAS 17208.
At this position, the CO(6--5) and CS(14--13) lines are detected in absorption.
In the surrounding region, CO emission is observed with a rotating pattern, but CS emission is not detected.

By comparing the observed intensity of CO(6--5) emission with non-LTE simulations performed with RADEX, the gas column density in the circumnuclear region ($r=40$--80\,pc) is estimated to be $\NHH=(0.6\text{--}2)\times10^{23}\,\persq{cm}$.
However, the gas parameters obtained in this region cannot simultaneously explain the intensity of CO absorption in the central beam ($r=20$\,pc).
The gas at the center is indicated to be dense ($\gtrsim1\times10^7\,\percb{cm}$) and warm ($\gtrsim200$\,K) by the detection of CS(14--13), and its column density is constrained to $\NHH>10^{23}\,\persq{cm}$ from RADEX calculations.
Under the assumption that the gas temperature is highest at the center, the central column density has an excess of $>$0.3\,dex over the circumnuclear region, indicating that there is a concentrated structure around the nucleus.
The map of CS(14--13) reinforces this picture.
Our results are consistent with the presence of a hot mid-IR core predicted from the bright HCN-vib emission observed at an angular resolution one order of magnitude larger than in our work \citep{Aalto+15_4ULIRGs}.

The \AKARI\ spectrum obtained after careful correction for hot-pixel effects has revealed deep and broad CO $v\!=\!1\!\leftarrow\!0$ $\Delta J\!=\!\pm1$ ro-vibrational absorption at 4.67\,\micron.
Its band profile is well represented by an LTE gas model with $\NHH=(0.7\text{--}2)\times10^{23}\,\persq{cm}$ and $\Tex\sim1000\,\mathrm{K}$.
The fact that LTE is established over a wide wavelength range (i.e., up to high $J$ levels) suggests a high gas density ($\gtrsim3\times10^6\,\percb{cm}$).

The properties of the gas observed through absorption in the sub-mm and near-IR are similar, but the temperature is higher in the latter.
This may mean that the regions probed in the two wavelength ranges are roughly coincident, but that in the near-IR, the hot region close to the nucleus is also included, because the effective background source should be hotter and smaller than in the sub-mm.

The most plausible heating source for the observed dense and warm gas of a large column density is X-rays.
The absorption region is considered to be located within the XDR formed around the AGN.
The abundance ratio of $[\mathrm{CS}]/[\mathrm{CO}]\sim10^{-2.5}$ found from the sub-mm absorption intensities is also consistent with the chemistry in dense XDRs.
The AGN power is evaluated from our observations to have a contribution of more than 30\% at 4--5\,\micron, based on the depth of the CO band.

Despite the suggested presence of an AGN, the \EWPAH, \tauice, \taudust, and continuum color measured from the \AKARI\ spectrum are not indicative of an (obscured) AGN.
This is probably because the AGN is severely attenuated at 2.5--4\,\micron\ and overwhelmed by the starburst contribution.
From a comparison of archival $\LHCNvib/\LIR$ and \tauice\ values, it is found that high \tauice\ indicative of an obscured AGN is rarely observed in HCN-vib bright galaxies, even where the presence of AGN has been confirmed by other means.
This result implies that the lack of AGN signatures at 2.5--4\,\micron\ is usual for AGNs in hot mid-IR cores.

It is also found that high optical depth of CO ro-vibrational absorption (\tauCO) is more frequently observed in HCN-vib bright galaxies than in faint ones, supporting the interpretation that the CO absorption is occurring in the hot mid-IR core.
Also, \tauCO\ is found to be diverse in HCN-vib faint galaxies, even those hosting obscured AGNs.
This can be interpreted as a geometrical effect of an obscuring structure that does not cover the entire solid angle.

Based on the morphology of the dust continuum emission, we conclude that the dust peak, and thus the CT AGN, belongs to the western one of the two nuclear disks revealed by \citet{Medling+14}.
The sub-mm CO(6--5) and CS(14--13) absorption lines appear to show a slight blueshift ($\sim-50$\,\kmps), and it might be a sign of an outflow if real, although the relation to known outflows is unclear.

\acknowledgments

We thank the anonymous referee for insightful comments, which helped us significantly improve the clarity of the arguments in this manuscript.
We thank Dr. A. Medling for providing details regarding the OSIRIS observation of IRAS 17208.
This paper makes use of the following ALMA data: ADS/JAO.ALMA\#2016.1.01223.S.
ALMA is a partnership of ESO (representing its member states), NSF (USA) and NINS (Japan), together with NRC (Canada), MOST and ASIAA (Taiwan), and KASI (Republic of Korea), in cooperation with the Republic of Chile.
The Joint ALMA Observatory is operated by ESO, AUI/NRAO and NAOJ.
This research is based on observations with AKARI,a JAXA project with the participation of ESA.
Data analysis was in part carried out on the Multi-wavelength Data Analysis System operated by the Astronomy Data Center (ADC), National Astronomical Observatory of Japan.
This work is supported by JSPS KAKENHI Grant Number JP19J00892.
S.B. is partially supported by a Grant-in-Aid for Scientific Research (A) of JSPS (KAKENHI 21H04496).
T.K. is supported from Fondecyt postdoctoral fellowship 3200470.


\vspace{5mm}
\facilities{ALMA, \AKARI(IRC)}


\software{
CASA \citep{CASA},
IRC Spectroscopy Toolkit \citep{Ohyama+07,Baba+16,Baba+19},
MPFIT \citep{MPFIT},
RADEX \citep{RADEX},
pyradex (\url{https://github.com/keflavich/pyradex}),
IPython \citep{IPython},
Jupyter Notebook \citep{JupyterNotebook},
NumPy \citep{NumPy},
SciPy \citep{SciPy},
Pandas \citep{Pandas},
Matplotlib \citep{Matplotlib},
Astropy \citep{Astropy,Astropy_v2},
APLpy \citep{APLpy,APLpy_v2}
}


\clearpage

\appendix

\section{Channel Maps}
\label{sec:cm}

Figure \ref{sec:cm} shows velocity channel maps of CO(6--5).
The line emission (red contours) shows a rotating pattern.
The line absorption (blue contours) is detected continuously over multiple channels only at the dust continuum peak (black plus sign).

\begin{figure}[t]
\plotone{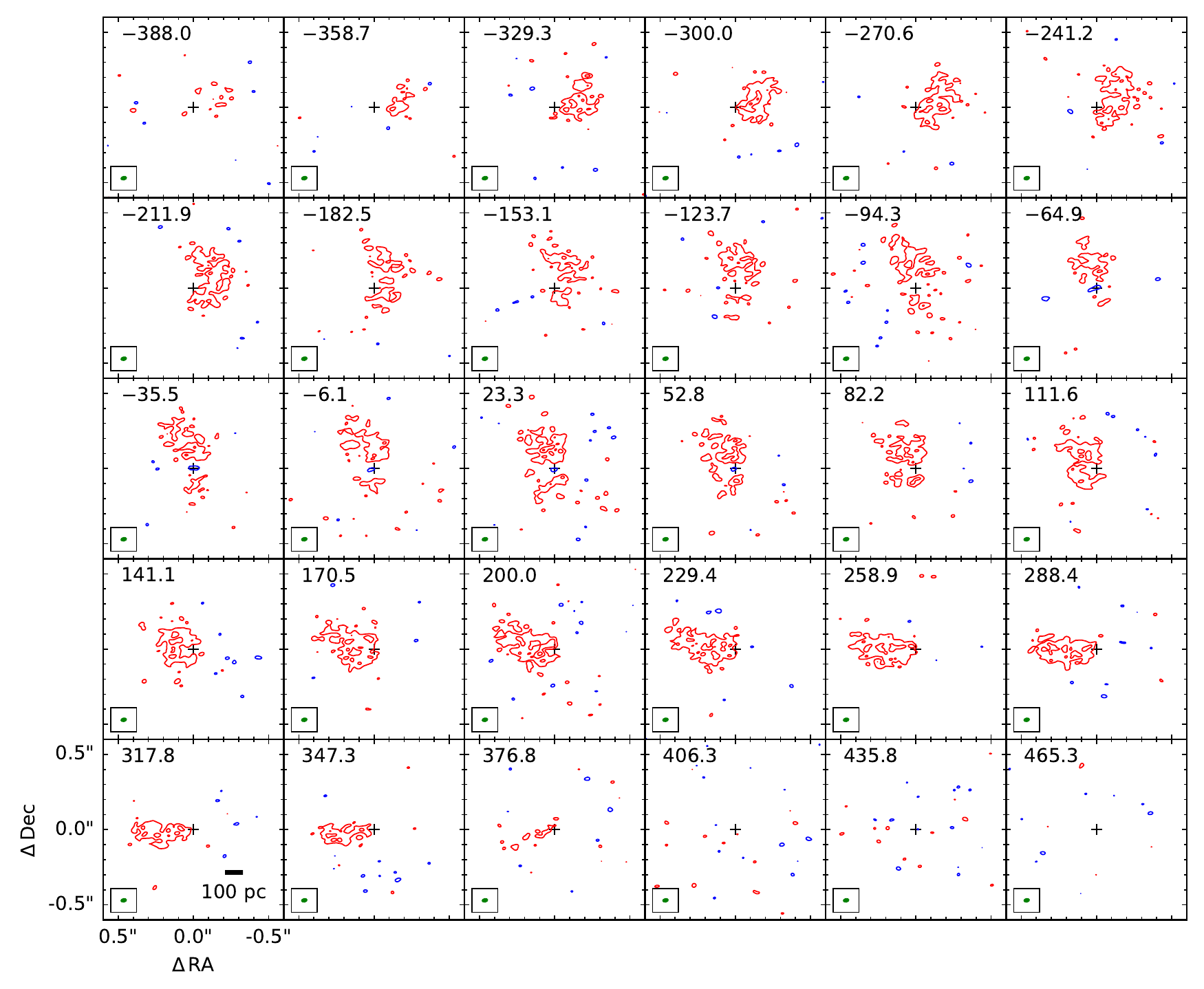}
\caption{Velocity channel maps of CO(6--5) in the central 1\farcs2$\times$1\farcs2 region.
The coordinates are expressed as offsets from the dust continuum peak, which is indicated by the black plus sign.
The velocity of each channel relative to \Vsys\ is written at the upper left corner of each panel in units of \kmps.
The red/blue contours show positive/negative intensities at $\pm$4, 8, and 12$\sigma$, where $\sigma$ is the blank sky rms noise of each channel, $1\sigma\sim4\,\mathrm{mJy\,\per{beam}}$.
The green ellipse at the lower left indicates the size of the synthesized beam.
Channels with velocities above 490\,\kmps\ are omitted because they do not show significant emission or absorption.
Significant absorption is observed at the dust peak continuously from $-$65 to 53\,\kmps.
\label{fig:cm}}
\end{figure}

\section{Position-Velocity Diagrams}
\label{sec:PV}

\begin{figure}[t]
\plotone{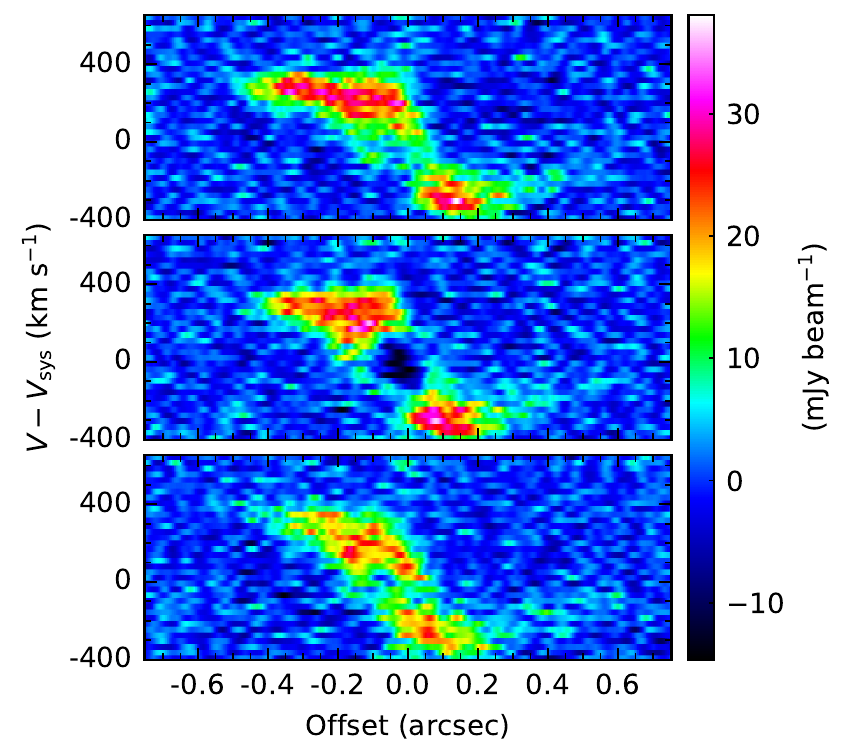}
\caption{PV diagrams for CO(6--5) created along the major axis ($\mathrm{PA}=95\arcdeg$; see Figure \ref{fig:velocity}) at three positions shifted from the center by $+0\farcs05$ (north; top panel), 0\arcsec\ (nucleus; middle panel), and $-0\farcs05$ (south; bottom panel).
The position origins are set on the minor axis.
For each panel, an averaging width of $0\farcs045$ was used.
The maximum and minimum intensities are 38 and $-$15\,mJy\,\per{beam}, respectively, and the rms noise in the blank sky is $\sim$3\,mJy\,\per{beam}.
\label{fig:PV}}
\end{figure}

In Figure \ref{fig:PV}, we present the PV diagrams of CO(6--5) created along the major axis ($\mathrm{PA}=95\arcdeg$) at the nuclear position and two off-nuclear positions shifted by 0\farcs05 to the north and south.
At each position, the CO data cube was averaged with a width of 0\farcs045 to increase the S/N.
We observe rotation patterns in all three PV diagrams.
Moreover, CO(6--5) is clearly seen in absorption at the nucleus.
The presence of the bright spots at $\pm0\farcs2$ and $\pm200\,\kmps$ relative to the systemic velocity (12824\,\kmps) is similar to those observed in the PV diagram of CO(2--1) presented by \citet{Garcia-Burillo+15}.
The pattern, in which the intensity is suppressed at the systemic velocity, resembles that in the PV diagrams of HCN(4--3) and HCO$^+$(4--3) presented by \citet{Aalto+15_4ULIRGs}.
Note that the position angles of the PV diagrams in these two works differ slightly from that in our work, namely, by $+8\arcdeg$ and $-5\arcdeg$, respectively.

\bibliography{reference}{}

\begin{thebibliography}{}
\expandafter\ifx\csname natexlab\endcsname\relax\def\natexlab#1{#1}\fi
\providecommand{\url}[1]{\href{#1}{#1}}
\providecommand{\dodoi}[1]{doi:~\href{http://doi.org/#1}{\nolinkurl{#1}}}
\providecommand{\doeprint}[1]{\href{http://ascl.net/#1}{\nolinkurl{http://ascl.net/#1}}}
\providecommand{\doarXiv}[1]{\href{https://arxiv.org/abs/#1}{\nolinkurl{https://arxiv.org/abs/#1}}}

\bibitem[{{Aalto} {et~al.}(2015{\natexlab{a}}){Aalto}, {Garcia-Burillo},
  {Muller}, {Winters}, {Gonzalez-Alfonso}, {van der Werf}, {Henkel},
  {Costagliola}, \& {Neri}}]{Aalto+15_Mrk231}
{Aalto}, S., {Garcia-Burillo}, S., {Muller}, S., {et~al.} 2015{\natexlab{a}},
  \aap, 574, A85, \dodoi{10.1051/0004-6361/201423987}

\bibitem[{{Aalto} {et~al.}(2015{\natexlab{b}}){Aalto}, {Mart{\'\i}n},
  {Costagliola}, {Gonz{\'a}lez-Alfonso}, {Muller}, {Sakamoto}, {Fuller},
  {Garc{\'\i}a-Burillo}, {van der Werf}, {Neri}, {Spaans}, {Combes}, {Viti},
  {M{\"u}hle}, {Armus}, {Evans}, {Sturm}, {Cernicharo}, {Henkel}, \&
  {Greve}}]{Aalto+15_4ULIRGs}
{Aalto}, S., {Mart{\'\i}n}, S., {Costagliola}, F., {et~al.} 2015{\natexlab{b}},
  \aap, 584, A42, \dodoi{10.1051/0004-6361/201526410}

\bibitem[{{Aalto} {et~al.}(2019){Aalto}, {Muller}, {K{\"o}nig}, {Falstad},
  {Mangum}, {Sakamoto}, {Privon}, {Gallagher}, {Combes}, {Garc{\'\i}a-Burillo},
  {Mart{\'\i}n}, {Viti}, {van der Werf}, {Evans}, {Black}, {Varenius},
  {Beswick}, {Fuller}, {Henkel}, {Kohno}, {Alatalo}, \& {M{\"u}hle}}]{Aalto+19}
{Aalto}, S., {Muller}, S., {K{\"o}nig}, S., {et~al.} 2019, \aap, 627, A147,
  \dodoi{10.1051/0004-6361/201935480}

\bibitem[{{Alonso-Herrero} {et~al.}(2016){Alonso-Herrero}, {Poulton}, {Roche},
  {Hern{\'a}n-Caballero}, {Aretxaga}, {Mart{\'\i}nez-Paredes}, {Ramos Almeida},
  {Pereira-Santaella}, {D{\'\i}az-Santos}, {Levenson}, {Packham}, {Colina},
  {Esquej}, {Gonz{\'a}lez-Mart{\'\i}n}, {Ichikawa}, {Imanishi}, {Rodr{\'\i}guez
  Espinosa}, \& {Telesco}}]{Alonso-Herrero+16}
{Alonso-Herrero}, A., {Poulton}, R., {Roche}, P.~F., {et~al.} 2016, \mnras,
  463, 2405, \dodoi{10.1093/mnras/stw2031}

\bibitem[{{Arribas} \& {Colina}(2003)}]{Arribas&Colina03}
{Arribas}, S., \& {Colina}, L. 2003, \apj, 591, 791, \dodoi{10.1086/375417}

\bibitem[{{Arribas} {et~al.}(2014){Arribas}, {Colina}, {Bellocchi}, {Maiolino},
  \& {Villar-Mart{\'\i}n}}]{Arribas+14}
{Arribas}, S., {Colina}, L., {Bellocchi}, E., {Maiolino}, R., \&
  {Villar-Mart{\'\i}n}, M. 2014, \aap, 568, A14,
  \dodoi{10.1051/0004-6361/201323324}

\bibitem[{{Astropy Collaboration} {et~al.}(2013){Astropy Collaboration},
  {Robitaille}, {Tollerud}, {Greenfield}, {Droettboom}, {Bray}, {Aldcroft},
  {Davis}, {Ginsburg}, {Price-Whelan}, {Kerzendorf}, {Conley}, {Crighton},
  {Barbary}, {Muna}, {Ferguson}, {Grollier}, {Parikh}, {Nair}, {Unther},
  {Deil}, {Woillez}, {Conseil}, {Kramer}, {Turner}, {Singer}, {Fox}, {Weaver},
  {Zabalza}, {Edwards}, {Azalee Bostroem}, {Burke}, {Casey}, {Crawford},
  {Dencheva}, {Ely}, {Jenness}, {Labrie}, {Lim}, {Pierfederici}, {Pontzen},
  {Ptak}, {Refsdal}, {Servillat}, \& {Streicher}}]{Astropy}
{Astropy Collaboration}, {Robitaille}, T.~P., {Tollerud}, E.~J., {et~al.} 2013,
  \aap, 558, A33, \dodoi{10.1051/0004-6361/201322068}

\bibitem[{{Astropy Collaboration} {et~al.}(2018){Astropy Collaboration},
  {Price-Whelan}, {Sip{\H{o}}cz}, {G{\"u}nther}, {Lim}, {Crawford}, {Conseil},
  {Shupe}, {Craig}, {Dencheva}, {Ginsburg}, {Vand erPlas}, {Bradley},
  {P{\'e}rez-Su{\'a}rez}, {de Val-Borro}, {Aldcroft}, {Cruz}, {Robitaille},
  {Tollerud}, {Ardelean}, {Babej}, {Bach}, {Bachetti}, {Bakanov}, {Bamford},
  {Barentsen}, {Barmby}, {Baumbach}, {Berry}, {Biscani}, {Boquien}, {Bostroem},
  {Bouma}, {Brammer}, {Bray}, {Breytenbach}, {Buddelmeijer}, {Burke},
  {Calderone}, {Cano Rodr{\'\i}guez}, {Cara}, {Cardoso}, {Cheedella}, {Copin},
  {Corrales}, {Crichton}, {D'Avella}, {Deil}, {Depagne}, {Dietrich}, {Donath},
  {Droettboom}, {Earl}, {Erben}, {Fabbro}, {Ferreira}, {Finethy}, {Fox},
  {Garrison}, {Gibbons}, {Goldstein}, {Gommers}, {Greco}, {Greenfield},
  {Groener}, {Grollier}, {Hagen}, {Hirst}, {Homeier}, {Horton}, {Hosseinzadeh},
  {Hu}, {Hunkeler}, {Ivezi{\'c}}, {Jain}, {Jenness}, {Kanarek}, {Kendrew},
  {Kern}, {Kerzendorf}, {Khvalko}, {King}, {Kirkby}, {Kulkarni}, {Kumar},
  {Lee}, {Lenz}, {Littlefair}, {Ma}, {Macleod}, {Mastropietro}, {McCully},
  {Montagnac}, {Morris}, {Mueller}, {Mumford}, {Muna}, {Murphy}, {Nelson},
  {Nguyen}, {Ninan}, {N{\"o}the}, {Ogaz}, {Oh}, {Parejko}, {Parley}, {Pascual},
  {Patil}, {Patil}, {Plunkett}, {Prochaska}, {Rastogi}, {Reddy Janga},
  {Sabater}, {Sakurikar}, {Seifert}, {Sherbert}, {Sherwood-Taylor}, {Shih},
  {Sick}, {Silbiger}, {Singanamalla}, {Singer}, {Sladen}, {Sooley},
  {Sornarajah}, {Streicher}, {Teuben}, {Thomas}, {Tremblay}, {Turner},
  {Terr{\'o}n}, {van Kerkwijk}, {de la Vega}, {Watkins}, {Weaver}, {Whitmore},
  {Woillez}, {Zabalza}, \& {Astropy Contributors}}]{Astropy_v2}
{Astropy Collaboration}, {Price-Whelan}, A.~M., {Sip{\H{o}}cz}, B.~M., {et~al.}
  2018, \aj, 156, 123, \dodoi{10.3847/1538-3881/aabc4f}

\bibitem[{{Baba} {et~al.}(2018){Baba}, {Nakagawa}, {Isobe}, \&
  {Shirahata}}]{Baba+18}
{Baba}, S., {Nakagawa}, T., {Isobe}, N., \& {Shirahata}, M. 2018, \apj, 852,
  83, \dodoi{10.3847/1538-4357/aa9f25}

\bibitem[{{Baba} {et~al.}(2019){Baba}, {Nakagawa}, {Usui}, {Yamagishi}, \&
  {Onaka}}]{Baba+19}
{Baba}, S., {Nakagawa}, T., {Usui}, F., {Yamagishi}, M., \& {Onaka}, T. 2019,
  \pasj, 71, 2, \dodoi{10.1093/pasj/psy131}

\bibitem[{{Baba} {et~al.}(2016){Baba}, {Nakagawa}, {Shirahata}, {Isobe},
  {Usui}, {Ohyama}, {Onaka}, {Yano}, \& {Kochi}}]{Baba+16}
{Baba}, S., {Nakagawa}, T., {Shirahata}, M., {et~al.} 2016, \pasj, 68, 27,
  \dodoi{10.1093/pasj/psw013}

\bibitem[{{Baker} \& {Menzel}(1938)}]{Baker&Menzel}
{Baker}, J.~G., \& {Menzel}, D.~H. 1938, \apj, 88, 52, \dodoi{10.1086/143959}

\bibitem[{{Barcos-Mu{\~n}oz} {et~al.}(2018){Barcos-Mu{\~n}oz}, {Aalto},
  {Thompson}, {Sakamoto}, {Mart{\'\i}n}, {Leroy}, {Privon}, {Evans}, \&
  {Kepley}}]{Barcos-Munoz+18}
{Barcos-Mu{\~n}oz}, L., {Aalto}, S., {Thompson}, T.~A., {et~al.} 2018, \apjl,
  853, L28, \dodoi{10.3847/2041-8213/aaa28d}

\bibitem[{{Barnes} \& {Hernquist}(1996)}]{Barnes&Hernquist96}
{Barnes}, J.~E., \& {Hernquist}, L. 1996, \apj, 471, 115,
  \dodoi{10.1086/177957}

\bibitem[{{Barnes} \& {Hernquist}(1991)}]{Barnes&Hernquist91}
{Barnes}, J.~E., \& {Hernquist}, L.~E. 1991, \apjl, 370, L65,
  \dodoi{10.1086/185978}

\bibitem[{{Borne} {et~al.}(2000){Borne}, {Bushouse}, {Lucas}, \&
  {Colina}}]{Borne+00}
{Borne}, K.~D., {Bushouse}, H., {Lucas}, R.~A., \& {Colina}, L. 2000, \apjl,
  529, L77, \dodoi{10.1086/312461}

\bibitem[{{Bushouse} {et~al.}(2002){Bushouse}, {Borne}, {Colina}, {Lucas},
  {Rowan-Robinson}, {Baker}, {Clements}, {Lawrence}, \& {Oliver}}]{Bushouse+02}
{Bushouse}, H.~A., {Borne}, K.~D., {Colina}, L., {et~al.} 2002, \apjs, 138, 1,
  \dodoi{10.1086/324019}

\bibitem[{{Cami}(2002)}]{Cami02}
{Cami}, J. 2002, PhD thesis, University of Amsterdam

\bibitem[{{Colina} {et~al.}(2001){Colina}, {Borne}, {Bushouse}, {Lucas},
  {Rowan-Robinson}, {Lawrence}, {Clements}, {Baker}, \& {Oliver}}]{Colina+01}
{Colina}, L., {Borne}, K., {Bushouse}, H., {et~al.} 2001, \apj, 563, 546,
  \dodoi{10.1086/323945}

\bibitem[{{Cui} {et~al.}(2001){Cui}, {Xia}, {Deng}, {Mao}, \& {Zou}}]{Cui+01}
{Cui}, J., {Xia}, X.~Y., {Deng}, Z.~G., {Mao}, S., \& {Zou}, Z.~L. 2001, \aj,
  122, 63, \dodoi{10.1086/321127}

\bibitem[{{Davies} {et~al.}(2015){Davies}, {Burtscher}, {Rosario},
  {Storchi-Bergmann}, {Contursi}, {Genzel}, {Graci{\'a}-Carpio}, {Hicks},
  {Janssen}, {Koss}, {Lin}, {Lutz}, {Maciejewski}, {M{\"u}ller-S{\'a}nchez},
  {Orban de Xivry}, {Ricci}, {Riffel}, {Riffel}, {Schartmann},
  {Schnorr-M{\"u}ller}, {Sternberg}, {Sturm}, {Tacconi}, \&
  {Veilleux}}]{Davies+15}
{Davies}, R.~I., {Burtscher}, L., {Rosario}, D., {et~al.} 2015, \apj, 806, 127,
  \dodoi{10.1088/0004-637X/806/1/127}

\bibitem[{{Di Matteo} {et~al.}(2005){Di Matteo}, {Springel}, \&
  {Hernquist}}]{Di-Matteo+05}
{Di Matteo}, T., {Springel}, V., \& {Hernquist}, L. 2005, \nat, 433, 604,
  \dodoi{10.1038/nature03335}

\bibitem[{{Doi} {et~al.}(2019){Doi}, {Nakagawa}, {Isobe}, {Baba}, {Yano}, \&
  {Yamagishi}}]{Doi+19}
{Doi}, R., {Nakagawa}, T., {Isobe}, N., {et~al.} 2019, \pasj, 71, 26,
  \dodoi{10.1093/pasj/psz019}

\bibitem[{{Duc} {et~al.}(1997){Duc}, {Mirabel}, \& {Maza}}]{Duc+97}
{Duc}, P.~A., {Mirabel}, I.~F., \& {Maza}, J. 1997, \aaps, 124, 533,
  \dodoi{10.1051/aas:1997205}

\bibitem[{{Fabian}(2012)}]{Fabian12}
{Fabian}, A.~C. 2012, \araa, 50, 455,
  \dodoi{10.1146/annurev-astro-081811-125521}

\bibitem[{{Falstad} {et~al.}(2019){Falstad}, {Hallqvist}, {Aalto}, {K{\"o}nig},
  {Muller}, {Aladro}, {Combes}, {Evans}, {Fuller}, {Gallagher},
  {Garc{\'\i}a-Burillo}, {Gonz{\'a}lez-Alfonso}, {Greve}, {Henkel}, {Imanishi},
  {Izumi}, {Mangum}, {Mart{\'\i}n}, {Privon}, {Sakamoto}, {Veilleux}, \& {van
  der Werf}}]{Falstad+19}
{Falstad}, N., {Hallqvist}, F., {Aalto}, S., {et~al.} 2019, \aap, 623, A29,
  \dodoi{10.1051/0004-6361/201834586}

\bibitem[{{Falstad} {et~al.}(2021){Falstad}, {Aalto}, {K{\"o}nig}, {Onishi},
  {Muller}, {Gorski}, {Sato}, {Stanley}, {Combes}, {Gonz{\'a}lez-Alfonso},
  {Mangum}, {Evans}, {Barcos-Mu{\~n}oz}, {Privon}, {Linden},
  {D{\'\i}az-Santos}, {Mart{\'\i}n}, {Sakamoto}, {Harada}, {Fuller},
  {Gallagher}, {van der Werf}, {Viti}, {Greve}, {Garc{\'\i}a-Burillo},
  {Henkel}, {Imanishi}, {Izumi}, {Nishimura}, {Ricci}, \&
  {M{\"u}hle}}]{Falstad+21}
{Falstad}, N., {Aalto}, S., {K{\"o}nig}, S., {et~al.} 2021, \aap, 649, A105,
  \dodoi{10.1051/0004-6361/202039291}

\bibitem[{{Franceschini} {et~al.}(2003){Franceschini}, {Braito}, {Persic},
  {Della Ceca}, {Bassani}, {Cappi}, {Malaguti}, {Palumbo}, {Risaliti},
  {Salvati}, \& {Severgnini}}]{Franceschini+03}
{Franceschini}, A., {Braito}, V., {Persic}, M., {et~al.} 2003, \mnras, 343,
  1181, \dodoi{10.1046/j.1365-8711.2003.06744.x}

\bibitem[{{Gao} {et~al.}(2020){Gao}, {Wang}, {Pearson}, {Gordon}, {Holwerda},
  {Hopkins}, {Brown}, {Bland -Hawthorn}, \& {Owers}}]{Gao+20}
{Gao}, F., {Wang}, L., {Pearson}, W.~J., {et~al.} 2020, \aap, 637, A94,
  \dodoi{10.1051/0004-6361/201937178}

\bibitem[{{Garc{\'\i}a-Burillo} {et~al.}(2015){Garc{\'\i}a-Burillo}, {Combes},
  {Usero}, {Aalto}, {Colina}, {Alonso-Herrero}, {Hunt}, {Arribas},
  {Costagliola}, {Labiano}, {Neri}, {Pereira-Santaella}, {Tacconi}, \& {van der
  Werf}}]{Garcia-Burillo+15}
{Garc{\'\i}a-Burillo}, S., {Combes}, F., {Usero}, A., {et~al.} 2015, \aap, 580,
  A35, \dodoi{10.1051/0004-6361/201526133}

\bibitem[{{Garc{\'\i}a-Burillo} {et~al.}(2019){Garc{\'\i}a-Burillo}, {Combes},
  {Ramos Almeida}, {Usero}, {Alonso-Herrero}, {Hunt}, {Rouan}, {Aalto},
  {Querejeta}, {Viti}, {van der Werf}, {Vives-Arias}, {Fuente}, {Colina},
  {Mart{\'\i}n-Pintado}, {Henkel}, {Mart{\'\i}n}, {Krips}, {Gratadour}, {Neri},
  \& {Tacconi}}]{Garcia-Burillo+19}
{Garc{\'\i}a-Burillo}, S., {Combes}, F., {Ramos Almeida}, C., {et~al.} 2019,
  \aap, 632, A61, \dodoi{10.1051/0004-6361/201936606}

\bibitem[{{Gonz{\'a}lez-Alfonso} \&
  {Sakamoto}(2019)}]{Gonzalez-Alfonso&Sakamoto19}
{Gonz{\'a}lez-Alfonso}, E., \& {Sakamoto}, K. 2019, \apj, 882, 153,
  \dodoi{10.3847/1538-4357/ab3a32}

\bibitem[{{Gonz{\'a}lez-Mart{\'\i}n} {et~al.}(2009){Gonz{\'a}lez-Mart{\'\i}n},
  {Masegosa}, {M{\'a}rquez}, \& {Guainazzi}}]{Gonzalez-Martin+09}
{Gonz{\'a}lez-Mart{\'\i}n}, O., {Masegosa}, J., {M{\'a}rquez}, I., \&
  {Guainazzi}, M. 2009, \apj, 704, 1570, \dodoi{10.1088/0004-637X/704/2/1570}

\bibitem[{{Gowardhan} {et~al.}(2018){Gowardhan}, {Spoon}, {Riechers},
  {Gonz{\'a}lez-Alfonso}, {Farrah}, {Fischer}, {Darling}, {Fergulio}, {Afonso},
  \& {Bizzocchi}}]{Gowardhan+18}
{Gowardhan}, A., {Spoon}, H., {Riechers}, D.~A., {et~al.} 2018, \apj, 859, 35,
  \dodoi{10.3847/1538-4357/aabccc}

\bibitem[{{Gravity Collaboration} {et~al.}(2020){Gravity Collaboration},
  {Pfuhl}, {Davies}, {Dexter}, {Netzer}, {H{\"o}nig}, {Lutz}, {Schartmann},
  {Sturm}, {Amorim}, {Brandner}, {Cl{\'e}net}, {de Zeeuw}, {Eckart},
  {Eisenhauer}, {F{\"o}rster Schreiber}, {Gao}, {Garcia}, {Genzel},
  {Gillessen}, {Gratadour}, {Kishimoto}, {Lacour}, {Millour}, {Ott}, {Paumard},
  {Perraut}, {Perrin}, {Peterson}, {Petrucci}, {Prieto}, {Rouan}, {Shangguan},
  {Shimizu}, {Sternberg}, {Straub}, {Straubmeier}, {Tacconi}, {Tristram},
  {Vermot}, {Waisberg}, {Widmann}, \& {Woillez}}]{GRAVITYcollab20}
{Gravity Collaboration}, {Pfuhl}, O., {Davies}, R., {et~al.} 2020, \aap, 634,
  A1, \dodoi{10.1051/0004-6361/201936255}

\bibitem[{{Haan} {et~al.}(2011){Haan}, {Surace}, {Armus}, {Evans}, {Howell},
  {Mazzarella}, {Kim}, {Vavilkin}, {Inami}, {Sanders}, {Petric}, {Bridge},
  {Melbourne}, {Charmandaris}, {Diaz-Santos}, {Murphy}, {U}, {Stierwalt}, \&
  {Marshall}}]{Haan+11}
{Haan}, S., {Surace}, J.~A., {Armus}, L., {et~al.} 2011, \aj, 141, 100,
  \dodoi{10.1088/0004-6256/141/3/100}

\bibitem[{{Harris} {et~al.}(2020){Harris}, {Millman}, {van der Walt},
  {Gommers}, {Virtanen}, {Cournapeau}, {Wieser}, {Taylor}, {Berg}, {Smith},
  {Kern}, {Picus}, {Hoyer}, {van Kerkwijk}, {Brett}, {Haldane}, {del R{\'\i}o},
  {Wiebe}, {Peterson}, {G{\'e}rard-Marchant}, {Sheppard}, {Reddy}, {Weckesser},
  {Abbasi}, {Gohlke}, \& {Oliphant}}]{NumPy}
{Harris}, C.~R., {Millman}, K.~J., {van der Walt}, S.~J., {et~al.} 2020, \nat,
  585, 357, \dodoi{10.1038/s41586-020-2649-2}

\bibitem[{{Hopkins} {et~al.}(2006){Hopkins}, {Hernquist}, {Cox}, {Di Matteo},
  {Robertson}, \& {Springel}}]{Hopkins+06}
{Hopkins}, P.~F., {Hernquist}, L., {Cox}, T.~J., {et~al.} 2006, \apjs, 163, 1,
  \dodoi{10.1086/499298}

\bibitem[{{Hopkins} {et~al.}(2008){Hopkins}, {Hernquist}, {Cox}, \&
  {Kere{\v{s}}}}]{Hopkins+08}
{Hopkins}, P.~F., {Hernquist}, L., {Cox}, T.~J., \& {Kere{\v{s}}}, D. 2008,
  \apjs, 175, 356, \dodoi{10.1086/524362}

\bibitem[{{Hunter}(2007)}]{Matplotlib}
{Hunter}, J.~D. 2007, Computing in Science and Engineering, 9, 90,
  \dodoi{10.1109/MCSE.2007.55}

\bibitem[{{Ichikawa} {et~al.}(2014){Ichikawa}, {Imanishi}, {Ueda}, {Nakagawa},
  {Shirahata}, {Kaneda}, \& {Oyabu}}]{Ichikawa+14}
{Ichikawa}, K., {Imanishi}, M., {Ueda}, Y., {et~al.} 2014, \apj, 794, 139,
  \dodoi{10.1088/0004-637X/794/2/139}

\bibitem[{{Ichikawa} {et~al.}(2019){Ichikawa}, {Ricci}, {Ueda}, {Bauer},
  {Kawamuro}, {Koss}, {Oh}, {Rosario}, {Shimizu}, {Stalevski}, {Fuller},
  {Packham}, \& {Trakhtenbrot}}]{Ichikawa+19}
{Ichikawa}, K., {Ricci}, C., {Ueda}, Y., {et~al.} 2019, \apj, 870, 31,
  \dodoi{10.3847/1538-4357/aaef8f}

\bibitem[{{Imanishi} {et~al.}(2006{\natexlab{a}}){Imanishi}, {Dudley}, \&
  {Maloney}}]{Imanishi+06_tau}
{Imanishi}, M., {Dudley}, C.~C., \& {Maloney}, P.~R. 2006{\natexlab{a}}, \apj,
  637, 114, \dodoi{10.1086/498391}

\bibitem[{{Imanishi} \& {Maloney}(2003)}]{Imanishi&Maloney03}
{Imanishi}, M., \& {Maloney}, P.~R. 2003, \apj, 588, 165,
  \dodoi{10.1086/368354}

\bibitem[{{Imanishi} {et~al.}(2008){Imanishi}, {Nakagawa}, {Ohyama},
  {Shirahata}, {Wada}, {Onaka}, \& {Oi}}]{Imanishi+08}
{Imanishi}, M., {Nakagawa}, T., {Ohyama}, Y., {et~al.} 2008, \pasj, 60, S489,
  \dodoi{10.1093/pasj/60.sp2.S489}

\bibitem[{{Imanishi} {et~al.}(2010){Imanishi}, {Nakagawa}, {Shirahata},
  {Ohyama}, \& {Onaka}}]{Imanishi+10}
{Imanishi}, M., {Nakagawa}, T., {Shirahata}, M., {Ohyama}, Y., \& {Onaka}, T.
  2010, \apj, 721, 1233, \dodoi{10.1088/0004-637X/721/2/1233}

\bibitem[{{Imanishi} \& {Nakanishi}(2013)}]{Imanishi&Nakanishi13}
{Imanishi}, M., \& {Nakanishi}, K. 2013, \aj, 146, 91,
  \dodoi{10.1088/0004-6256/146/4/91}

\bibitem[{{Imanishi} {et~al.}(2016{\natexlab{a}}){Imanishi}, {Nakanishi}, \&
  {Izumi}}]{Imanishi+16_IRAS20551}
{Imanishi}, M., {Nakanishi}, K., \& {Izumi}, T. 2016{\natexlab{a}}, \apj, 825,
  44, \dodoi{10.3847/0004-637X/825/1/44}

\bibitem[{{Imanishi} {et~al.}(2016{\natexlab{b}}){Imanishi}, {Nakanishi}, \&
  {Izumi}}]{Imanishi+16_14galaxies}
---. 2016{\natexlab{b}}, \aj, 152, 218, \dodoi{10.3847/0004-6256/152/6/218}

\bibitem[{{Imanishi} {et~al.}(2018){Imanishi}, {Nakanishi}, {Izumi}, \&
  {Wada}}]{Imanishi+18}
{Imanishi}, M., {Nakanishi}, K., {Izumi}, T., \& {Wada}, K. 2018, \apjl, 853,
  L25, \dodoi{10.3847/2041-8213/aaa8df}

\bibitem[{{Imanishi} {et~al.}(2006{\natexlab{b}}){Imanishi}, {Nakanishi}, \&
  {Kohno}}]{Imanishi+06_HCN}
{Imanishi}, M., {Nakanishi}, K., \& {Kohno}, K. 2006{\natexlab{b}}, \aj, 131,
  2888, \dodoi{10.1086/503527}

\bibitem[{{Imanishi} {et~al.}(2020){Imanishi}, {Nguyen}, {Wada}, {Hagiwara},
  {Iguchi}, {Izumi}, {Kawakatu}, {Nakanishi}, \& {Onishi}}]{Imanishi+20}
{Imanishi}, M., {Nguyen}, D.~D., {Wada}, K., {et~al.} 2020, \apj, 902, 99,
  \dodoi{10.3847/1538-4357/abaf50}

\bibitem[{{Inami} {et~al.}(2018){Inami}, {Armus}, {Matsuhara}, {Charmandaris},
  {D{\'\i}az-Santos}, {Surace}, {Stierwalt}, {Ohyama}, {Howell}, {Marshall},
  {Evans}, {Linden}, \& {Mazzarella}}]{Inami+18}
{Inami}, H., {Armus}, L., {Matsuhara}, H., {et~al.} 2018, \aap, 617, A130,
  \dodoi{10.1051/0004-6361/201833053}

\bibitem[{{Iwasawa} {et~al.}(2011){Iwasawa}, {Sanders}, {Teng}, {U}, {Armus},
  {Evans}, {Howell}, {Komossa}, {Mazzarella}, {Petric}, {Surace}, {Vavilkin},
  {Veilleux}, \& {Trentham}}]{Iwasawa+11}
{Iwasawa}, K., {Sanders}, D.~B., {Teng}, S.~H., {et~al.} 2011, \aap, 529, A106,
  \dodoi{10.1051/0004-6361/201015264}

\bibitem[{{Izumi} {et~al.}(2018){Izumi}, {Wada}, {Fukushige}, {Hamamura}, \&
  {Kohno}}]{Izumi+18}
{Izumi}, T., {Wada}, K., {Fukushige}, R., {Hamamura}, S., \& {Kohno}, K. 2018,
  \apj, 867, 48, \dodoi{10.3847/1538-4357/aae20b}

\bibitem[{{Izumi} {et~al.}(2016){Izumi}, {Kohno}, {Aalto}, {Espada}, {Fathi},
  {Harada}, {Hatsukade}, {Hsieh}, {Imanishi}, {Krips}, {Mart{\'\i}n},
  {Matsushita}, {Meier}, {Nakai}, {Nakanishi}, {Schinnerer}, {Sheth},
  {Terashima}, \& {Turner}}]{Izumi+16}
{Izumi}, T., {Kohno}, K., {Aalto}, S., {et~al.} 2016, \apj, 818, 42,
  \dodoi{10.3847/0004-637X/818/1/42}

\bibitem[{{Izumi} {et~al.}(2020){Izumi}, {Nguyen}, {Imanishi}, {Kawamuro},
  {Baba}, {Nakano}, {Kohno}, {Matsushita}, {Meier}, {Turner}, {Michiyama},
  {Harada}, {Mart{\'\i}n}, {Nakanishi}, {Takano}, {Wiklind}, {Nakai}, \&
  {Hsieh}}]{Izumi+20}
{Izumi}, T., {Nguyen}, D.~D., {Imanishi}, M., {et~al.} 2020, \apj, 898, 75,
  \dodoi{10.3847/1538-4357/ab9cb1}

\bibitem[{{Kamenetzky} {et~al.}(2014){Kamenetzky}, {Rangwala}, {Glenn},
  {Maloney}, \& {Conley}}]{Kamenetzky+14}
{Kamenetzky}, J., {Rangwala}, N., {Glenn}, J., {Maloney}, P.~R., \& {Conley},
  A. 2014, \apj, 795, 174, \dodoi{10.1088/0004-637X/795/2/174}

\bibitem[{{Kameno} {et~al.}(2020){Kameno}, {Sawada-Satoh}, {Impellizzeri},
  {Espada}, {Nakai}, {Sugai}, {Terashima}, {Kohno}, {Lee}, \&
  {Mart{\'\i}n}}]{Kameno+20}
{Kameno}, S., {Sawada-Satoh}, S., {Impellizzeri}, C.~M.~V., {et~al.} 2020,
  \apj, 895, 73, \dodoi{10.3847/1538-4357/ab8bd6}

\bibitem[{{Kawakatu} {et~al.}(2020){Kawakatu}, {Wada}, \&
  {Ichikawa}}]{Kawakatu+20}
{Kawakatu}, N., {Wada}, K., \& {Ichikawa}, K. 2020, \apj, 889, 84,
  \dodoi{10.3847/1538-4357/ab5f60}

\bibitem[{{Kim} {et~al.}(2015){Kim}, {Im}, {Kim}, {Jun}, {Woo}, {Lee}, {Lee},
  {Nakagawa}, {Matsuhara}, {Wada}, {Oyabu}, {Takagi}, {Ohyama}, \&
  {Lee}}]{Kim+15}
{Kim}, D., {Im}, M., {Kim}, J.~H., {et~al.} 2015, \apjs, 216, 17,
  \dodoi{10.1088/0067-0049/216/1/17}

\bibitem[{{Kluyver} {et~al.}(2016){Kluyver}, {Ragan-Kelley}, {P{\'e}rez},
  {Granger}, {Bussonnier}, {Frederic}, {Kelley}, {Hamrick}, {Grout}, {Corlay},
  {Inanov}, {Avila}, {Abdalla}, {Willing}, \& {Jupyter Development
  Team}}]{JupyterNotebook}
{Kluyver}, T., {Ragan-Kelley}, B., {P{\'e}rez}, F., {et~al.} 2016, in
  Positioning and Power in Academic Publishing: Players, Agents and Agendas,
  ed. L.~{Loizides} \& B.~{Schmidt} (IOP Press), 87--90,
  \dodoi{10.3233/978-1-61499-649-1-87}

\bibitem[{{Koss} {et~al.}(2018){Koss}, {Blecha}, {Bernhard}, {Hung}, {Lu},
  {Trakhtenbrot}, {Treister}, {Weigel}, {Sartori}, {Mushotzky}, {Schawinski},
  {Ricci}, {Veilleux}, \& {Sanders}}]{Koss+18}
{Koss}, M.~J., {Blecha}, L., {Bernhard}, P., {et~al.} 2018, \nat, 563, 214,
  \dodoi{10.1038/s41586-018-0652-7}

\bibitem[{{Lebouteiller} {et~al.}(2011){Lebouteiller}, {Barry}, {Spoon},
  {Bernard-Salas}, {Sloan}, {Houck}, \& {Weedman}}]{CASSIS}
{Lebouteiller}, V., {Barry}, D.~J., {Spoon}, H.~W.~W., {et~al.} 2011, \apjs,
  196, 8, \dodoi{10.1088/0067-0049/196/1/8}

\bibitem[{{Leja} {et~al.}(2018){Leja}, {Johnson}, {Conroy}, \& {van
  Dokkum}}]{Leja+18}
{Leja}, J., {Johnson}, B.~D., {Conroy}, C., \& {van Dokkum}, P. 2018, \apj,
  854, 62, \dodoi{10.3847/1538-4357/aaa8db}

\bibitem[{{Lutz} {et~al.}(2004){Lutz}, {Sturm}, {Genzel}, {Spoon}, \&
  {Stacey}}]{Lutz+04}
{Lutz}, D., {Sturm}, E., {Genzel}, R., {Spoon}, H.~W.~W., \& {Stacey}, G.~J.
  2004, \aap, 426, L5, \dodoi{10.1051/0004-6361:200400072}

\bibitem[{{Lutz} {et~al.}(1999){Lutz}, {Veilleux}, \& {Genzel}}]{Lutz+99}
{Lutz}, D., {Veilleux}, S., \& {Genzel}, R. 1999, \apjl, 517, L13,
  \dodoi{10.1086/312014}

\bibitem[{{Markwardt}(2009)}]{MPFIT}
{Markwardt}, C.~B. 2009, Astronomical Society of the Pacific Conference Series,
  Vol. 411, {Non-linear Least-squares Fitting in IDL with MPFIT}, ed. D.~A.
  {Bohlender}, D.~{Durand}, \& P.~{Dowler} ({Astronomical Society of the
  Pacific}), 251

\bibitem[{{Mart{\'\i}n} {et~al.}(2016){Mart{\'\i}n}, {Aalto}, {Sakamoto},
  {Gonz{\'a}lez-Alfonso}, {Muller}, {Henkel}, {Garc{\'\i}a-Burillo}, {Aladro},
  {Costagliola}, {Harada}, {Krips}, {Mart{\'\i}n-Pintado}, {M{\"u}hle}, {van
  der Werf}, \& {Viti}}]{Martin+16}
{Mart{\'\i}n}, S., {Aalto}, S., {Sakamoto}, K., {et~al.} 2016, \aap, 590, A25,
  \dodoi{10.1051/0004-6361/201528064}

\bibitem[{{McKinney}(2010)}]{Pandas}
{McKinney}, W. 2010, in Proc. 9th Python in Science Conf., ed. S.~{van der
  Walt} \& J.~{Millman}, Vol. 445, 51--56

\bibitem[{{McMullin} {et~al.}(2007){McMullin}, {Waters}, {Schiebel}, {Young},
  \& {Golap}}]{CASA}
{McMullin}, J.~P., {Waters}, B., {Schiebel}, D., {Young}, W., \& {Golap}, K.
  2007, in Astronomical Society of the Pacific Conference Series, Vol. 376,
  Astronomical Data Analysis Software and Systems XVI, ed. R.~A. {Shaw},
  F.~{Hill}, \& D.~J. {Bell}, 127

\bibitem[{{Medling} {et~al.}(2014){Medling}, {U}, {Guedes}, {Max}, {Mayer},
  {Armus}, {Holden}, {Ro{\v{s}}kar}, \& {Sanders}}]{Medling+14}
{Medling}, A.~M., {U}, V., {Guedes}, J., {et~al.} 2014, \apj, 784, 70,
  \dodoi{10.1088/0004-637X/784/1/70}

\bibitem[{{Medling} {et~al.}(2015){Medling}, {U}, {Rich}, {Kewley}, {Armus},
  {Dopita}, {Max}, {Sanders}, \& {Sutherland}}]{Medling+15}
{Medling}, A.~M., {U}, V., {Rich}, J.~A., {et~al.} 2015, \mnras, 448, 2301,
  \dodoi{10.1093/mnras/stv081}

\bibitem[{{Meijerink} \& {Spaans}(2005)}]{Meijerink&Spaans05}
{Meijerink}, R., \& {Spaans}, M. 2005, \aap, 436, 397,
  \dodoi{10.1051/0004-6361:20042398}

\bibitem[{{Meijerink} {et~al.}(2006){Meijerink}, {Spaans}, \&
  {Israel}}]{Meijerink+06}
{Meijerink}, R., {Spaans}, M., \& {Israel}, F.~P. 2006, \apjl, 650, L103,
  \dodoi{10.1086/508938}

\bibitem[{{Melnick} \& {Mirabel}(1990)}]{Melnick&Mirabel90}
{Melnick}, J., \& {Mirabel}, I.~F. 1990, \aap, 231, L19

\bibitem[{{Merloni} {et~al.}(2014){Merloni}, {Bongiorno}, {Brusa}, {Iwasawa},
  {Mainieri}, {Magnelli}, {Salvato}, {Berta}, {Cappelluti}, {Comastri},
  {Fiore}, {Gilli}, {Koekemoer}, {Le Floc'h}, {Lusso}, {Lutz}, {Miyaji},
  {Pozzi}, {Riguccini}, {Rosario}, {Silverman}, {Symeonidis}, {Treister},
  {Vignali}, \& {Zamorani}}]{Merloni+14}
{Merloni}, A., {Bongiorno}, A., {Brusa}, M., {et~al.} 2014, \mnras, 437, 3550,
  \dodoi{10.1093/mnras/stt2149}

\bibitem[{{Mihos} \& {Hernquist}(1996)}]{Mihos&Hernquist96}
{Mihos}, J.~C., \& {Hernquist}, L. 1996, \apj, 464, 641, \dodoi{10.1086/177353}

\bibitem[{{Momjian} {et~al.}(2003){Momjian}, {Romney}, {Carilli}, {Troland}, \&
  {Taylor}}]{Momjian+03}
{Momjian}, E., {Romney}, J.~D., {Carilli}, C.~L., {Troland}, T.~H., \&
  {Taylor}, G.~B. 2003, \apj, 587, 160, \dodoi{10.1086/367722}

\bibitem[{{Murphy} {et~al.}(1996){Murphy}, {Armus}, {Matthews}, {Soifer},
  {Mazzarella}, {Shupe}, {Strauss}, \& {Neugebauer}}]{Murphy+96}
{Murphy}, T.~W., J., {Armus}, L., {Matthews}, K., {et~al.} 1996, \aj, 111,
  1025, \dodoi{10.1086/117849}

\bibitem[{{Nardini} {et~al.}(2009){Nardini}, {Risaliti}, {Salvati}, {Sani},
  {Watabe}, {Marconi}, \& {Maiolino}}]{Nardini+09}
{Nardini}, E., {Risaliti}, G., {Salvati}, M., {et~al.} 2009, \mnras, 399, 1373,
  \dodoi{10.1111/j.1365-2966.2009.15357.x}

\bibitem[{{Nardini} {et~al.}(2010){Nardini}, {Risaliti}, {Watabe}, {Salvati},
  \& {Sani}}]{Nardini+10}
{Nardini}, E., {Risaliti}, G., {Watabe}, Y., {Salvati}, M., \& {Sani}, E. 2010,
  \mnras, 405, 2505, \dodoi{10.1111/j.1365-2966.2010.16618.x}

\bibitem[{{Ohyama} {et~al.}(2007){Ohyama}, {Onaka}, {Matsuhara}, {Wada}, {Kim},
  {Fujishiro}, {Uemizu}, {Sakon}, {Cohen}, {Ishigaki}, {Ishihara}, {Ita},
  {Kataza}, {Matsumoto}, {Murakami}, {Oyabu}, {Tanab{\'e}}, {Takagi}, {Ueno},
  {Usui}, {Watarai}, {Pearson}, {Takeyama}, {Yamamuro}, \& {Ikeda}}]{Ohyama+07}
{Ohyama}, Y., {Onaka}, T., {Matsuhara}, H., {et~al.} 2007, \pasj, 59, S411,
  \dodoi{10.1093/pasj/59.sp2.S411}

\bibitem[{{Onaka} {et~al.}(2010){Onaka}, {Matsuhara}, {Wada}, {Ishihara},
  {Ita}, {Ohyama}, {Ootsubo}, {Oyabu}, {Sakon}, {Shimonishi}, {Takita},
  {Tanab{\`e}}, {Usui}, \& {Murakami}}]{Onaka+10}
{Onaka}, T., {Matsuhara}, H., {Wada}, T., {et~al.} 2010, in Society of
  Photo-Optical Instrumentation Engineers (SPIE) Conference Series, Vol. 7731,
  Space Telescopes and Instrumentation 2010: Optical, Infrared, and Millimeter
  Wave, ed. J.~{Oschmann}, Jacobus~M., M.~C. {Clampin}, \& H.~A. {MacEwen},
  77310M, \dodoi{10.1117/12.856568}

\bibitem[{{Onishi} {et~al.}(2021){Onishi}, {Nakagawa}, {Baba}, {Matsumoto},
  {Isobe}, {Shirahata}, {Terada}, {Usuda}, \& {Oyabu}}]{Onishi+21}
{Onishi}, S., {Nakagawa}, T., {Baba}, S., {et~al.} 2021, \apj, 921, 141,
  \dodoi{10.3847/1538-4357/ac1c6d}

\bibitem[{{Pearson} {et~al.}(2016){Pearson}, {Rigopoulou}, {Hurley}, {Farrah},
  {Afonso}, {Bernard-Salas}, {Borys}, {Clements}, {Cormier}, {Efstathiou},
  {Gonzalez-Alfonso}, {Lebouteiller}, \& {Spoon}}]{Pearson+16}
{Pearson}, C., {Rigopoulou}, D., {Hurley}, P., {et~al.} 2016, \apjs, 227, 9,
  \dodoi{10.3847/0067-0049/227/1/9}

\bibitem[{{Perez} \& {Granger}(2007)}]{IPython}
{Perez}, F., \& {Granger}, B.~E. 2007, Computing in Science and Engineering, 9,
  21, \dodoi{10.1109/MCSE.2007.53}

\bibitem[{{Rangwala} {et~al.}(2015){Rangwala}, {Maloney}, {Wilson}, {Glenn},
  {Kamenetzky}, \& {Spinoglio}}]{Rangwala+15}
{Rangwala}, N., {Maloney}, P.~R., {Wilson}, C.~D., {et~al.} 2015, \apj, 806,
  17, \dodoi{10.1088/0004-637X/806/1/17}

\bibitem[{{Ricci} {et~al.}(2017){Ricci}, {Bauer}, {Treister}, {Schawinski},
  {Privon}, {Blecha}, {Arevalo}, {Armus}, {Harrison}, {Ho}, {Iwasawa},
  {Sanders}, \& {Stern}}]{Ricci+17_fraction}
{Ricci}, C., {Bauer}, F.~E., {Treister}, E., {et~al.} 2017, \mnras, 468, 1273,
  \dodoi{10.1093/mnras/stx173}

\bibitem[{{Risaliti} {et~al.}(2010){Risaliti}, {Imanishi}, \&
  {Sani}}]{Risaliti+10}
{Risaliti}, G., {Imanishi}, M., \& {Sani}, E. 2010, \mnras, 401, 197,
  \dodoi{10.1111/j.1365-2966.2009.15622.x}

\bibitem[{{Risaliti} {et~al.}(2006){Risaliti}, {Maiolino}, {Marconi}, {Sani},
  {Berta}, {Braito}, {Della Ceca}, {Franceschini}, \& {Salvati}}]{Risaliti+06}
{Risaliti}, G., {Maiolino}, R., {Marconi}, A., {et~al.} 2006, \mnras, 365, 303,
  \dodoi{10.1111/j.1365-2966.2005.09715.x}

\bibitem[{{Robitaille}(2019)}]{APLpy_v2}
{Robitaille}, T. 2019, {APLpy v2.0: The Astronomical Plotting Library in
  Python}, 2.0,  Zenodo, \dodoi{10.5281/zenodo.2567476}

\bibitem[{{Robitaille} \& {Bressert}(2012)}]{APLpy}
{Robitaille}, T., \& {Bressert}, E. 2012, {APLpy: Astronomical Plotting Library
  in Python}.
\newblock \doeprint{1208.017}

\bibitem[{{Rose} {et~al.}(2020){Rose}, {Edge}, {Combes}, {Hamer}, {McNamara},
  {Russell}, {Gaspari}, {Salom{\'e}}, {Sarazin}, {Tremblay}, {Baum}, {Bremer},
  {Donahue}, {Fabian}, {Ferland}, {Nesvadba}, {O'Dea}, {Oonk}, \&
  {Peck}}]{Rose+20}
{Rose}, T., {Edge}, A.~C., {Combes}, F., {et~al.} 2020, \mnras, 496, 364,
  \dodoi{10.1093/mnras/staa1474}

\bibitem[{{Ruffa} {et~al.}(2018){Ruffa}, {Vignali}, {Mignano}, {Paladino}, \&
  {Iwasawa}}]{Ruffa+18}
{Ruffa}, I., {Vignali}, C., {Mignano}, A., {Paladino}, R., \& {Iwasawa}, K.
  2018, \aap, 616, A127, \dodoi{10.1051/0004-6361/201732268}

\bibitem[{{Rupke} \& {Veilleux}(2013)}]{Rupke&Veilleux13}
{Rupke}, D. S.~N., \& {Veilleux}, S. 2013, \apj, 768, 75,
  \dodoi{10.1088/0004-637X/768/1/75}

\bibitem[{{Sakamoto} {et~al.}(2010){Sakamoto}, {Aalto}, {Evans}, {Wiedner}, \&
  {Wilner}}]{Sakamoto+10}
{Sakamoto}, K., {Aalto}, S., {Evans}, A.~S., {Wiedner}, M.~C., \& {Wilner},
  D.~J. 2010, \apjl, 725, L228, \dodoi{10.1088/2041-8205/725/2/L228}

\bibitem[{{Sakamoto} {et~al.}(2021){Sakamoto}, {Mart{\'\i}n}, {Wilner},
  {Aalto}, {Evans}, \& {Harada}}]{Sakamoto+21}
{Sakamoto}, K., {Mart{\'\i}n}, S., {Wilner}, D.~J., {et~al.} 2021, \apj, 923,
  240, \dodoi{10.3847/1538-4357/ac29bf}

\bibitem[{{Sakamoto} {et~al.}(2009){Sakamoto}, {Aalto}, {Wilner}, {Black},
  {Conway}, {Costagliola}, {Peck}, {Spaans}, {Wang}, \&
  {Wiedner}}]{Sakamoto+09}
{Sakamoto}, K., {Aalto}, S., {Wilner}, D.~J., {et~al.} 2009, \apjl, 700, L104,
  \dodoi{10.1088/0004-637X/700/2/L104}

\bibitem[{{Sanders} {et~al.}(2003){Sanders}, {Mazzarella}, {Kim}, {Surace}, \&
  {Soifer}}]{Sanders+03}
{Sanders}, D.~B., {Mazzarella}, J.~M., {Kim}, D.~C., {Surace}, J.~A., \&
  {Soifer}, B.~T. 2003, \aj, 126, 1607, \dodoi{10.1086/376841}

\bibitem[{{Sanders} \& {Mirabel}(1996)}]{Sanders&Mirabel96}
{Sanders}, D.~B., \& {Mirabel}, I.~F. 1996, \araa, 34, 749,
  \dodoi{10.1146/annurev.astro.34.1.749}

\bibitem[{{Saunders} {et~al.}(2000){Saunders}, {Sutherland}, {Maddox},
  {Keeble}, {Oliver}, {Rowan-Robinson}, {McMahon}, {Efstathiou}, {Tadros},
  {White}, {Frenk}, {Carrami{\~n}ana}, \& {Hawkins}}]{Saunders+00}
{Saunders}, W., {Sutherland}, W.~J., {Maddox}, S.~J., {et~al.} 2000, \mnras,
  317, 55, \dodoi{10.1046/j.1365-8711.2000.03528.x}

\bibitem[{{Sch{\"o}ier} {et~al.}(2005){Sch{\"o}ier}, {van der Tak}, {van
  Dishoeck}, \& {Black}}]{LAMDA}
{Sch{\"o}ier}, F.~L., {van der Tak}, F.~F.~S., {van Dishoeck}, E.~F., \&
  {Black}, J.~H. 2005, \aap, 432, 369, \dodoi{10.1051/0004-6361:20041729}

\bibitem[{{Scoville} {et~al.}(2017){Scoville}, {Murchikova}, {Walter},
  {Vlahakis}, {Koda}, {Vanden Bout}, {Barnes}, {Hernquist}, {Sheth}, {Yun},
  {Sanders}, {Armus}, {Cox}, {Thompson}, {Robertson}, {Zschaechner}, {Tacconi},
  {Torrey}, {Hayward}, {Genzel}, {Hopkins}, {van der Werf}, \&
  {Decarli}}]{Scoville+17}
{Scoville}, N., {Murchikova}, L., {Walter}, F., {et~al.} 2017, \apj, 836, 66,
  \dodoi{10.3847/1538-4357/836/1/66}

\bibitem[{{Scoville} {et~al.}(2000){Scoville}, {Evans}, {Thompson}, {Rieke},
  {Hines}, {Low}, {Dinshaw}, {Surace}, \& {Armus}}]{Scoville+00}
{Scoville}, N.~Z., {Evans}, A.~S., {Thompson}, R., {et~al.} 2000, \aj, 119,
  991, \dodoi{10.1086/301248}

\bibitem[{{Secrest} {et~al.}(2020){Secrest}, {Ellison}, {Satyapal}, \&
  {Blecha}}]{Secrest+20}
{Secrest}, N.~J., {Ellison}, S.~L., {Satyapal}, S., \& {Blecha}, L. 2020,
  \mnras, 499, 2380, \dodoi{10.1093/mnras/staa1692}

\bibitem[{{Shirahata} {et~al.}(2013){Shirahata}, {Nakagawa}, {Usuda}, {Goto},
  {Suto}, \& {Geballe}}]{Shirahata+13}
{Shirahata}, M., {Nakagawa}, T., {Usuda}, T., {et~al.} 2013, \pasj, 65, 5,
  \dodoi{10.1093/pasj/65.1.5}

\bibitem[{{Smith} {et~al.}(2007){Smith}, {Draine}, {Dale}, {Moustakas},
  {Kennicutt}, {Helou}, {Armus}, {Roussel}, {Sheth}, {Bendo}, {Buckalew},
  {Calzetti}, {Engelbracht}, {Gordon}, {Hollenbach}, {Li}, {Malhotra},
  {Murphy}, \& {Walter}}]{Smith+07}
{Smith}, J.~D.~T., {Draine}, B.~T., {Dale}, D.~A., {et~al.} 2007, \apj, 656,
  770, \dodoi{10.1086/510549}

\bibitem[{{Spoon} {et~al.}(2005){Spoon}, {Keane}, {Cami}, {Lahuis}, {Tielens},
  {Armus}, \& {Charmandaris}}]{Spoon+05}
{Spoon}, H.~W.~W., {Keane}, J.~V., {Cami}, J., {et~al.} 2005, in IAU Symposium,
  Vol. 231, Astrochemistry: Recent Successes and Current Challenges, ed. D.~C.
  {Lis}, G.~A. {Blake}, \& E.~{Herbst}, 281--290,
  \dodoi{10.1017/S1743921306007277}

\bibitem[{{Spoon} {et~al.}(2004){Spoon}, {Armus}, {Cami}, {Tielens}, {Chiar},
  {Peeters}, {Keane}, {Charmandaris}, {Appleton}, {Teplitz}, \&
  {Burgdorf}}]{Spoon+04}
{Spoon}, H.~W.~W., {Armus}, L., {Cami}, J., {et~al.} 2004, \apjs, 154, 184,
  \dodoi{10.1086/422813}

\bibitem[{{Stierwalt} {et~al.}(2013){Stierwalt}, {Armus}, {Surace}, {Inami},
  {Petric}, {Diaz-Santos}, {Haan}, {Charmand aris}, {Howell}, {Kim},
  {Marshall}, {Mazzarella}, {Spoon}, {Veilleux}, {Evans}, {Sanders},
  {Appleton}, {Bothun}, {Bridge}, {Chan}, {Frayer}, {Iwasawa}, {Kewley},
  {Lord}, {Madore}, {Melbourne}, {Murphy}, {Rich}, {Schulz}, {Sturm},
  {Vavilkin}, \& {Xu}}]{Stierwalt+13}
{Stierwalt}, S., {Armus}, L., {Surace}, J.~A., {et~al.} 2013, \apjs, 206, 1,
  \dodoi{10.1088/0067-0049/206/1/1}

\bibitem[{{Storey} \& {Hummer}(1995)}]{Strey&Hummer95}
{Storey}, P.~J., \& {Hummer}, D.~G. 1995, \mnras, 272, 41,
  \dodoi{10.1093/mnras/272.1.41}

\bibitem[{{Tanimoto} {et~al.}(2020){Tanimoto}, {Ueda}, {Odaka}, {Ogawa},
  {Yamada}, {Kawaguchi}, \& {Ichikawa}}]{Tanimoto+20}
{Tanimoto}, A., {Ueda}, Y., {Odaka}, H., {et~al.} 2020, \apj, 897, 2,
  \dodoi{10.3847/1538-4357/ab96bc}

\bibitem[{{U} {et~al.}(2019){U}, {Medling}, {Inami}, {Armus},
  {D{\'\i}az-Santos}, {Charmandaris}, {Howell}, {Stierwalt}, {Privon},
  {Linden}, {Sanders}, {Max}, {Evans}, {Barcos-Mu{\~n}oz}, {Chiang},
  {Appleton}, {Canalizo}, {Fazio}, {Iwasawa}, {Larson}, {Mazzarella}, {Murphy},
  {Rich}, \& {Surace}}]{U+19}
{U}, V., {Medling}, A.~M., {Inami}, H., {et~al.} 2019, \apj, 871, 166,
  \dodoi{10.3847/1538-4357/aaf1c2}

\bibitem[{{van der Tak} {et~al.}(2007){van der Tak}, {Black}, {Sch{\"o}ier},
  {Jansen}, \& {van Dishoeck}}]{RADEX}
{van der Tak}, F.~F.~S., {Black}, J.~H., {Sch{\"o}ier}, F.~L., {Jansen}, D.~J.,
  \& {van Dishoeck}, E.~F. 2007, \aap, 468, 627,
  \dodoi{10.1051/0004-6361:20066820}

\bibitem[{{Veilleux} {et~al.}(2002){Veilleux}, {Kim}, \&
  {Sanders}}]{Veilleux+02}
{Veilleux}, S., {Kim}, D.~C., \& {Sanders}, D.~B. 2002, \apjs, 143, 315,
  \dodoi{10.1086/343844}

\bibitem[{{Veilleux} {et~al.}(1995){Veilleux}, {Kim}, {Sanders}, {Mazzarella},
  \& {Soifer}}]{Veilleux+95}
{Veilleux}, S., {Kim}, D.~C., {Sanders}, D.~B., {Mazzarella}, J.~M., \&
  {Soifer}, B.~T. 1995, \apjs, 98, 171, \dodoi{10.1086/192158}

\bibitem[{{Veilleux} {et~al.}(2006){Veilleux}, {Kim}, {Peng}, {Ho}, {Tacconi},
  {Dasyra}, {Genzel}, {Lutz}, \& {Sanders}}]{Veilleux+06}
{Veilleux}, S., {Kim}, D.~C., {Peng}, C.~Y., {et~al.} 2006, \apj, 643, 707,
  \dodoi{10.1086/503188}

\bibitem[{{Veilleux} {et~al.}(2009){Veilleux}, {Rupke}, {Kim}, {Genzel},
  {Sturm}, {Lutz}, {Contursi}, {Schweitzer}, {Tacconi}, {Netzer}, {Sternberg},
  {Mihos}, {Baker}, {Mazzarella}, {Lord}, {Sanders}, {Stockton}, {Joseph}, \&
  {Barnes}}]{Veilleux+09}
{Veilleux}, S., {Rupke}, D.~S.~N., {Kim}, D.~C., {et~al.} 2009, \apjs, 182,
  628, \dodoi{10.1088/0067-0049/182/2/628}

\bibitem[{{Veilleux} {et~al.}(2013){Veilleux}, {Mel{\'e}ndez}, {Sturm},
  {Gracia-Carpio}, {Fischer}, {Gonz{\'a}lez-Alfonso}, {Contursi}, {Lutz},
  {Poglitsch}, {Davies}, {Genzel}, {Tacconi}, {de Jong}, {Sternberg}, {Netzer},
  {Hailey-Dunsheath}, {Verma}, {Rupke}, {Maiolino}, {Teng}, \&
  {Polisensky}}]{Veilleux+13}
{Veilleux}, S., {Mel{\'e}ndez}, M., {Sturm}, E., {et~al.} 2013, \apj, 776, 27,
  \dodoi{10.1088/0004-637X/776/1/27}

\bibitem[{{Virtanen} {et~al.}(2020){Virtanen}, {Gommers}, {Oliphant},
  {Haberland}, {Reddy}, {Cournapeau}, {Burovski}, {Peterson}, {Weckesser},
  {Bright}, {van der Walt}, {Brett}, {Wilson}, {Millman}, {Mayorov}, {Nelson},
  {Jones}, {Kern}, {Larson}, {Carey}, {Polat}, {Feng}, {Moore}, {Vand erPlas},
  {Laxalde}, {Perktold}, {Cimrman}, {Henriksen}, {Quintero}, {Harris},
  {Archibald}, {Ribeiro}, {Pedregosa}, {van Mulbregt}, \& {SciPy 1. 0
  Contributors}}]{SciPy}
{Virtanen}, P., {Gommers}, R., {Oliphant}, T.~E., {et~al.} 2020, Nature
  Methods, 17, 261, \dodoi{10.1038/s41592-019-0686-2}

\bibitem[{{Wada} {et~al.}(2018){Wada}, {Fukushige}, {Izumi}, \&
  {Tomisaka}}]{Wada+18}
{Wada}, K., {Fukushige}, R., {Izumi}, T., \& {Tomisaka}, K. 2018, \apj, 852,
  88, \dodoi{10.3847/1538-4357/aa9e53}

\bibitem[{{Wada} {et~al.}(2016){Wada}, {Schartmann}, \& {Meijerink}}]{Wada+16}
{Wada}, K., {Schartmann}, M., \& {Meijerink}, R. 2016, \apjl, 828, L19,
  \dodoi{10.3847/2041-8205/828/2/L19}

\bibitem[{{Wheeler} {et~al.}(2020){Wheeler}, {Glenn}, {Rangwala}, \&
  {Fyhrie}}]{Wheeler+20}
{Wheeler}, J., {Glenn}, J., {Rangwala}, N., \& {Fyhrie}, A. 2020, \apj, 896,
  43, \dodoi{10.3847/1538-4357/ab8f32}

\bibitem[{{Winter} {et~al.}(2012){Winter}, {Veilleux}, {McKernan}, \&
  {Kallman}}]{Winter+12}
{Winter}, L.~M., {Veilleux}, S., {McKernan}, B., \& {Kallman}, T.~R. 2012,
  \apj, 745, 107, \dodoi{10.1088/0004-637X/745/2/107}

\bibitem[{{Wright} {et~al.}(2010){Wright}, {Eisenhardt}, {Mainzer}, {Ressler},
  {Cutri}, {Jarrett}, {Kirkpatrick}, {Padgett}, {McMillan}, {Skrutskie},
  {Stanford}, {Cohen}, {Walker}, {Mather}, {Leisawitz}, {Gautier}, {McLean},
  {Benford}, {Lonsdale}, {Blain}, {Mendez}, {Irace}, {Duval}, {Liu}, {Royer},
  {Heinrichsen}, {Howard}, {Shannon}, {Kendall}, {Walsh}, {Larsen}, {Cardon},
  {Schick}, {Schwalm}, {Abid}, {Fabinsky}, {Naes}, \& {Tsai}}]{Wright+10}
{Wright}, E.~L., {Eisenhardt}, P. R.~M., {Mainzer}, A.~K., {et~al.} 2010, \aj,
  140, 1868, \dodoi{10.1088/0004-6256/140/6/1868}

\bibitem[{{Yamada} {et~al.}(2013){Yamada}, {Oyabu}, {Kaneda}, {Yamagishi},
  {Ishihara}, {Kim}, \& {Im}}]{Yamada+13}
{Yamada}, R., {Oyabu}, S., {Kaneda}, H., {et~al.} 2013, \pasj, 65, 103,
  \dodoi{10.1093/pasj/65.5.103}

\bibitem[{{Yamashita} {et~al.}(2017){Yamashita}, {Komugi}, {Matsuhara},
  {Armus}, {Inami}, {Ueda}, {Iono}, {Kohno}, {Evans}, \&
  {Arimatsu}}]{Yamashita+17}
{Yamashita}, T., {Komugi}, S., {Matsuhara}, H., {et~al.} 2017, \apj, 844, 96,
  \dodoi{10.3847/1538-4357/aa7af1}

\bibitem[{{Yang} {et~al.}(2010){Yang}, {Stancil}, {Balakrishnan}, \&
  {Forrey}}]{Yang+10}
{Yang}, B., {Stancil}, P.~C., {Balakrishnan}, N., \& {Forrey}, R.~C. 2010,
  \apj, 718, 1062, \dodoi{10.1088/0004-637X/718/2/1062}

\bibitem[{{Yuan} {et~al.}(2010){Yuan}, {Kewley}, \& {Sanders}}]{Yuan+10}
{Yuan}, T.~T., {Kewley}, L.~J., \& {Sanders}, D.~B. 2010, \apj, 709, 884,
  \dodoi{10.1088/0004-637X/709/2/884}

\end{thebibliography}
\bibliographystyle{aasjournal}



\end{document}